\documentclass[longauth]{aa} 

\usepackage{xcolor}  
\usepackage{natbib}  
\bibpunct{(}{)}{;}{a}{}{,} 

\usepackage[colorlinks = true,
            linkcolor = blue,
            urlcolor  = blue,
            citecolor = blue]{hyperref} 

\usepackage{graphicx}
\usepackage{txfonts}
\usepackage{tikz}
\usepackage{graphicx}
\usepackage{txfonts}
\usepackage{lipsum}
\usepackage{subcaption}         
\usepackage{lscape}             
\usepackage{placeins}           
                                
\usepackage[
    starfontserif 
    ]{starfont}
\usepackage{amsmath}

\begin{document}

   \title{Two neighbours of the ultra-short-period Earth-sized planet K2-157~b in the warm Neptunian savanna\thanks{Based on observations collected at the European Southern Observatory (ESO) under the programmes with IDs 1102.C-0744, 1102.C-0958, 1104.C-0350, and 106.21M2.004.}}

   \author{A. Castro-González\inst{ \ref{CAB_villafranca}}
        \and F.~Bouchy\inst{ \ref{obs_geneva}, \ref{geneva_departement}}
        \and  A.~C.~M.~Correia\inst{\ref{cfisuc-coimbra},\ref{obs_paris}}
        \and
        A.~Sozzetti\inst{\ref{INAF_torin}}
        \and J.~Lillo-Box\inst{\ref{CAB_villafranca}} \and P.~Figueira\inst{\ref{obs_geneva}, \ref{CAUP}} 
        \and
        B.~Lavie\inst{\ref{obs_geneva}}
        \and
        C.~Lovis\inst{\ref{obs_geneva}}
        \and
        M.~J.~Hobson\inst{\ref{obs_geneva}}
        \and
        S.~G.~Sousa\inst{\ref{CAUP}, \ref{dep_astro_porto}}
        \and
        V.~Adibekyan\inst{\ref{CAUP}, \ref{dep_astro_porto}}
        \and
        M.~R.~Standing\inst{\ref{ESAC}}
        \and
        N.~C.~Hara\inst{\ref{geneva_departement},\ref{LAM}}
        \and
        D.~Barrado\inst{\ref{CAB_villafranca}}
        \and
        A.~M.~Silva\inst{\ref{CAUP},\ref{dep_astro_porto},\ref{lisboa}}
        \and 
        V.~Bourrier\inst{\ref{obs_geneva}}
        \and
        J.~Korth\inst{\ref{lund}}
        \and 
        N.~C.~Santos\inst{\ref{CAUP},\ref{dep_astro_porto}}
        \and
        M.~Damasso\inst{\ref{INAF_torin}}
        \and
        M.~R.~Zapatero~Osorio\inst{\ref{CAB_villafranca}}
        \and \\
        J.~Rodrigues\inst{\ref{CAUP},\ref{dep_astro_porto},\ref{OFXB}}
        \and 
        Y.~Alibert\inst{\ref{space_hab_switz},\ref{physikalisches_inst}}
        \and
        S.~C.~C. Barros\inst{\ref{CAUP}, \ref{dep_astro_porto}}
        \and
        S.~Cristiani\inst{\ref{INAF_trieste}}
        \and P.~Di~Marcantonio\inst{\ref{INAF_trieste}}
        \and  \\
        J.~I.~Gonz\'alez Hern\'andez\inst{\ref{IAC},\ref{uni_laguna}}
        \and 
        G.~Lo Curto\inst{\ref{ESO_chile}}
        \and
        C.~J.~A.~P.~Martins\inst{\ref{CAUP},\ref{centro_CAUP}}
        \and 
        N.~J.~Nunes\inst{\ref{lisboa}}
        \and \\
        E.~Palle\inst{\ref{IAC},\ref{uni_laguna}}
        \and 
        F.~Pepe\inst{\ref{uni_geneva}}
        \and 
        A.~Su\'{a}rez~Mascare\~{n}o\inst{\ref{IAC},\ref{uni_laguna}}
        \and
        H.~M.~Tabernero\inst{\ref{ice_csic},\ref{ieec}}
        }

   \institute{Centro de Astrobiolog\'{i}a, CSIC-INTA, Camino Bajo del Castillo s/n, 28692 Villanueva de la Ca\~{n}ada, Madrid, Spain\label{CAB_villafranca} \\
    \email{acastro@cab.inta-csic.es}
    \and
    Observatoire de l'Universit\'e de Gen\`eve, 51 chemin Pegasi, 1290 Sauverny, Switzerland\label{obs_geneva}
    \and
    Département d’Astronomie, Université de Genève, Chemin Pegasi 51, 1290 Versoix, Switzerland\label{geneva_departement} 
    \and
    CFisUC, Departamento de F\'isica, Universidade de Coimbra, 3004-516 Coimbra, Portugal\label{cfisuc-coimbra}
    \and
    IMCCE, UMR8028 CNRS, Observatoire de Paris, PSL Universit\'{e}, 77 Av. Denfert-Rochereau, 75014 Paris, France\label{obs_paris}
    \and
    INAF - Osservatorio Astrofisico di Torin, Via Osservatorio 20, I-10025 Pino Torinese, Italy\label{INAF_torin}
    \and
    Instituto de Astrof\'isica e Ci\^encias do Espa\c{c}o, Universidade do Porto, CAUP, Rua das Estrelas, 4150-762 Porto, Portugal\label{CAUP}
    \and
    Departamento de Fisica e Astronomia, Universidade do Porto, Rua do Campo Alegre, 4169-007 Porto, Portugal\label{dep_astro_porto}
    \and
    European Space Agency (ESA), European Space Astronomy Centre (ESAC), Camino Bajo del Castillo s/n, 28692 Villanueva de la Cañada, Madrid, Spain\label{ESAC}
    \and
    Université Aix Marseille, CNRS, CNES, LAM, Marseille, France\label{LAM}
    \and
    Instituto de Astrofísica e Ciências do Espaço, Universidade de Lisboa, Campo Grande, 1749-016 Lisboa\label{lisboa}
    \and
    Lund Observatory, Division of Astrophysics, Department of Physics, Lund University, Box 118, 22100 Lund, Sweden\label{lund}
    \and
     Observatoire François-Xavier Bagnoud -- OFXB, 3961 Saint-Luc, Switzerland\label{OFXB}
     \and
    Center for Space and Habitability, University of Bern, Gesellschaftsstrasse 6, CH-3012 Bern, Switzerland \label{space_hab_switz}
    \and
    Physics Institute of University of Bern, Gesellschaftsstrasse 6, CH-3012 Bern, Switzerland\label{physikalisches_inst}
    \and
     INAF – Osservatorio Astronomico di Trieste, via G. B. Tiepolo 11, I-34143, Trieste, Italy\label{INAF_trieste}
    \and
    Instituto de Astrof{\'\i}sica de Canarias, E-38205 La Laguna, Tenerife, Spain\label{IAC}
    \and
    Universidad de La Laguna, Dept. Astrof{\'\i}sica, E-38206 La Laguna, Tenerife, Spain\label{uni_laguna}
    \and
    European Southern Observatory, Av. Alonso de Cordova 3107, Casilla 19001, Santiago de Chile, Chile\label{ESO_chile}
    \and
     Centro de Astrof\'{\i}sica da Universidade do Porto, Rua das Estrelas, 4150-762 Porto, Portugal\label{centro_CAUP}
     \and
     Département d’Astronomie, Université de Genève, Ch. des Maillettes 51, 1290 Versoix, Switzerland\label{uni_geneva}
     \and
     Institut de Ciències de l’Espai (ICE, CSIC), Campus UAB, c/ de Can Magrans s/n, 08193 Cerdanyola del Vallès, Barcelona, Spain\label{ice_csic}
     \and
     Institut d'Estudis Espacials de Catalunya (IEEC), Edifici RDIT, Campus UPC, 08860 Castelldefels (Barcelona), Spain\label{ieec}
    \\  }

   \date{Received 25 March 2025 / Accepted 25 April 2025}

 
  \abstract
   {The formation and evolution of ultra-short-period (USP) rocky planets is poorly understood. However, it is widely thought that these planets could not have formed at their present-day close-in orbits, but instead migrated inwards through interactions with outer neighbours.}
   {We aim to confirm and characterise the USP Earth-sized validated planet K2-157~b ($P_{\rm orb}$ = 8.8 h) and constrain the presence of additional companions in the system through radial velocity (RV) measurements.}
   {We measured 49 RVs with the ESPRESSO spectrograph and tested different planetary and non-planetary configurations to infer the model that best represents our data set. We derived the orbital and physical properties of the system through a global RV and transit model.}
  {We detected two additional super-Neptune-mass planets located within the warm Neptunian savanna, K2-157~c ($P_{\rm orb, c}$ = $25.942^{+0.045}_{-0.044}$~d,  $M_{\rm p, c} \, \textrm{sin} \, i$ = $30.8 \pm 1.9$ $\rm M_{\oplus}$) and K2-157~d ($P_{\rm orb, d}$ = $66.50^{+0.71}_{-0.59}$ d,  $M_{\rm p, d} \, \textrm{sin} \, i$ = $23.3 \pm 2.5$ $\rm M_{\oplus}$). The joint analysis constrains the mass of K2-157~b at the 2.7$\sigma$ level, $M_{\rm p, b}$ = $1.14^{+0.41}_{-0.42}$ $\rm M_{\oplus}$ ($<$ 2.4 $\rm M_{\oplus}$ at 3$\sigma$),  which, together with the inferred radius, $R_{\rm p, b}$ = 0.935 $\pm$ 0.090~$\rm R_{\oplus}$, make the planet compatible with a rocky composition with a likely ($68\%$ confidence) higher iron-to-silicate mass fraction than Earth.  K2 data discard non-grazing transit configurations for K2-157~c ($i_{\rm c}$ $<$ 88.4$^{\circ}$ at 3$\sigma$), and ESPRESSO data constrain the eccentricities of K2-157 c and K2-157 d to $e_{\rm c}$ $<$ 0.2 and $e_{\rm d}$ $<$ 0.5 at 3$\sigma$. Our dynamical analysis indicates that the system is stable for eccentricities up to $e_{\rm c}$, $e_{\rm d}$ $\sim$ 0.3 and mutual inclinations up to $\sim$~60$^{\circ}$. At a population level, we find that the trend that the closest USP planets tend to orbit late-type stars does not hold when scaling the orbital separation to the Roche limit, which suggests that the orbital distribution of the closest planets across spectral types is primarily determined by tidal disruption. }
  {The orbital architecture of K2-157 is unusual in the known exoplanet plethora, with only one similar case reported to date: 55 Cnc. The USP planets of these systems, being accompanied by massive, long-period, relatively spaced, and possibly misaligned neighbours, could have migrated inwards through eccentricity-based mechanisms triggered by secular interactions.}

   \maketitle

\section{Introduction}

Ultra-short-period (USP) planets are defined as planets that orbit their host stars in less than one day. Appearing in 0.51 ± 0.07$\%$ of G-dwarf stars and 0.83 ± 0.18$\%$ of K-dwarf stars \citep{2014ApJ...787...47S}, these planets are an unusual outcome of planet formation and evolution. Interestingly, while USP planets have been detected in a wide range of radii and masses, from sub-Earths to super-Jupiters, they generally show terrestrial sizes (i.e. $R_{\rm p}$ $<$ 2 $\rm R_{\oplus}$). Hence, their extremely short orbits offer unique observational advantages that facilitate the characterisation of small planets with Earth and sub-Earth masses. Being subjected to extreme irradiation and gravitational conditions, USPs also allow us to explore physical and chemical phenomena for which there are no analogues in the Solar System. 

An Earth-mass USP planet with a half-day orbit around a Sun-like star induces a radial velocity (RV) signal of 80 $\rm cm \, s^{-1}$. While apparently small, 0.5-to-1 $\rm m \, s^{-1}$ planetary signals are within the reach of state-of-the-art spectrographs. In particular, the high-resolution ESPRESSO spectrograph \citep{2021A&A...645A..96P}, with an on-sky RV precision better than 10 $\rm cm \, s^{-1}$ on short timescales ($<$ 1 h) and of 40 $\rm cm \, s^{-1}$ in the long term \citep{figueira2025}, offers an excellent opportunity to characterize the population of small USP planets. 

Small USP planets receive insolations thousands of times larger than that received by the Earth, so that any primordial H/He-dominated atmosphere is expected to have been removed through photo-evaporation \citep[e.g.][]{2014ApJ...787...47S,2016NatCo...711201L,2017MNRAS.472..245L,2019AREPS..47...67O}. The masses and radii of small USP planets have thus been directly used to probe their internal structures \citep[e.g.][]{2019ApJ...883...79D,2024MNRAS.529..141G,2024A&A...684A..83M,2025A&A...693A.144G}. Most rocky USP planets have equilibrium temperatures larger than 1800 K, which implies that their rocky mantles are partially molten, forming extensive ‘magma oceans' on the side facing the star \citep[e.g.][]{2017RSPTA.37550394T,2021ApJ...922L...4D,2024ApJ...962L...8S,2025TrGeo...7...51L}. While magma oceans have a negligible impact on the inferred compositions of rocky planets \citep[e.g.][]{2023ApJ...954..202B}, molten rocks have a great storage capacity of volatiles, which can be outgassed, possibly generating secondary high mean-molecular weight atmospheres \citep[e.g.][]{2020PNAS..11718264K,2024ApJ...963..157T}. Interestingly, USP planets lie inside the stellar Alfvénic sphere, so they are thought to undergo strong magnetic interactions with their host stars. Today, signs of these interactions have been detected in systems with giant planets \citep[e.g.][]{2005ApJ...622.1075S,2019NatAs...3.1128C,2024A&A...684A.160C}, and next-generation optical and radio instrumentation may open the door to similar detections in low-mass planets \citep[e.g.][]{2018A&A...615A.117B,2018haex.bookE..20S}. Small USP planets are also amenable to observations of phase curves and secondary eclipses with the James Webb Space Telescope \citep[JWST;][]{2006SSRv..123..485G}, which might unveil the presence of secondary atmospheres or the surface mineralogy in bare rocky surfaces \citep[e.g.][]{2012ApJ...752....7H,2016Natur.532..207D,2024ApJ...961L..44Z}.

USP planets are known to be affected by strong tidal forces exerted by their host stars, which gradually shrink their orbits, potentially bringing them within the tidal disruption limit. In this region, rocky material can be detached from the planet, even to the point of completely disrupting the whole planet \citep[]{2013ApJ...773L..15R}. Orbital decay towards the disruption limit may occur over timescales ranging from millions of years to several gigayears \citep[e.g.][]{2017MNRAS.465..149J,2018NewAR..83...37W,2024AJ....168..101D}, depending primarily on the orbital separation of the USP planet. In this regard, some authors point out that the disintegrated material would be engulfed by the star, potentially altering its photospheric compositions by a measurable amount \citep[e.g.][]{2015ApJ...808...13R,2018ApJ...854..138O,2023MNRAS.521.2969B,2024Natur.627..501L,2025A&A...693A..47S}, and others have managed to detect the possible disintegration process through strong asymmetries in the transit signals \citep[e.g.][]{2012ApJ...752....1R,2014ApJ...784...40R,2015ApJ...812..112S,2025ApJ...984L...3H}. 

While the end-state of the life of small USP planets is fairly well understood, many doubts remain about their origins. The fundamental problem is that it is unlikely that these planets formed at their present-day locations due to the extreme temperatures in these close-in regions \citep[e.g.][]{1998AREPS..26...53B} and the likely truncation of the proto-planetary disc at much larger distances \citep[e.g.][]{2017ApJ...842...40L}. In this regard, while tides can shrink a planetary orbit once a close-in location is reached, they are probably not efficient enough to bring planets to the USP regime within the age of the Universe \citep[e.g.][]{1966Icar....5..375G,2018NewAR..83...37W}. An early theory proposed that USP planets are the remnant cores of tidally disrupted migrated giant planets \citep{2013ApJ...779..165J,2016CeMDA.126..227J}. However, this possibility has been discarded as the predominant mechanism given that hot Jupiters \citep[e.g.][]{2001A&A...373.1019S} and hot Neptunes \citep[e.g.][]{2025AJ....169..117V} are found to preferentially orbit metal-rich stars, while the hosts of rocky USP planets show a more solar-like metallicity distribution \citep[e.g.][]{2017AJ....154...60W}. 

The most promising theories of USP planet formation involve high-eccentricity tidal migration (HEM) triggered by secular chaos \citep{2019AJ....157..180P}, low-eccentricity migration due to secular forcing \citep{2019MNRAS.488.3568P}, and obliquity-driven tidal migration \citep{2020ApJ...905...71M}. While fundamentally different, these theories converge into a common prediction: small USP planets are brought to their present-day orbits through interactions with outer planetary companions. In this line, the \textit{Kepler} mission \citep{2010Sci...327..977B,2014PASP..126..398H} showed that USP planets in multi-planetary systems are common. Based on a homogeneous analysis of ten campaigns of the extended K2 mission, \citet[][]{2021PSJ.....2..152A} estimated that all USP planets are expected to have planetary neighbours. In addition, thanks to their transiting nature, \textit{Kepler}-based works enabled the measurement of their mutual inclinations \citep[e.g.][]{2012ApJ...761...92F,2012A&A...541A.139F,2012AJ....143...94T,2014ApJ...790..146F,2018ApJ...864L..38D,2021PSJ.....2..152A}, which is crucial to test formation theories. These large-scale photometry-based surveys are of great value given the wealth of constraints achieved in a short time span. However, their conclusions have to be taken with caution, since they are highly biased towards systems with small mutual inclinations and short-orbit companions due to the loss of non-transiting configurations and typically short temporal observing baselines. In this regard, RV follow-up monitoring of USP-hosting systems is crucial to explore relevant regions of the period-inclination parameter space missed by transit surveys. 

In this work, we observed the G9 V star K2-157 ($V$ = 12.792 $\pm$ 0.057 mag) with the ESPRESSO spectrograph to confirm and characterize the statistically validated USP Earth-sized planet K2-157~b \citep[$P_{\rm orb}$ = 8.8 h;][]{2018AJ....155..136M} and constrain the presence of additional companions. These observations allowed us to detect two super-Neptune-mass planets in the warm Neptunian savanna and measure the mass of K2-157~b. In Sect.~\ref{sec:obs}, we describe the observations analysed in this work. In Sect.~\ref{sec:stellar_charact}, we present our stellar characterisation of K2-157 based on a high-resolution, high-S/N ESPRESSO spectrum.  In Sect.~\ref{sec:analysis}, we describe our analyses and results. In Sect.~\ref{sec:discussion}, we discuss the results, and we summarise and conclude in Sect.~\ref{sec:conclusions}.

\section{Observations}
\label{sec:obs}
\subsection{K2 high-precision photometry}
\label{sec:K2_data}

K2-157 (EPIC 201130233) was observed by K2 \citep{2014PASP..126..398H} in Campaign 10 (C10) from 13 July 2016 (JD 2457582.61) to 20 September 2016 (JD 2457651.69) in the so-called long-cadence mode (i.e. 29.4 min cadence). We accessed and downloaded the data through the Mikulski Archive for Space Telescopes (MAST)\footnote{\url{https://archive.stsci.edu}}.   Between JD 2457589.3 and JD 2457604.3, there was a failure in a detector (module 4) that caused a data gap of 15~d, resulting in a lower than usual observing time of 54~d. The loss of two reaction wheels caused the spacecraft to undergo continuous drifts that translated into larger systematic errors than those observed in the primary mission. We thus used the \texttt{everest} pipeline\footnote{Available at \url{https://github.com/rodluger/everest}} \citep{2016AJ....152..100L,2018AJ....156...99L} to correct for instrumental systematics based on the pixel-level decorrelation technique \citep[PLD;][]{2015ApJ...805..132D}. In addition, we downloaded the K2SFF photometry based on the self-flat fielding technique \citep[SFF;][]{2014PASP..126..948V}. We analysed the K2 photometry (Sect.~\ref{sec:joint_fit}) by considering both corrections and found consistent results within 1$\sigma$. We opted to present the results based on the \texttt{everest} correction, but we highlight that the properties of this system do not depend on the correction pipeline. \citet{2018AJ....155..136M} analysed WIYN/NESSI speckle and archival images to ensure that no nearby sources are contaminating the photometric aperture. We extended this analysis by searching for nearby \textit{Gaia} DR3 \citep{2023A&A...674A...1G} sources and found that the nearest star is located 20{\arcsec} away from K2-157 with a $G$ magnitude of 20.4 mag (Gaia DR3 3596379973268959744), which ensures a negligible contamination, in agreement with \citet{2018AJ....155..136M}. Therefore, no additional dilution correction was necessary. In Table \ref{tab:K2_data}, we present the \texttt{everest} photometry used in this work.  

We computed the transit least squares periodogram \citep[\texttt{TLS};][]{2019A&A...623A..39H}\footnote{Available at \url{https://github.com/hippke/tls}} of K2 C10 data to determine the significance of K2-157~b and potentially unveil additional transits. We first detrended the photometry through the bi-weight technique as implemented in \texttt{wotan}\footnote{Available at \url{https://github.com/hippke/wotan}} \citep{2019AJ....158..143H}. We generated 14 light curves considering a wide range of window lengths, between 0.2~d and 1.5~d (step size of 0.1~d), to ensure that we avoided possible detrending systematics related to the K2-157~b transit duration or orbital period, which could bias the \texttt{TLS} analysis. The resulting time series show standard deviations between 134 and 147 ppm (parts per million). The maximum power peaks correspond to the periodicity of K2-157~b (0.365~d) and emerge with signal detection efficiencies (SDE) between 26.0 and 28.1, which are well above the commonly considered thresholds for a significant transit detection \citep[i.e. SDEs from 6 to 10; ][]{2012ApJ...761..123S,2015ApJ...807...45D,2018AJ....156...78L,2018MNRAS.473L.131W}. In Fig.~\ref{fig:TLS}, we show the periodogram of the detrended time series with the largest SDE for K2-157~b (corresponding to a window length of 0.5~d). We then masked the eclipses of K2-157~b and re-computed the \texttt{TLS} periodograms, finding no additional significant peaks.  
\subsection{TESS high-precision photometry}
\label{sec:TESS_data}

The Transiting Exoplanets Satellite Survey \citep[TESS;][]{2014SPIE.9143E..20R} observed K2-157 (TIC 349445372) in Sector 46 (S46) from 3 December 2021 (JD 2459551.56) to 30 December 2021 (JD 2459578.70) with a 2-min cadence. We downloaded the available data products from MAST. At the mid-sector, there is a gap of 3 d due to the satellite repointing towards the Earth to downlink the data, resulting in a total observing time of 25 d. We used the PDCSAP photometry, which is the pixel-calibrated simple aperture photometry (SAP) corrected from instrumental systematics by the PDC algorithm \citep{2012PASP..124.1000S, 2012PASP..124..985S,2016SPIE.9913E..3EJ}. Given the larger TESS pixel scale than Kepler's (i.e. 21{\arcsec} vs 3.98{\arcsec}) and the consequent larger probability of flux contamination, we used the \texttt{TESS-cont} algorithm\footnote{Available at \url{https://github.com/castro-gzlz/TESS-cont}} \citep{2024A&A...691A.233C} to quantify the flux fraction from nearby sources falling within the photometric aperture. We find that 99.7$\%$ of the flux comes from K2-157, ensuring negligible contributions from nearby sources. We present the PDCSAP of K2-157 in Table~\ref{tab:TESS_data}.

We computed the \texttt{TLS} periodogram of the PDCSAP, which we detrended similarly to the K2 data (i.e. bi-weight method with a window length of 0.5~d). The resulting time series has a standard deviation of 3574 ppm (or 982 ppm in 30-min bins), which is seven times larger than that in K2's data. We show the periodogram in Fig.~\ref{fig:TLS}. No peak is found with SDE~$>$~7, indicating that the TESS data is not precise enough to detect K2-157~b and that no additional transit signals are detected. 

\subsection{ESPRESSO high-resolution spectroscopy}
\label{sec:ESPRESSO_data}

We observed K2-157 with the echelle spectrograph for rocky exoplanets and stable spectroscopic observations \citep[ESPRESSO;][]{2021A&A...645A..96P}, which is mounted on the Very Large Telescope (VLT) located at ESO’s Paranal Observatory. These observations were carried out in the context of the Guaranteed Time Observations (GTOs) of the instrument, which previously allowed us to characterize dozens of small rocky planets detected by TESS \citep[e.g.][]{2022A&A...665A.154B,2020A&A...642A..31D,2021A&A...648A..75S,2021A&A...653A..41D,2023A&A...675A..52C,2023A&A...679A..33D,2023A&A...673A..69L,2024A&A...685A..56S,2024A&A...688A.216H,2025A&A...695A.237R} and K2 \citep[e.g.][]{2020MNRAS.499.5004M,2020A&A...641A..92T,2024A&A...684A..22P}. We acquired 49 spectra between 19 February 2019 and 19 April 2021 (programme IDs 1102.C-0744, 1102.C-0958, 1104.C-0350, and 106.21M2.004) with a typical cadence of 2-3~d and exposure times between 900 and 1200 s. Given the moderate brightness of the star, we conducted all observations through a single VLT unit in the slow-readout high-resolution mode (HR21), which allowed us to get high-S/N spectra with spectral resolutions of 140\,000 across the 380-788 nm wavelength coverage. During the observations, simultaneous calibration was conducted by illuminating a second fibre connected to the instrument with a Fabry-Pérot interferometer \citep[][]{1899ApJ.....9...87P} to control environmental changes \citep{2010SPIE.7735E..4XW}.

We used the ESPRESSO Data Reduction Software (DRS; version 3.3.0)\footnote{The latest version of the ESPRESSO Data Reduction Software can be found at \url{https://www.eso.org/sci/software/pipelines}} to extract the calibrated spectra and associated RVs. The RV extraction is based on the technique presented by \citet{1996A&AS..119..373B}, whereby the average line profile (or cross-correlation function; CCF) is extracted using a mask weighted according to the RV content \citep[i.e. line depth and sharpness;][]{2001A&A...374..733B,2002A&A...388..632P}. For K2-157, we chose the G9 pipeline mask to obtain the CCFs, which were fitted to a Gaussian profile to compute the wavelength shifts. The relative RVs between the instrument and K2-157 were finally converted into stellar RVs by subtracting the barycentric Earth radial velocity (BERV) and correcting for secular acceleration. The RVs of K2-157 have a median uncertainty of 1.5 $\rm m\,s^{-1}$ and a standard deviation of 6.2 $\rm m \, s^{-1}$. This RV scatter widely exceeds the exquisite ESPRESSO on-sky RV performance \citep[e.g.][]{2020A&A...639A..77S,2022A&A...658A.115F,2024A&A...690A..79G,figueira2025} and cannot be explained through the small measured uncertainties, suggesting the existence of planetary or stellar signals. The DRS also computes different activity indicators based on the average line profile (i.e. FWHM, BIS, and Contrast of the CCF) and the emission of individual lines particularly sensitive to stellar activity (i.e. $\rm H_{\alpha}$, Ca-index, S-index, Na-index, and log $\rm R'_{HK}$). We additionally used the novel template-matching (TM) algorithm \texttt{sbart}\footnote{Available at \url{https://github.com/iastro-pt/sBART}} \citep{2022A&A...663A.143S} to get an independent RV extraction, which we note that it is not affected by the recently identified systematic bias in TM algorithms given the large baseline of the observations \citep{silva2025}. In this work, we opted to present the
results based on the CCF RVs computed by the DRS and ensured that they are independent of the extraction pipeline (i.e. consistent at 1$\sigma$ and similar uncertainties). In Table~\ref{tab:ESPRESSO_data}, we present the ESPRESSO data set.

\begin{figure}
    \centering
    \includegraphics[width=0.48\textwidth]{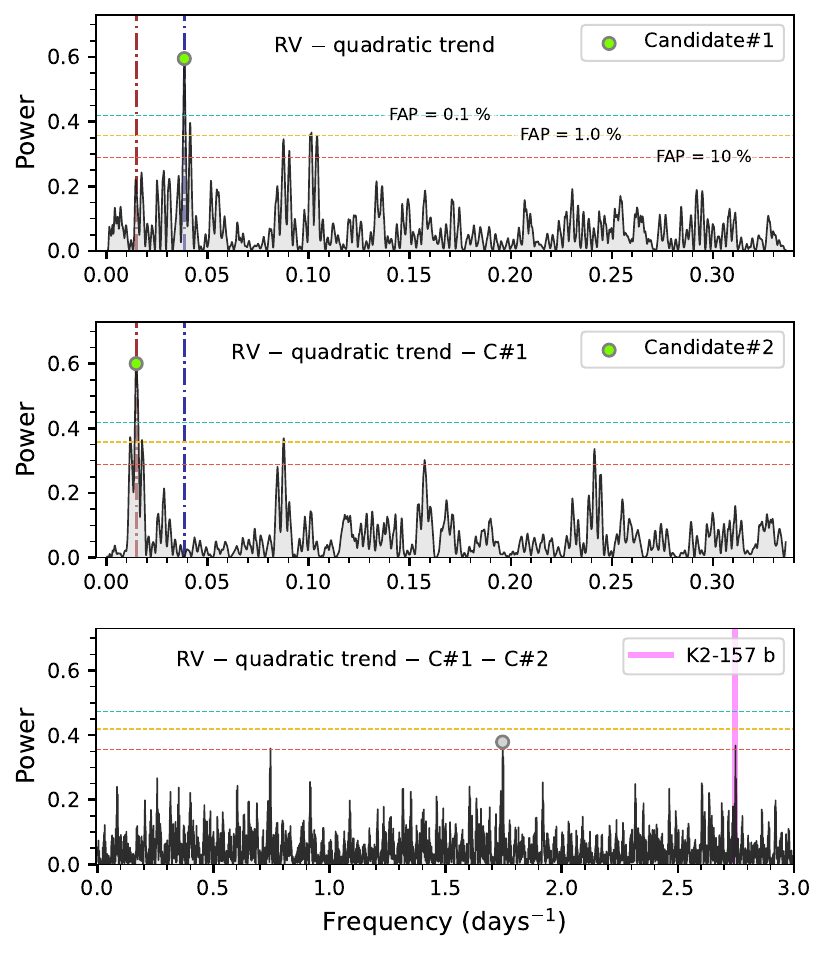}
    \caption{Generalized Lomb-Scargle (\texttt{GLS}) periodograms of the ESPRESSO RVs of K2-157 after subtracting a quadratic trend (upper panel) and the RV signals of Candidate$\#$1 (C$\#$1; middle panel) and Candidate$\#$2 (C$\#$2; lower panel). The circles indicate the maximum power frequencies and are highlighted in green when they correspond to a significant signal (FAP $<$ 0.1~$\%$). }
    \label{fig:gls_to_RVs}
\end{figure}

We computed the generalised Lomb-Scargle periodogram \citep[\texttt{GLS};][]{2009A&A...496..577Z}\footnote{We used the \texttt{PyAstronomy} implementation \citep[][]{pya}, which is available at \url{https://github.com/sczesla/PyAstronomy}.} of the ESPRESSO RVs and activity indicators to identify sinusoidal signals that could be attributed to planets or stellar activity. The periodogram of the RVs shows its maximum peak at a frequency of 0.03852 $\rm d^{-1}$ (periodicity of 25.96 d) with a False Alarm Probability (FAP) of 1.8 $\times$ $10^{-3}$ $\%$, which is below the commonly used threshold for considering that a signal is significant (FAP < 0.1 $\%$). When subtracting a quadratic trend previously fit to the data through least squares, the maximum power peak remains at this periodicity and its FAP decreases down to 2.5 $\times$ $10^{-5}$~$\%$. In the upper panel of Fig.~\ref{fig:gls_to_RVs}, we show this periodogram. In the middle panel of the same figure, we show the periodogram of the quadratically detrended RVs subtracted from the 25.96-d sinusoidal signal. A new significant peak pops up at a frequency of 0.014725 $\rm d^{-1}$ (67.91 d) and a FAP of 1.8 $\times$ $10^{-5}$~$\%$. After subtracting this second signal from the RV data set, no additional significant peaks appear (see the lower panel of Fig.~\ref{fig:gls_to_RVs}). However, we can appreciate three peaks that reach the 10~$\%$ FAP level\footnote{In this periodogram, we forced the algorithm to explore signals above the Nyquist frequency to unveil possible hints of K2-157~b.}. One of these peaks has a frequency of 2.7478~$\rm d^{-1}$ (0.36~d), which perfectly coincides with the orbital period of K2-157~b. The other peaks have frequencies of 1.7455 $\rm d^{-1}$ (0.57~d) and 0.7449 (1.34~d). Overall, while not significant, the appearance of a tentative peak (FAP < 10 $\%$) with a 0.36~d periodicity after subtracting the two significant signals suggests that the RV signature of K2-157~b is embedded in the ESPRESSO data. 

To assess whether the two significant RV signals could have a planetary origin or be contaminated by the activity of the host star, we computed the periodograms of the six activity indicators extracted by the DRS (Fig.~\ref{fig:gls_to_indicators}). No statistically significant peaks (FAP < 0.1~$\%$) appear in any of the indicators, either at the RV frequencies or at any other frequency. To try to unveil possible dependencies between the RVs and activity indicators, we searched for linear correlations through the Pearson correlation coefficient \citep{doi:10.1080/00031305.1988.10475524}. We find coefficients $r < 0.3$ for all indicators (Fig.~\ref{fig:person_r}), which is considered to reflect negligible correlations. This absence of evident stellar signals suggests that the prominent RV signals correspond to two additional planets in the system. We thus hereinafter refer to the RV signals as planet candidates: Candidate$\#$1 (25.96~d) and Candidate$\#$2 (67.91~d). We note that, while not significant, the periodogram of the $\rm H_{\alpha}$ indicator shows a relatively prominent peak at 36.9~d with a FAP of 0.26$\%$, making it an interesting candidate to reflect the stellar rotation period, which we further explore in Sect.~\ref{sec:stellar_activity}. 

We also used the $\ell_1$ periodogram technique by \citet{hara2017} and checked that the phase, amplitude, and frequency of Candidate$\#$1 and Candidate$\#$2 are consistent. The $\ell_1$ periodogram technique is based on a sparse recovery technique called the basis pursuit algorithm~\citep{chen1998}. 
It aims to find a representation of the RV time series as a sum of a small number of sinusoids whose frequencies are in the input grid.  It outputs a figure which has a similar aspect as a regular periodogram, but with fewer peaks due to aliasing. The peaks can be assigned a FAP, whose interpretation is close to the FAP of a regular periodogram peak, but which takes into account the correlated noise following~\cite{delisle2020}. First, the $\ell_1$ periodogram takes several parameters as input, in particular a list of vectors, or predictor, which are fitted linearly along with the search for periodic signals. Second, it needs an assumed covariance model, which can be non-diagonal to account for correlated noise. For the base model, we chose an offset and a linear and quadratic trend. We selected the covariance model following~\cite{hara2020}: we considered a grid of values for the noise model, and ranked them with cross-validation and Bayesian evidence, computed with the Laplace approximation. In Fig.~\ref{fig:l1per}, we show the $\ell_1$ periodogram corresponding to the highest ranked model. The peaks at 25.9~d and 69.3~d are significant, with FAP 2.7 $\times$$ 10^{-4}$ $\%$ and 1.9 $\times$ $10^{-4}$ $\%$. The other signals exhibit much higher FAPs ($>$ 90~$\%$).

To check that the phase, amplitude, and frequency of the RV planet candidates are consistent, we used an apodized sine periodogram~\cite[ASP;][]{hara2022} consisting of fitting the coefficients $A$ and $B$ of an apodized sine function, 
\begin{equation}
    e^{-\frac{(t-t_0)^2}{2\tau^2}}  (A \cos \omega t + B \sin \omega t ) \label{eq:asp},
\end{equation}
for a grid of values of frequency, $\omega$, signal lifetime, $\tau$, and signal centre, $t_0$. We applied the ASP iteratively: the signal with the best fit is removed from the data, and then the ASP is computed on the residuals. We show the first three iterations in Fig.~\ref{fig:ASP}, and find that Candidate$\#$1 and Candidate$\#$2 are consistent with being strictly periodic (i.e. the value of $\tau$ is compatible with +$\infty$). Overall, the \texttt{GLS} and $\ell_1$ periodograms, correlation analyses, and signal consistency strongly suggest that the two planet candidates are fiducial planets, which we further study in Sect.~\ref{sec:blind_search}.

\section{Stellar Characterization}
\label{sec:stellar_charact}

K2-157 is a late G-type star \citep[$V$ = 12.792 $\pm$ 0.057 mag, $B$-$V$ = 0.805 mag;][]{2016yCat.2336....0H} located in the solar neighbourhood \citep[$\pi$ = 3.446 $\pm$ 0.015 mas; ][]{2023A&A...674A...1G}. In Table~\ref{tab:stellar_prop}, we compile its main astrometric and photometric properties from the literature. \citet{2018AJ....155..136M} used one spectrum acquired with the TRES spectrograph to derive the spectroscopic parameters $T_{\rm eff}$, log $g$, and [Fe/H] through the Stellar Parameter Classification (SPC) tool \citep{2012Natur.486..375B}. The authors obtained $T_{\rm eff}$ = 5456 $\pm$ 50 K, [Fe/H] = 0.13 $\pm$ 0.08 dex, and log $g$ = 4.55 $\pm$ 0.10, and input them into the \texttt{isochrones} package \citep{2015ascl.soft03010M} to derive a stellar mass and radius of  $M_{\star}$ = $0.940^{+0.023}_{-0.027}$ $\rm M_{\odot}$ and $R_{\star}$ = $0.876^{+0.045}_{-0.025}$ $\rm R_{\odot}$, respectively. The photometry-based TESS Input Catalog \citep[TIC v8.2;][]{2018AJ....156..102S} obtains compatible values: $T_{\rm eff}$ = $5337^{+147}_{-159}$ K, log $g$ = $4.478^{+0.092}_{-0.069}$, $M_{\star}$ = $0.93^{+0.12}_{-0.10}$ $\rm M_{\odot}$, and $R_{\star}$ = $0.921^{+0.043}_{-0.059}$ $\rm R_{\odot}$. In the following sections, we describe our stellar characterisation based on a high-resolution, high-S/N ESPRESSO spectrum obtained from the co-adding of the 49 individual observations. 


\begin{table}[]
\renewcommand{\arraystretch}{1.193}
\setlength{\tabcolsep}{3pt}
\caption{Stellar properties of K2-157 compiled from the literature and derived in this work (Sect.~\ref{sec:stellar_charact}).}
\label{tab:stellar_prop}
\begin{tabular}{llc}
\hline \hline
Parameter                               & Value                                   & Reference            \\ \hline 
\multicolumn{3}{l}{Identifiers}                                                                          \\ \hline
K2                                      & 157                                     & (1)                  \\
EPIC                                    & 201130233                               & (2)                  \\
TIC                                     & 349445372                               & (3)                  \\
UCAC4                                   & 422-055378                              & (4)                  \\
Gaia DR3                                & 3596333042162818304                     & (5)                  \\ \hline
Astrometric properties                  &                                         &                      \\ \hline
RA, Dec                                 & 12:15:00.36, -5:46:55.27                & (5)                  \\
$\rm \mu_{\alpha}$ ($\rm mas\,yr^{-1}$) & 36.683 $\pm$ 0.017                      & (5)                  \\
$\rm \mu_{\delta}$ ($\rm mas\,yr^{-1}$) & -7.522 $\pm$ 0.012                      & (5)                  \\
Parallax (mas)                          & 3.446 $\pm$ 0.015                       & (5)                  \\
RV ($\rm km\,s^{-1}$)                   & 44.5 $\pm$ 1.4                          & (5)                  \\ \hline
\multicolumn{3}{l}{Atmospheric parameters and spectral type}                                             \\ \hline
$T_{\rm eff}$ (K)                       & $5334 \pm 64$                          & (6)                  \\
log $g$ (cgs)                           & $4.45 \pm 0.04$                         & (6)                  \\
{[}Fe/H{]} (dex)                        & $0.02 \pm 0.04$                       & (6)                  \\
$\xi_{\mathrm{t}}$ ($\rm km\,s^{-1}$)   & $0.73 \pm 0.04$                       & (6)                  \\
SpT                                     & G9 V            & (6)                  \\ \hline
Physical parameters                     &                                         & \multicolumn{1}{l}{} \\ \hline
$R_{\star}$ $(\rm R_{\odot})$           & $0.860 \pm 0.019$                       & (6)                  \\
$M_{\star}$ $(\rm M_{\odot})$           & $0.890 \pm 0.029$                       & (6)                  \\
$\rm Age$ (Ga)                        & $8.8 \pm 4.0$                       & (6)                  \\

$v \, \textrm{sin} \, i_{\star}$ ($\rm km\,s^{-1}$) & $<$ 3                & (6)   \\

$\textrm{log} \, R'_{HK}$ (dex)         & -5.077 $\pm$ 0.049            & (6)                  \\
$P_{\rm rot, H_{\alpha}}$$^{(\dagger)}$  (d)                     & $36.76^{+0.40}_{-0.44}$           & (6)                  \\ \hline
Chemical abundances                     &                                         & \multicolumn{1}{l}{} \\ \hline
{[}Mg/H{]} (dex)                       & $0.07 \pm 0.05$ & (6)                  \\
{[}Si/H{]} (dex)                        & $0.05 \pm 0.05$ & (6)                  \\
\hline
Photometric properties                  &                                         &                      \\ \hline
V (mag)                                 & 12.792 $\pm$ 0.057                      & (7)                  \\
B (mag)                                 & 13.597 $\pm$ 0.049                      & (7)                  \\
g' (mag)                                & 13.160 $\pm$ 0.006                      & (7)                  \\
r' (mag)                                & 12.579 $\pm$ 0.012                      & (7)                  \\
i' (mag)                                & 12.387 $\pm$ 0.018                      & (7)                  \\
z' (mag)                                & 12.367 $\pm$ 0.061                      & (7)                  \\
J (mag)                                 & 11.398 $\pm$ 0.022                      & (8)                  \\
H (mag)                                 & 11.023 $\pm$ 0.022                      & (8)                  \\
$\rm K_{s}$ (mag)                       & 10.980 $\pm$ 0.023                      & (8)                  \\ \hline
\multicolumn{3}{l}{Galactic space velocities and membership}                                      \\ \hline
$U_{\rm LSR}$$^{(\dagger\dagger)}$ ($\rm km \, s^{-1}$)           & $-66.89 \pm 0.31$  & (6)    \\
$V_{\rm LSR}$$^{(\dagger\dagger)}$ ($\rm km \, s^{-1}$)           & $5.15 \pm 0.74$  & (6)    \\
$W_{\rm LSR}$$^{(\dagger\dagger)}$ ($\rm km \, s^{-1}$)           & $45.65 \pm 1.14$   & (6)    \\
Gal. population$^{(\dagger\dagger)}$                  & Thin disc (54$\%$)         & (6)   \\ \hline
\end{tabular}
\tablefoot{($\dagger$) $P_{\rm rot, H_{\alpha}}$ was obtained by fitting a sinusoidal model to the $H_{\rm \alpha}$ activity index ($\Delta \mathcal{Z}$ = +2.2; Sect.~\ref{sec:stellar_activity}), and it is compatible with the $P_{\rm rot}$ estimated from empirical $\textrm{log}(R'_{\rm HK }$)-$P_{\rm rot}$ relations (Sect.~\ref{subsec:rotation_proxies}). ($\dagger\dagger$) $UVW$ are referred to the local standard of rest from \citet{2010MNRAS.403.1829S}, and the probability is based on the prescription by \citet{2014A&A...562A..71B}.   (1) \citealp{2018AJ....155..136M}; (2) \citealp{2016ApJS..224....2H}; (3) \citealp{2018AJ....156..102S}; (4) \citealp{2013AJ....145...44Z}; (5) \citealp{2023A&A...674A...1G}; (6) This work; (7) \citet{2016yCat.2336....0H};    (8) \citet{2006AJ....131.1163S}.}
\end{table}

\subsection{Stellar parameters and chemical abundances}
\label{sec:stellar_params_abund}

We used the spectral analysis technique ARES+MOOG to derive the stellar atmospheric parameters ($T_{\mathrm{eff}}$, $\log g$, micro-turbulence, and [Fe/H]), following the methodology described in \citet[][]{Sousa-21, Sousa-14, Santos-13}. The latest version of ARES\footnote{Available at \url{https://github.com/sousasag/ARES}} \citep{Sousa-07, Sousa-15} was used to consistently measure the equivalent widths (EW) of a list of iron lines presented in \citet[][]{Sousa-08}. We used a minimisation process to find the ionisation and excitation equilibrium and converge to the best spectroscopic parameters. We used a grid of Kurucz model atmospheres \citep{Kurucz-93} and the radiative transfer code MOOG \citep{Sneden-73}. The derived parameters are listed in Table \ref{tab:stellar_prop}. We also derived the trigonometric surface gravity using \textit{Gaia} DR3 data following the methodology described in \citet[][]{Sousa-21}. The mass and radius of the star were inferred using PARAM 1.3\footnote{Available at \url{http://stev.oapd.inaf.it/cgi-bin/param\_1.3}} \citep[][]{daSilva-06}: $M_{\star}$ = $0.890 \pm 0.029$~$\rm M_{\odot}$ and $R_{\star}$ = $0.860 \pm 0.019$ $\rm R_{\odot}$. 

Chemical abundances are relevant to study the internal composition of highly irradiated rocky planets such as K2-157~b, since they are considered to be a good proxy of the original composition of the proto-planetary discs \citep[e.g.][]{2021Sci...374..330A}. We derived the abundances of Mg and Si using the same tools and models as for stellar parameter determination, as well as the classical curve-of-growth analysis method, assuming local thermodynamic equilibrium \citep[e.g.][]{Adibekyan-12, Adibekyan-15}. Although the EWs of the spectral lines were automatically measured with ARES, for Mg, which has only three lines available, we performed a careful visual inspection of the EWs. We list the inferred abundances in Table \ref{tab:stellar_prop}.

\subsection{Rotation period through the $\textrm{log}(R'_{\rm HK }$) activity index}
\label{subsec:rotation_proxies}

\begin{figure*}
    \centering
    \includegraphics[width=\textwidth]{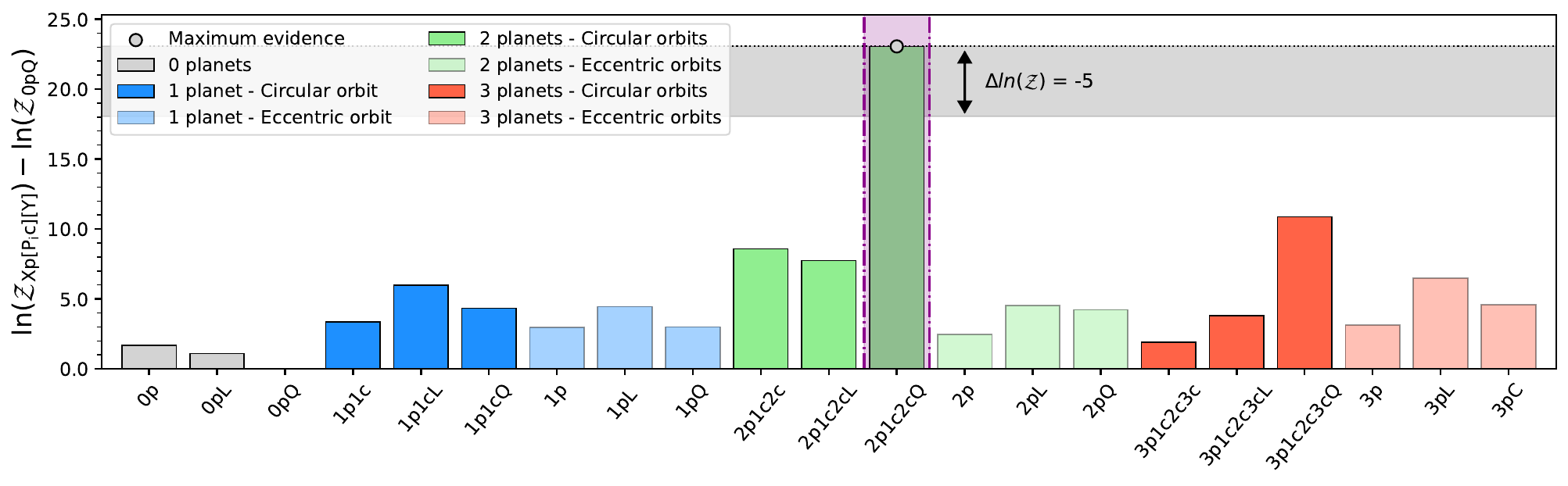}
    \caption{Bar chart showing the differences of the log-evidences of the 21 tested models, which are labelled on the $X$ axis. The grey, blue, green, and red bars represent models with zero, one, two, and three planets, respectively. The vertical magenta shade highlights the simplest model that best represents our data set (2p1c2cQ). The horizontal grey shade indicates the 0 $\geq$ $\rm \Delta ln(\mathcal{Z})$ $\geq$ -6 region from the model with the largest evidence.}
    \label{fig:ev_blind_search}
\end{figure*}

The $\textrm{log}(R'_{\rm HK }$) values measured by the ESPRESSO DRS range from  -5.214 $\pm$ 0.019 to -4.911 $\pm$ 0.015, indicating that K2-157 is a chromospherically inactive star \citep[i.e. $\textrm{log} \, R'_{\rm HK }$ $<$ -4.75;][]{1980PASP...92..385V,1982A&A...107...31M,1996AJ....111..439H,2020A&A...641A.110G}. We adopted a $\textrm{log}(R'_{\rm HK }$) of -5.077 $\pm$ 0.049, which we obtained as the median value of the 49 individual measurements. The error bar corresponds to the largest value between the median uncertainty (i.e. 0.0067) and the standard deviation of the measurements (i.e. 0.049). The $\textrm{log}(R'_{\rm HK }$) index has been shown to correlate with the stellar rotation period through different age–rotation–activity relations, which also depend on the $B-V$ colour \citep{1984ApJ...279..763N,2008ApJ...687.1264M} or spectral type \citep{2016A&A...595A..12S}. The UCAC4 survey \citep{2013AJ....145...44Z} measures $B$ = 13.609 $\pm$ 0.010 mag and $V$ = 12.823 $\pm$ 0.020 mag ($B-V$ = 0.786 $\pm$ 0.022) for K2-157. The APASS survey \citep{2016yCat.2336....0H} DR10 \citep{2019JAVSO..47..130H} measures $B$ = 13.597 $\pm$ 0.049 mag and $V$ = 12.792 $\pm$ 0.057 mag ($B-V$ = 0.805 $\pm$ 0.075 mag). The resulting $B-V$ colours of both surveys are compatible with a t-statistic of 0.09, so we selected the more precise value from UCAC4. We used the \texttt{pyrhk} code\footnote{Available at \url{https://github.com/gomesdasilva/pyrhk}} to estimate the rotation period of K2-157 by following the relations from \citet{1984ApJ...279..763N} and \citet{2008ApJ...687.1264M}, and obtain $P_{\rm rot}$ = 42.6 $\pm$ 7.8 d and $P_{\rm rot}$ = 46.4 $\pm$ 4.2~d, respectively. From the latter, we also obtain an estimated age of 8.8 $\pm$ 4.0 Ga. We also used the updated relation for GKM stars from \citet{2016A&A...595A..12S}, obtaining $P_{\rm rot}$ = 39.0 $\pm$ 3.5~d. Interestingly, these $P_{\rm rot}$ estimations are in quite good agreement with the prominent peak at $\simeq$37~d found in the $H_{\rm \alpha}$ indicator. In Sect.~\ref{sec:stellar_activity}, we analyse in more detail the time series of the activity indicators to try to measure the true $P_{\rm rot}$.

\section{Analysis and results}
\label{sec:analysis}

We aim to make a comprehensive characterisation of the K2-157 planetary system. Given the signal complexity shown in the RV periodograms and the large amount of data provided by K2 and TESS we split the analysis into different steps. 

\subsection{Model inference and parameter determination}
\label{sec:model_inference}

We inferred the models that best describe our data sets and determined the associated parameters through Bayesian inference \citep{1763RSPT...53..370B}. In brief, we first built Gaussian likelihood functions assuming that our data ($D$) is normally distributed, and then obtained both the posterior distributions of the parameters of the models ($M_{i}$) and the Bayesian evidences $\mathcal{Z}_{i} = P(D|M_{i})$ through dynamic nested sampling \citep[see][for a detailed description of the algorithm]{2014A&A...564A.125B,2019S&C....29..891H}. Dynamic nested sampling has been proven to tackle high-dimensional problems better than traditional nestled sampling algorithms thanks to the inclusion of dynamically changing live-points, allowing more exhaustive explorations of the parameter space. We used the nested sampling implementation in \texttt{dynesty}\footnote{Available at \url{https://github.com/joshspeagle/dynesty}} \citep{2020MNRAS.493.3132S} and considered 5000 live-points with a conservative stopping criterion $\Delta \mathcal{Z}$ $<$ $10^{-5}$, $\Delta \mathcal{Z}$ being the Bayesian evidence update during each iteration. 

We searched for the best models describing our data through model comparison. To do so, we followed Occam's razor principle, so that we always selected the simplest possible model unless there was a more complex one with significantly larger evidence. \citet{jeffreys1961theory} empirically estimated that differences of the logarithmic evidences $\mathcal{B}$ = $\textrm{ln} (\mathcal{Z}_{\rm M_{1}}) - \textrm{ln} (\mathcal{Z}_{\rm M_{2}})$ between 1 and 2.5 indicate weak evidence in favour of the higher evidence model, differences between 2.5 and 5 indicate a moderate evidence, while differences larger than 5 reflect strong evidence. We adopt this criterion and hence only consider that a complex model better describes our data than a simpler one if $\mathcal{B}$ $>$ 5.

\subsection{Blind search for RV planetary signals}
\label{sec:blind_search}

We followed the procedure described in Sect.~\ref{sec:model_inference} to search for planetary signals in the ESPRESSO data with no prior assumptions on their existence (i.e. blind search). We built a total of 21 models involving one, two, and three circular and eccentric orbits with and without long-term linear and quadratic trends. 

We built the planetary models through Keplerians as implemented in the \texttt{radvel} package\footnote{Available at \url{https://radvel.readthedocs.io/en/latest}} \citep{2018PASP..130d4504F} by adopting the parametrization $ P_{\rm orb}, T_{0}, K, \sqrt{e} \cos(\omega), \textrm{and} \sqrt{e} \sin(\omega)$, where $P_{\rm orb}$ is the planetary orbital period, $T_{0}$ is the time of inferior conjunction, $K$ is the semi-amplitude, $e$ is the orbital eccentricity, and $\omega$ is the argument of the periastron of the orbit. Models with linear drifts, quadratic trends, or without long-term trends, were implemented as  $\delta \, \times \, (t-t_{\rm init})^{2} +  \gamma \,  \times \,  (t-t_{\rm init}) + v_{\rm sys}$, where $t$ is the Julian date and $t_{\rm init}$ corresponds to the first observing time ($t_{\rm init}$ = JD 2458533.735). We note that $v_{\rm sys}$ corresponds to the systemic velocity of the star as measured by ESPRESSO at $t$ = $t_{\rm init}$. In addition, we incorporated a white noise component (i.e. a jitter term; $\sigma_{\rm jit}$) which we added quadratically to the formal RV uncertainties ($\sigma_{i}$): $(\sigma_{i}^{2} + \sigma_{\rm jit}^{2})^{0.5}$.

\begin{figure*}
    \centering
    \includegraphics[width=\textwidth]{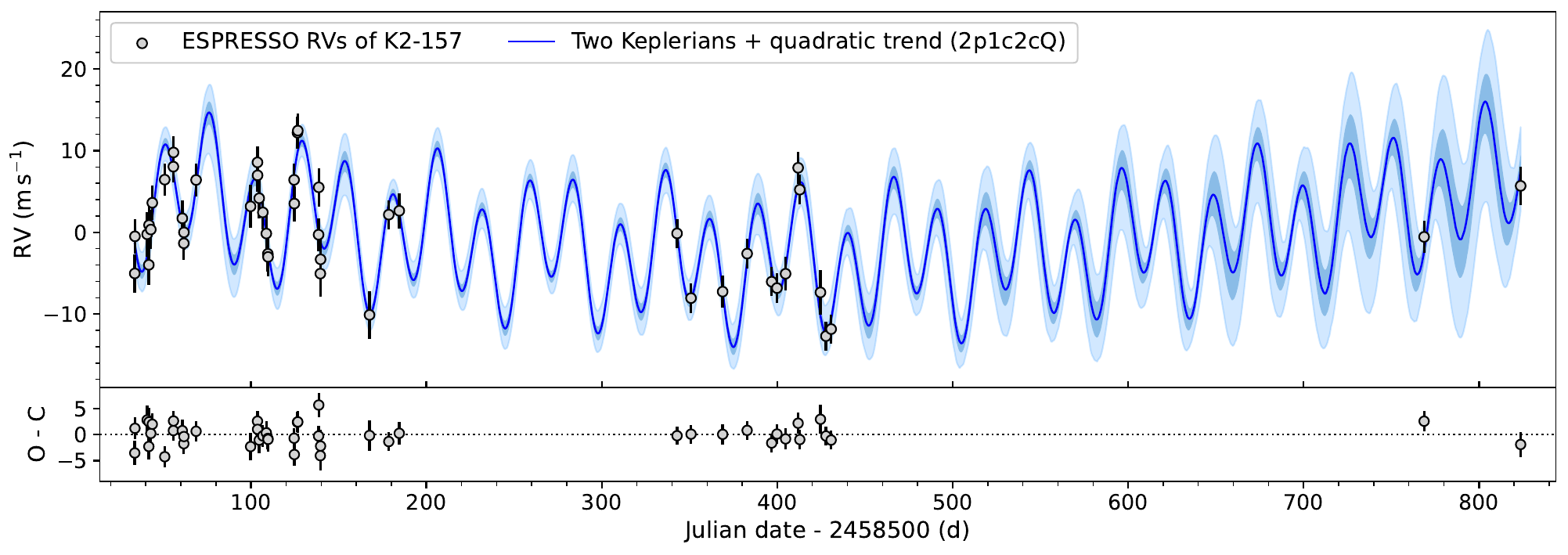}
    \caption{RVs of K2-157. The solid blue line indicates the median posterior model that best represents our data set (i.e. two circular Keplerians plus a quadratic trend), and the dark and light blue shades correspond to the 1$\sigma$ and 3$\sigma$ confidence intervals, respectively.}
    \label{fig:rv_complete_blind}
\end{figure*}

We considered wide and uninformative priors for all the parameters involved in the models to not biasing the search, parameter determination, and evidence computation. In particular, we adopted $\mathcal{U} (0,400)$ d for $P_{\rm orb}$ (i.e. half the total ESPRESSO observing baseline), $\mathcal{U} (t_{\rm init}, t_{\rm init}+400)$ d for $T_{0}$, $\mathcal{U} (0,20)$ $\rm m\,s^{-1}$ for $K$, $\mathcal{U} (0,10)$ $\rm m\,s^{-1}$ for $\sigma_{\rm jit}$, and $\mathcal{U} (-1,1)$ for $\sqrt{e} \cos(\omega)$ and $\sqrt{e} \sin(\omega)$ in the eccentric cases. In Fig.~\ref{fig:ev_blind_search}, we show the log-evidences of each model subtracted from the lowest-evidence model (i.e. the zero-planet model with a quadratic trend), as in \citet{2020A&A...642A.121L}. We label the models following the notation  $\rm Xp \left[ P_{i}c \right] \left[ Y \right]$, where X is the total number of Keplerians, $\rm P_{i}$ indicates which planets are considered to have a circular orbit, and Y indicates whether the model has a linear (L) or quadratic (Q) trend. We find that the Bayesian evidence is boosted for the two-planet model with circular orbits and a quadratic trend, 2p1c2cQ. This model has the largest evidence, and meets the criterion for strong evidence ($\mathcal{B}$ $>$ 5) when compared to simpler models, thus positioning itself as the best description of the ESPRESSO RVs. The considerable evidence difference between this model and the other two-planet models highlights the importance of the quadratic trend, which exhibits a long-term high amplitude comparable to the Keplerians. 

The analyses presented above indicate that the ESPRESSO RVs of K2-157 are best described when including two planetary signals, which coincide with the periodicities of Candidate$\#$1 and Candidate$\#$2 (Sect.~\ref{sec:ESPRESSO_data} and Fig.~\ref{fig:gls_to_RVs}). The ESPRESSO activity indicators show no hints of sinusoidal signals at these periodicities (Sect.~\ref{sec:ESPRESSO_data} and Fig.~\ref{fig:gls_to_indicators}), strongly supporting a planetary origin. We note, however, that before considering these RV planetary signals as confirmed planets, we conducted more dedicated analyses of the stellar activity of K2-157 (Sect.~\ref{sec:stellar_activity}), which definitively allowed us to refer to them as K2-157~c ($P_{\rm orb}$ = 25.97~d) and K2-157~d ($P_{\rm orb}$ = 66.58 d). In Table~\ref{tab:parameters_ESPRESSO}, we show the median and 1$\sigma$ (i.e. 68.3$\%$ credible intervals) of the posterior distributions of the fitted parameters of the 2p1c2cQ model. In Fig.~\ref{fig:rv_complete_blind}, we plot the RV ESPRESSO data set together with the model evaluated on the median parameters. In Fig.~\ref{fig:corner_ESPRESSO}, we show a corner plot with the posterior distributions of the parameters.  

\begin{figure}
    \centering
    \includegraphics[width=0.48\textwidth]{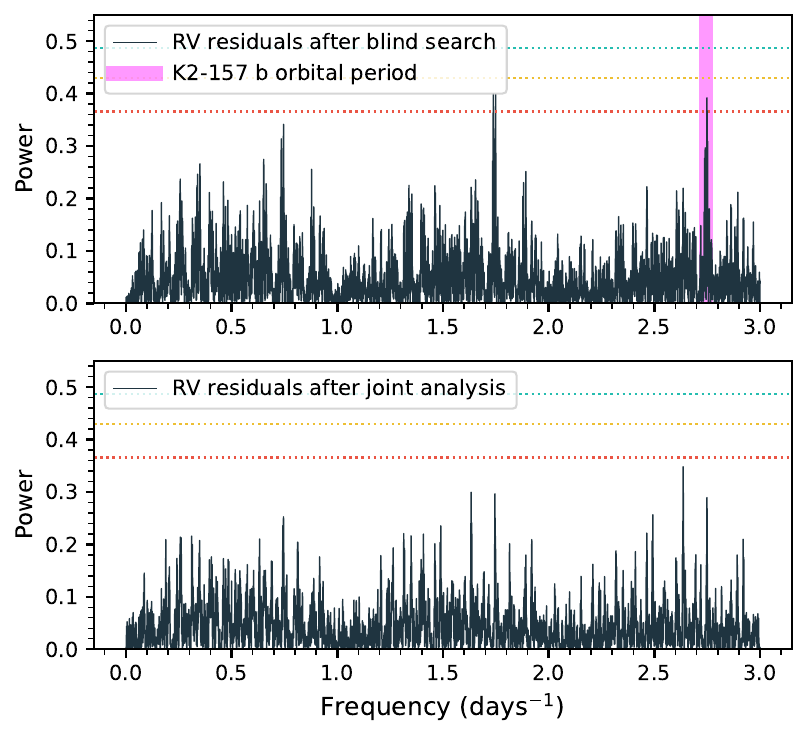}
    \caption{\texttt{GLS} periodogram of the residual RVs after subtracting the two-planet model from the blind-search analysis (Sect.~\ref{sec:blind_search}) and the three-planet model from the joint analysis (Sect.~\ref{sec:joint_fit}). The horizontal dashed red, orange, and green lines indicate the 10$\%$, 1.0$\%$, and 0.1$\%$ FAP levels, respectively.}
    \label{fig:gls_residuals}
\end{figure}

In addition to the detection of two additional planet signals, we can draw two main conclusions from the blind-search analysis. First, the significant quadratic long-term trend suggests the existence of an additional long-period massive companion in the system, or, alternatively, the imprint of the magnetic cycle of the star (we can confidently discard an instrumental origin given the well-known ESPRESSO long-term stability). \citet{2011arXiv1107.5325L} found that long-term RV trends caused by the stellar magnetic cycle would be similarly detected in the FWHM indicator, and the opposite trend would be detected in the Contrast indicator. For K2-157, the FWHM shows a tentative upward trend, and the Contrast shows a clear linear downward trend with no hints of a parabolic behaviour. Hence, we cannot infer a preferred origin for the RV trend based on the indicators. The second conclusion is that the RV signal of K2-157~b is not detected in ESPRESSO data when considering wide, uninformative priors (i.e. through blind search). However, a tentative peak (FAP $<$ 10 $\%$) with the exact periodicity of K2-157~b  appears in the periodogram of the residuals of the two-planet model (Fig.~\ref{fig:gls_residuals}, upper panel). This periodogram resembles the periodogram of the RVs after subtracting a quadratic trend plus the two most significant sinusoidal signals (Fig.~\ref{fig:gls_to_RVs}, lower panel), hinting again that the RV signature of K2-157~b is embedded in our data. 

\subsection{Stellar activity and rotation period}
\label{sec:stellar_activity}

\begin{figure}
    \centering
    \includegraphics[width=0.48\textwidth]{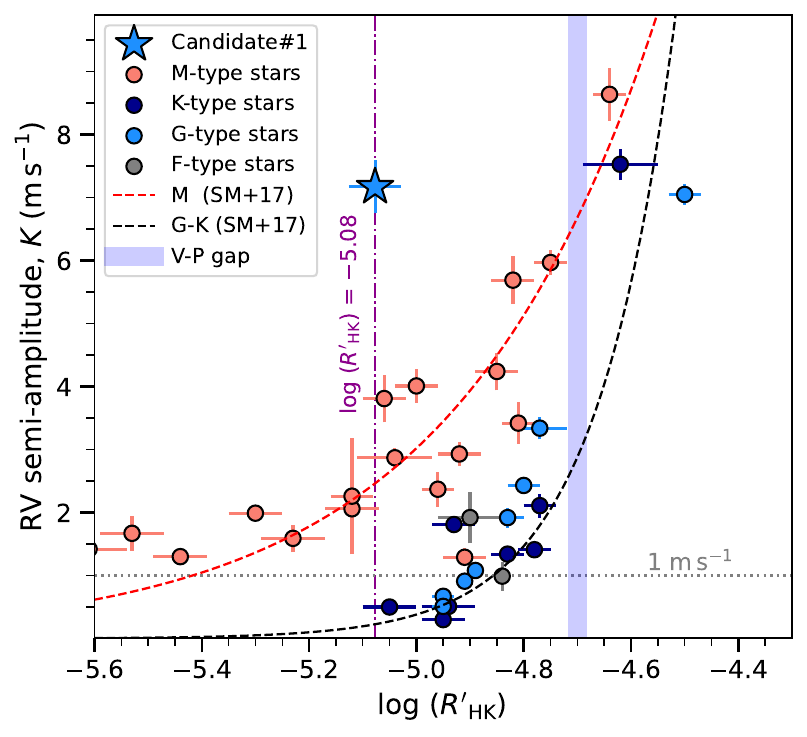}
    \caption{RV semi-amplitude ($K$) of rotation-induced signals versus the chromospheric activity level $\textrm{log}(R'_{\rm HK }$) of 37 stars studied in \citealt{2017MNRAS.468.4772S} (SM+17). The blue star corresponds to the semi-amplitude of Candidate$\#1$. The vertical band represents the Vaughan-Preston (V-P) gap \citep[][]{1980PASP...92..385V} separating inactive and active stars at $\textrm{log}(R'_{\rm HK }$) $\simeq$ -4.7. }
    \label{fig:SM+17}
\end{figure}

The low $\textrm{log}(R'_{\rm HK }$) activity index (i.e. -5.077 $\pm$ 0.049; Sect.~\ref{subsec:rotation_proxies}) together with the absence of significant sinusoidal activity signals (Sect.~\ref{sec:ESPRESSO_data}) indicates that K2-157 is chromospherically inactive. While apparently challenging, the main goal of this section is to try to measure the rotation period of K2-157 by further exploring its activity indexes through a specific activity model.

Given that activity signals are not necessarily periodic \citep[e.g.][]{Ioannidis_2016} or sinusoidal \citep[e.g.][]{2014ApJ...796..132D}, we modelled the activity indicators with a Gaussian Process regression \citep[GP;][]{2006gpml.book.....R,2012RSPTA.37110550R} defined by a quasiperiodic kernel \citep{2015ITPAM..38..252A}. This GP is physically motivated (i.e. it depends on a set of hyperparameters interpretable through different spot/plage properties) and can be implemented through \texttt{george}\footnote{Available at \url{https://github.com/dfm/george}} as:
\begin{equation}
    K_{QP} (\tau) = \eta_{1}^{2} \rm{exp} \left[ - \frac{\tau^{2}}{2\eta_{2}^{2}} - \frac{2sin^{2} \left(  \frac{\pi \tau}{\eta_{3}} \right)}{\eta_{4}^{2}}  \right].
\end{equation}

\noindent The hyperparameter $\eta_{1}$ scales with the amplitude of the stellar activity signal. $\eta_{3}$ corresponds to the main periodicity of the signal, and it is considered to reflect the stellar rotation period \citep[i.e. $\eta_{3}$~=~$P_{\rm rot}$;][]{2018MNRAS.474.2094A}. $\eta_{2}$ is the length-scale of exponential decay, and thus it is considered a measure of the timescale of growth and decline of the active regions. $\eta_{4}$ controls the relative importance between the long-term decay and the periodic variability \citep[see][for detailed descriptions of these parameters]{2014MNRAS.443.2517H,2016A&A...588A..31F,2018MNRAS.474.2094A}. 

Similarly to the blind-search RV analysis, we followed the procedure described in Sect.~\ref{sec:model_inference} to search for activity signals in the ESPRESSO indicators with no prior assumptions on their existence: $\eta_{1}$ $\in$ $\mathcal{U}$(0,1000), $\eta_{3}$ $\in$ $\mathcal{U}$(0,100), $\eta_{2}$ $\in$ $\mathcal{LU}$(0.001,1000), and  $\eta_{4}$ $\in$ $\mathcal{LU}$(0.001,100). In Fig.~\ref{fig:eta3_posteriors}, we show the posterior distributions of the $\eta_{3}$ (or $P_{\rm rot}$) hyperparameter for each time series. $\eta_{1}$ either converged to a non-zero amplitude or showed a well-constrained zero-truncated distribution, and both $\eta_{2}$ and $\eta_{4}$ always showed unconstrained posterior distributions. The $\eta_{3}$ distribution converged at $\eta_{\rm 3, H_{\alpha}}$ = $37.03^{+1.25}_{- 0.92}$  d for the $H_{\rm \alpha}$ time series, where we had previously identified a prominent \texttt{GLS} peak at 36.9~d (Sect.~\ref{sec:ESPRESSO_data}). In contrast, the other five indicators show unconstrained $\eta_{\rm 3}$ distributions. We note, however, over-density regions of compatible solutions at  $\eta_{\rm 3}$ $\simeq$ 37-38 d in the Na indicator, and at $\eta_{\rm 3}$ $\simeq$ 35-36 d in the FWHM indicator, which could be reflecting the same activity signal. Interestingly, this periodicity falls within the 1$\sigma$ confidence interval of the expected $P_{\rm rot}$ from the $\textrm{log}(R'_{\rm HK }$)-$P_{\rm rot}$ empirical relations from \citealt{2016A&A...595A..12S} (Sect.~\ref{subsec:rotation_proxies}), which further supports a stellar origin for the signal. We note that the log-evidences do not favour the activity model against the null hypothesis, with  $\Delta \mathcal{Z}$ ranging from -11 to -2. Getting negative evidence indicates that the considered activity model is too complex for the studied data sets. In particular, for the case of the $H_{\rm \alpha}$ indicator, where only $\eta_{\rm 1}$ and $\eta_{\rm 3}$ converged, the unconstrained $\eta_{\rm 2}$ and $\eta_{\rm 4}$ hyperparameters are useless and so can be safely discarded. 

We repeated the analysis by considering a simple sinusoid plus a linear trend as the activity model. The log-evidence of this model compared to the linear trend alone is  $\Delta \mathcal{Z}$ = +2.2, and its periodicity converges at $P_{\rm rot, H_{\alpha}}$ = $36.76^{+0.40}_{-0.44}$ d (i.e. compatible with  $\eta_{\rm 3, H_{\alpha}}$ within 1$\sigma$). We note that this log-evidence is not high enough to claim this $P_{\rm rot}$ as statistically significant, but being positive and near the moderate evidence threshold (i.e. $\Delta \mathcal{Z}$ $>$ 2.5) we consider it as a good candidate to be the true $P_{\rm rot}$, especially when also taking into account that it matches the empirical predictions obtained from the $\textrm{log}(R'_{\rm HK }$) index. We thus report this value in Table \ref{tab:stellar_prop} while including a cautionary footnote highlighting its still not significant Bayesian Evidence. In Fig.~\ref{fig:ha_timeseries} (left panel), we show the complete $H_{\alpha}$ time series together with the median posterior model. In the right panel, we show the $H_{\rm \alpha}$ data folded to the inferred $P_{\rm rot, H_{\alpha}}$.

We also tested the activity-unrelated nature of Candidate$\#$1 by taking into account its large RV amplitude (i.e. $K_{\#1}$ = 7.10 $\pm$ 0.49 $\rm m \, s^{-1}$). \citealt{2017MNRAS.468.4772S} (SM+17) used HARPS RVs of 37 stars to find that the $\textrm{log}(R'_{\rm HK }$) index and the rotation-induced RV semi-amplitude are correlated. In Fig.~\ref{fig:SM+17}, we contextualise Candidate$\#$1 within their sample and empirical relations. A GK star with a $\textrm{log}(R'_{\rm HK }$) of -5.08 is expected to generate a RV semi-amplitude of about 0.2 $\rm m\,s^{-1}$, which is well below the observed 7 $\rm m \, s^{-1}$ for Candidate$\#$1. We note that this prediction comes from an extrapolation, given that SM+17 relations do not contain stars with such a low $\textrm{log}(R'_{\rm HK }$), so the exact value should be taken with care. Still, stars with  $\textrm{log}(R'_{\rm HK }$) $\lessapprox$ -4.9 show semi-amplitudes $\lessapprox$ 1 $\rm m\,s^{-1}$, which are still well below the observed signal semi-amplitude. 

From the analyses presented above, we can safely claim that Candidate$\#$1 is not generated by the stellar rotation of the star, and instead, $P_{\rm rot, H_{\alpha}}$ = $36.76^{+0.40}_{-0.44}$ d positions as the most likely rotation period of K2-157. This, together with the blind-search analysis (Sect.~\ref{sec:blind_search}), allows us to confirm the planetary nature of Candidate$\#$1 ($P_{\rm orb}$ = 25.97~d) and Candidate$\#$2 ($P_{\rm orb}$ = 66.62~d), which we refer to as K2-157~c and K2-157~d, respectively. 

\subsection{Joint RV and transit analysis}
\label{sec:joint_fit}

\begin{figure*}
    \centering
    \includegraphics[width=0.99\textwidth]{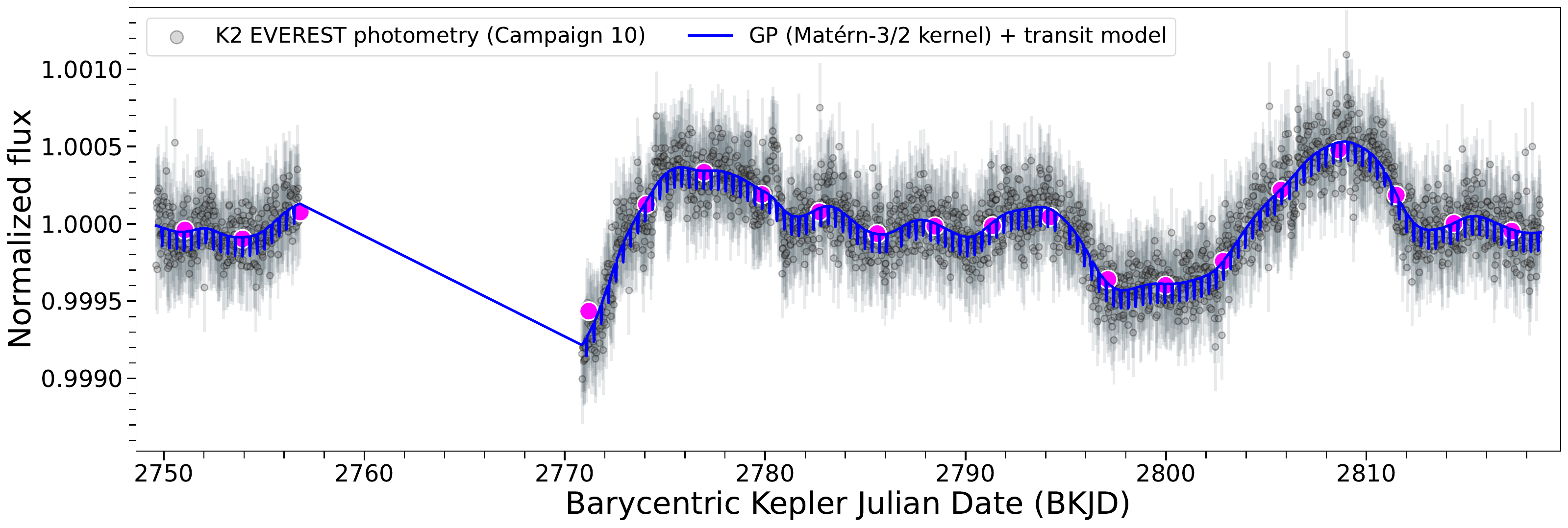}
    \caption{K2 EVEREST photometry of K2-157 (Campaign 10) together with the median posterior model (transit + GP) inferred in Sect.~\ref{sec:joint_fit}. The magenta data points correspond to bins of 3~d. BKJD equals Barycentric Julian Date (BJD) $-$ 2454833~d.}
    \label{fig:K2_C10}
\end{figure*}

We inferred the final parameters of the K2-157 system by jointly modelling the ESPRESSO RVs and K2 photometry through a joint RV (3p1c2c3cQ) and transit model.

\begin{figure*}
    \centering
    \includegraphics[width=0.47\textwidth]{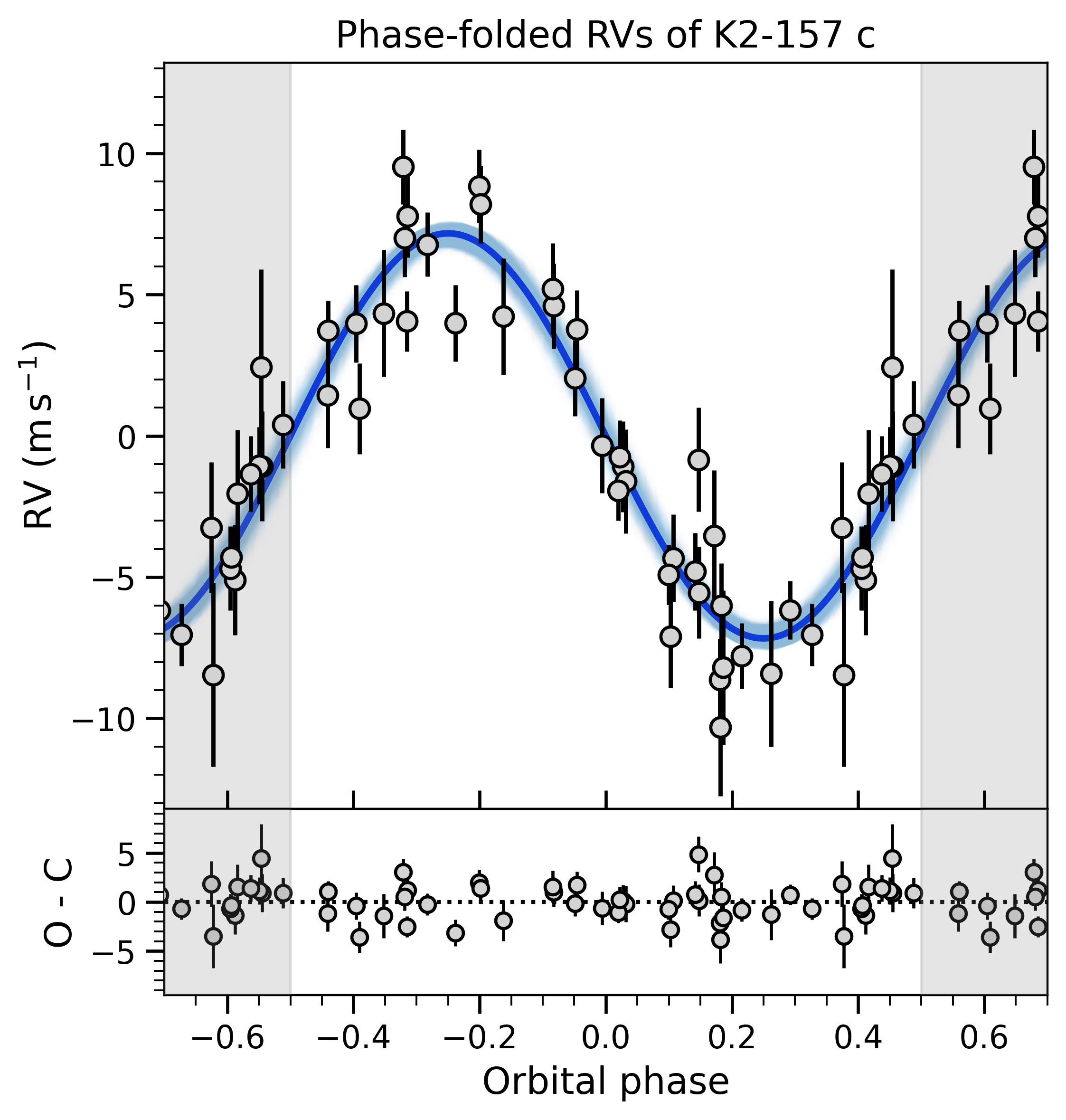}
    \includegraphics[width=0.47\textwidth]{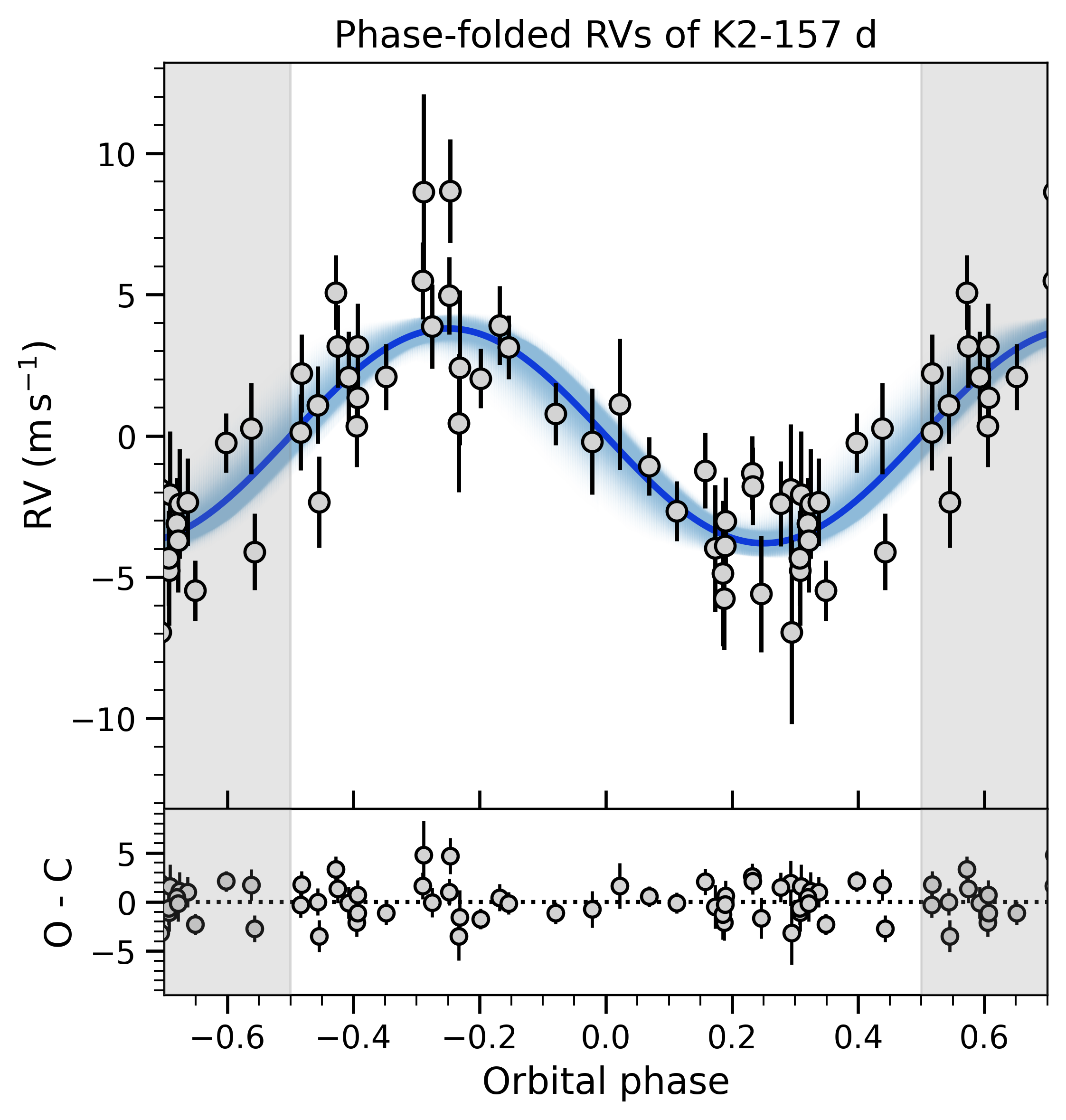}
    \caption{ESPRESSO RVs of K2-157~c (left) and K2-157 d (right) folded to their respective orbital periods inferred in Sect.~\ref{sec:joint_fit}. The solid blue lines indicate the median posterior models, and the shades indicate the 1$\sigma$ confidence intervals.}
    \label{fig:phase_K2-157cd}
\end{figure*}

\begin{figure*}
    \centering
    \includegraphics[width=0.474\textwidth]{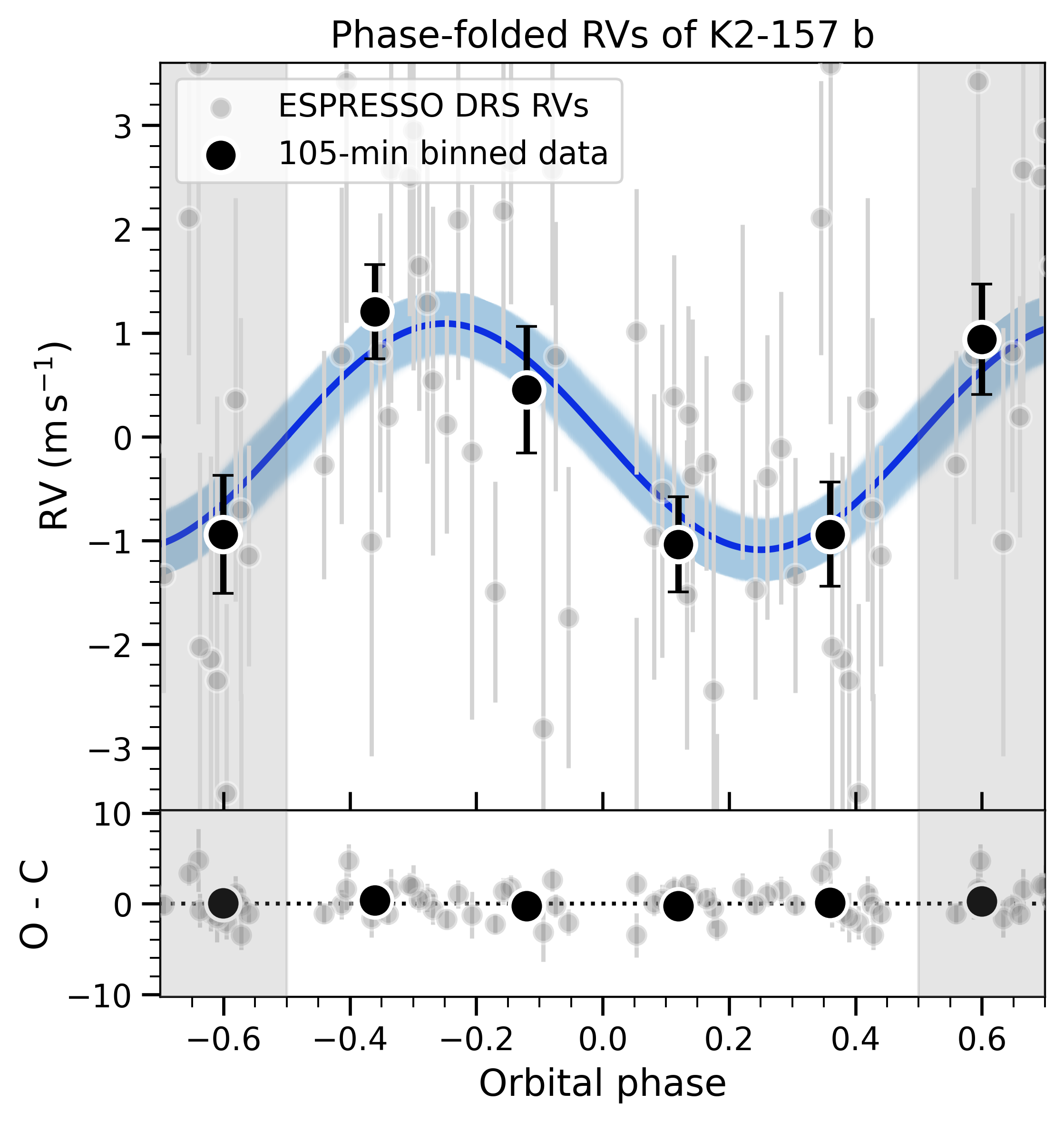}
    \includegraphics[width=0.479\textwidth]{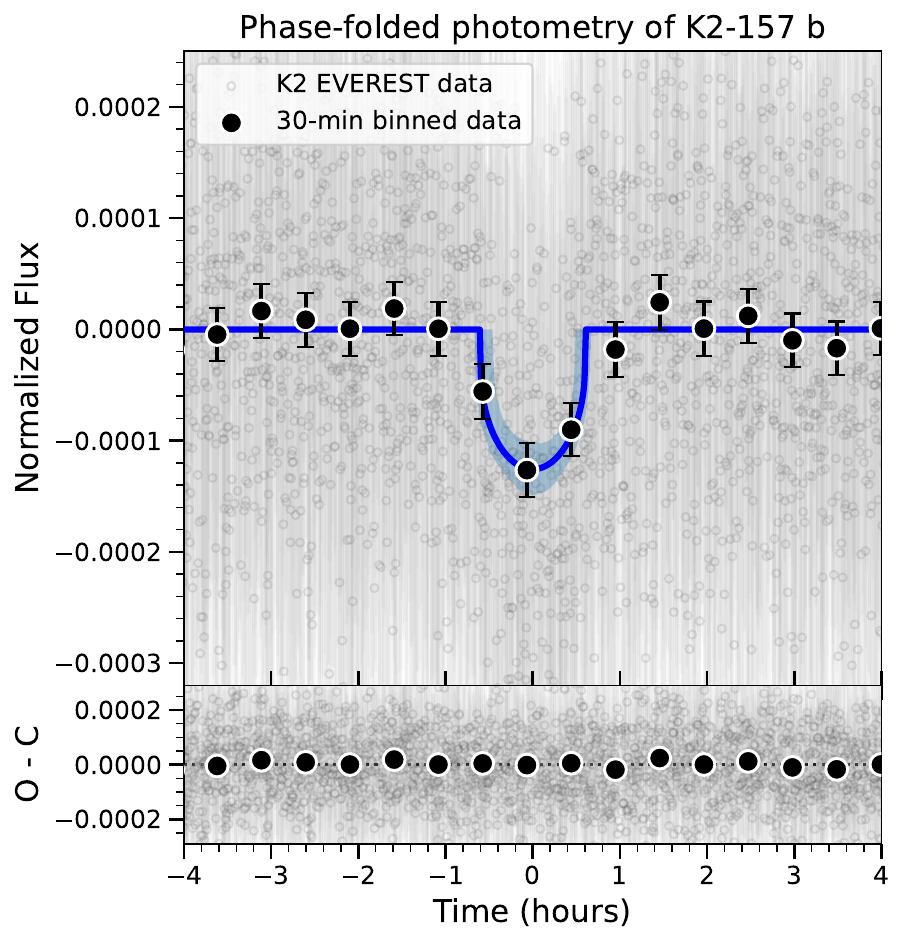}
    \caption{ESPRESSO RVs (left) and K2 photometry (right) of K2-157~b folded to its orbital period inferred in Sect.~\ref{sec:joint_fit}. In both panels, the solid blue line indicates the median posterior model, and the shade indicates the 1$\sigma$ confidence interval.}
    \label{fig:phase_K2-157b}
\end{figure*}

We used the quadratic limb-darkened transit model from \citet{2002ApJ...580L.171M} as implemented in \texttt{batman}\footnote{Available at \url{https://github.com/lkreidberg/batman}} \citep{2015PASP..127.1161K}. This model is defined by seven parameters: the orbital period of the transiting planet ($P_{\rm orb}$), its time of inferior conjunction (or time of mid-transit, $T_{0}$), the inclination of the planetary orbit ($i$), the planet-to-star radius ratio ($R_{\rm p} / R_{\star}$), the quadratic limb darkening (LD) coefficients $u_{1}$ and $u_{2}$, and the semimajor axis scaled to the stellar radius ($a / R_{\star}$). We parametrized the LD parameters through the prescription for effective uninformative sampling proposed by \citet{2013MNRAS.435.2152K}: $q_{1}$ = $\left(u_{1} + u_{2} \right)^{2}$, and  $q_{2}$ = 0.5 $u_{1} \left(u_{1} + u_{2} \right)^{-1}$, and also parametrized $a / R_{\star}$ through the stellar
mass ($M_{\star}$) and radius ($R_{\star}$) in order to
better constrain the model parameters \citep[][]{2007ApJ...664.1190S}.

As we can see in Fig.~\ref{fig:K2_C10}, K2 data is affected by considerable correlated noise provoking low-frequency trends, which is a well-known behaviour caused by the spacecraft drift (potentially mixed with some degree of stellar variability). In order to better preserve the shape and depth of the transits, we modelled jointly the transit signal and correlated noise through a GP. This procedure has been proven successful to model shallow transit signals \citep[e.g.][]{2022MNRAS.515.1328D,2023A&A...679A..33D} and ensures a proper propagation of the uncertainties of the model parameters \citep[e.g.][]{2021A&A...649A..26L}. We chose a GP defined by a Matérn-3/2 kernel, which, given its flexibility, is one of the preferred kernels to model photometric satellite data with an unknown mixture of instrumental and stellar noise \citep[e.g.][]{2021A&A...656A.124K,2022A&A...658A.138G,2023A&A...677A.182M,2023A&A...673A..32M,2024A&A...692A.238L}. This kernel can be written as:

\begin{equation}
    k(x_{i}, x_{j}) = \eta_{\sigma}^{2} \left[ \left(1 + \frac{1}{\epsilon} \right) e^{-(1-\epsilon) \sqrt{3} \tau / \eta_{\rho}} \cdot \left(1 - \frac{1}{\epsilon} \right) e^{-(1+\epsilon) \sqrt{3} \tau / \eta_{\rho}} \right],
\end{equation}where $\tau = x_{i} - x_{j}$ is the temporal separation between two time stamps, and $\eta_{\sigma}$ and $\eta_{\rho}$ are two hyperparameters that represent the characteristic amplitude and timescale of the correlated variations, respectively. The parameter $\epsilon$ controls the approximation to the exact kernel, which we fixed to its default value of $10^{-2}$ \citep{2017AJ....154..220F}.

We ran the joint fit analysis through the procedure described in Sect.~\ref{sec:model_inference} and using the priors specified in Table~\ref{tab:parameters_joint}. Most priors are uniform (either uninformed or poorly informed), except the stellar mass and radius, for which we used Gaussian priors based on the characterisation described in Sect.~\ref{sec:stellar_charact}. For the quadratic LD coefficients, we tested unconstrained uniform priors, and also constrained Gaussian priors based on the \texttt{ldtk} package\footnote{Available at \url{https://github.com/hpparvi/ldtk}} \citep{2015MNRAS.453.3821P}. This package uses the spectroscopic atmospheric parameters $T_{\rm eff}$, log $g$, and [Fe/H], together with the instrumental transmission curve (\textit{Kepler} in this case) to infer the coefficients of any LD law (quadratic in this case) based on the synthetic spectra library from \citet{2013A&A...553A...6H}. We obtain $u_{1}$ = $0.5672 \pm 0.0015$ and $u_{2}$ = $0.1154 \pm 0.0022$ (i.e. $q_{1}$ = $0.4660 \pm 0.0037$ and $q_{2}$ = $0.4154 \pm 0.0014$). We note that the uncertainties of theoretical LD coefficients have been commonly found to be misestimated \citep[e.g.][]{2022AJ....163..228P}, which can induce biases in the transit modelling. To try to avoid this situation, we performed three different modellings: considering the formal \texttt{ldtk} uncertainties, arbitrarily and conservatively enlarging them up to 0.3, and uniformly sampling the parameter space with $\mathcal{U} (0,1)$ priors. We find that in the unconstrained case, the posterior $q_{1}$ and $q_{2}$ distributions remain quite unconstrained, and that in the more constrained scenarios, the resulting planetary parameters are compatible with the unconstrained case at the 1$\sigma$ level. Therefore, we can conclude that the a priori constraining level of the LD coefficients does not have an impact on the characterisation of this system, and we thus arbitrarily chose to report the unconstrained case. In Table~\ref{tab:parameters_joint}, we show the median and $1\sigma$ intervals of the posterior parameter distributions.

We draw three conclusions from the results of the joint analysis. First, the RV semi-amplitudes of planets K2-157~c and K2-157~d are compatible at 1$\sigma$ with those inferred from the blind-search analysis. Second, our derived orbital and physical parameters for K2-157~b are also compatible with those from previous K2-based works. In particular, the planet-to-star radius ratio ($R_{\rm p, b}/ R_{\star}$ = 
$0.00996 \pm 0.00094$) agrees with that of \citealt{2018AJ....155..136M} ($R_{\rm p, b}/ R_{\star}$ = 
$0.0111^{+0.0014}_{-0.0010}$) with a t-statistic of 0.83 and with that of \citealt{2021PSJ.....2..152A} ($R_{\rm p, b}/ R_{\star}$ = 
$0.0107^{+0.0011}_{-0.0006}$) with a t-statistic of 0.66. The third conclusion is that we were able to measure the RV semi-amplitude of K2-157~b at a $\simeq$ 2.7$\sigma$ level, $K_{\rm b} = 1.10^{+0.39}_{-0.41}$ $\rm m \, s^{-1}$, which translates into a mass of $M_{\rm p, b}$ = $1.14^{+0.41}_{-0.42}$ $\rm M_{\oplus}$ ($<$ 2.4 $\rm M_{\oplus}$ at 3$\sigma$)\footnote{We note that some authors opt to report the 3$\sigma$ mass upper limit when the semi-amplitude significance is below 3$\sigma$ \citep[e.g.][]{2023A&A...677A..33B,2023MNRAS.523.3069O,2023Natur.623..932L,2024A&A...685A..19C}, while others prefer to report the median and 1$\sigma$ confidence intervals for significances about $\simeq$2-3$\sigma$  \citep[e.g.][]{2013ApJ...768...14W,2022A&A...665A.154B,2023Natur.617..701P,2024ApJS..272...32P}. For K2-157~b, given the closeness to the 3$\sigma$ threshold, we decide to report both the $1\sigma$ uncertainties and the upper limit (in brackets).}.

In Fig.~\ref{fig:phase_K2-157cd}, we show the ESPRESSO RVs caused by K2-157~c and K2-157~d subtracted from the other components in the model and folded to their inferred orbital periods. In Fig.~\ref{fig:K2_C10}, we show the C10 K2 light curve together with the transit + GP model for K2-157 b. In Fig.~\ref{fig:phase_K2-157b}, we show the ESPRESSO RVs and K2 photometry folded to the orbital period of K2-157 b. In Fig.~\ref{fig:gls_residuals}, we show the periodogram of the RV residuals. In Fig.~\ref{fig:corner_joint}, we show a corner plot with the posterior distributions of the fit parameters. 

Finally, while not preferred by the RV data (Sect.~\ref{sec:blind_search}), we performed an additional joint analysis by letting the orbital eccentricities vary uniformly between 0 and 1 to infer the upper limits compatible with our data set. As a result, the posterior of $e_{\rm b}$ resulted in a quasi-flat, unconstrained distribution. However, from the posteriors of  $e_{\rm c}$ and $e_{\rm d}$, truncated at zero, we were able to derive 3$\sigma$ upper limits of $e_{\rm c}$ $<$ 0.2 and $e_{\rm d}$ $<$ 0.5. Regarding the remaining parameters, we find values compatible at 1$\sigma$ with those from the circular model. These eccentricities are poorly constrained, but still allow us to discard highly eccentric orbits for K2-157~c and K2-157~d. In Sect.~\ref{sec:alexandre}, we conduct an orbital stability analysis of the system to try to better constrain the eccentricities of these planets.

\subsection{Possible transits and inclinations of K2-157 c and d}
\label{sec:transits_inclinations}

We explored the possibility that K2-157 c and K2-157 d transit their host star and studied the range of orbital inclinations compatible with the K2 data. To do so, we first considered the measured minimum masses and used the empirical mass-radius relations from \citet{2017ApJ...834...17C} to estimate the minimum radii for both planets, $R_{\rm p, c}$ = $6.03^{+2.48}_{-1.78}$ $\rm R_{\oplus}$ and $R_{\rm p, d}$ = $5.21^{+1.94}_{-1.56}$ $\rm R_{\oplus}$, which correspond to minimum transit depths of $\delta_{\rm c}$ = 4.1 $\pm$ 2.4 ppt and  $\delta_{\rm d}$ = 3.1 $\pm$ 1.9 ppt. These transit signals would have been clearly detected both within K2 and TESS data (see Sects.~\ref{sec:K2_data} and \ref{sec:TESS_data}). The non-detections can be explained through three possible scenarios: 1) the planets transit but the transit times do not fall within the photometric temporal baseline, 2) the planets transit with a grazing configuration that reduces the observed transit depth, or 3) the planets do not transit at all. We tested scenario 1 by propagating the RV-derived ephemeris to the K2 and TESS observing windows. In Fig.~\ref{fig:tr_simul}, we show the expected transit locations with their associated 1$\sigma$ uncertainties. We see that at least one transit of K2-157~c would have been observed in the K2 data. In contrast, the larger periodicity and ephemeris uncertainty of K2-157~d prevent us from determining whether it would have been observed to transit within this data set. We tested scenarios 2 and 3 for K2-157~c by repeating the joint fit from Sect.~\ref{sec:joint_fit} but including a transit component for this planet\footnote{We did not use TESS data for this analysis, given the much lower constraining capacity than K2's.}. We set a uniform prior for the orbital inclination $\mathcal{U_{\rm c}}$(0,90) and a Gaussian prior for the minimum mass, which we translated into $R_{\rm p} / R_{\star}$ through the sampled inclinations and the mass-radius relation. We highlight that this approach accounts for the uncertainties in our mass determination and the scatter in the empirical relations, and it also considers the dependency of the true mass on the orbital inclination. The posterior distribution results in a 3$\sigma$ upper limit of $i_{\rm c}$ $<$ 88.4$^{\circ}$, which translates into a 3$\sigma$ lower limit for the impact parameter of $b_{\rm c}$ $>$ 1.15. We note that this value corresponds to a configuration where the planet is located outside the sky-projected stellar disc, and, in order to transit, it requires a planetary radius of at least $\simeq$14~$\rm R_{\oplus}$, which coincides with the 3$\sigma$ upper limit of the minimum radius estimated for K2-157~c.

\subsection{Sensitivity limits and constraints on additional planets}
\label{sec:sensitivity_limits}

We computed the sensitivity limits of the ESPRESSO data by following the procedure described in \cite{Standing2022} as used in previous works
\citep{Sairam2022,Grieves2022,Standing2023,Baycroft2023,John2023,Sairam2024,Balsalobre-Ruza2025}. 
We began by running \texttt{kima}\footnote{Available at \url{https://github.com/kima-org/kima}} \citep{Faria2018} on the ESPRESSO RVs after removing the signals of K2-157~b, K2-157~c, and K2-157~d. \texttt{kima} utilises a diffusive nested sampling algorithm \citep{Brewer2011} and allows the number of planetary signals ($N_p$) to be fit as a free parameter. This initial run ensures that there are no remaining signals in the residual data set. 

\begin{figure}
    \centering
    \includegraphics[width=0.48\textwidth]{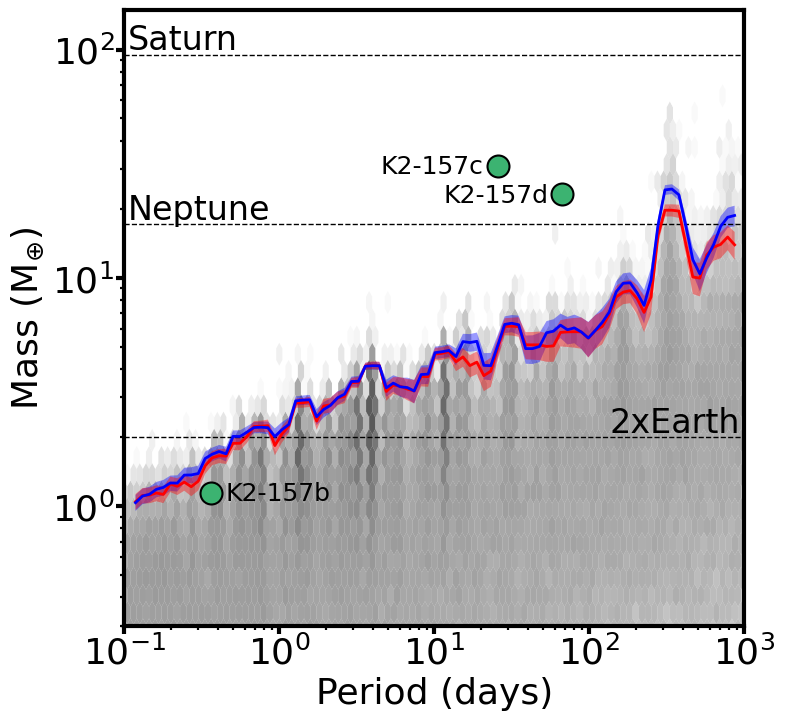}
    \caption{Hexbin plot of posterior samples obtained from \texttt{kima} runs on the residual RV data with $N_p$ fixed to 1. The blue line shows the 3$\sigma$ detection limit, whereas the red line shows the same limit computed on a subset of posterior samples with eccentricity $<0.1$. The uncertainties on these lines are illustrated by the faded lines of the associated colour.
}
\label{fig:senslim}
\end{figure}

We ran \texttt{kima} again with the number of planetary signals fixed to one. The posterior samples obtained from this run are compatible with the data, but their signals are not formally detected. For our analysis, we obtained over $520\,000$ samples. The 3$\sigma$ upper limit is then calculated in period bins, yielding the detection limit, which can be seen as the blue line in Fig.~\ref{fig:senslim}. The uncertainty on this limit is calculated as described in \cite{Standing2023}.
Any planetary signal above the blue detection limit would have been detected in the blind search on the ESPRESSO data. This was the case for planets K2-157 c and K2-157~d, which are well above the detection limit in Fig.~\ref{fig:senslim}. The signal expected from K2-157~b is below the detection limit, which explains why the blind search was unable to detect it. Overall, from Fig.~\ref{fig:senslim} we conclude that we can discard the presence of planets with $M_{\rm p}$ $\gtrsim$ 2 $\rm M_{\oplus}$ and $P_{\rm orb}$ $\lesssim$ 1 d, $M_{\rm p}$ $\gtrsim$ 4 $\rm M_{\oplus}$ and $P_{\rm orb}$ $\lesssim$ 10~d, and $M_{\rm p}$ $\gtrsim$ 10~$\rm M_{\oplus}$ and $P_{\rm orb}$ $\lesssim$ 200 d.

\section{Discussion}
\label{sec:discussion}

The mass determination ($M_{\rm p, b}$ = $1.14^{+0.41}_{-0.42}$ $\rm M_{\oplus}$; $<$ 2.4 $\rm M_{\oplus}$ at 3$\sigma$) and subsequent RV confirmation of the USP Earth-sized planet K2-157~b ($P_{\rm orb, b}$ $\simeq$ 8.8 h; $R_{\rm p, b}$ = 0.935 $\pm$ 0.090 $\rm R_{\oplus}$), together with the detection of two additional warm super-Neptune-mass planets, can provide insightful clues into the formation and evolution of this planetary system. USP planets such as K2-157~b are scarce in exoplanet catalogues, and those with relatively massive and warm companions are even rarer, which positions K2-157 as a relevant USP-hosting system to test planet formation and evolution theories. In Sect.~\ref{sec:K2-157b}, we contextualise and discuss the main properties of K2-157~b in the known population of USP planets (i.e. within the period-radius and mass-radius parameter spaces), and explore the recently identified dependency between the semi-major axis of USP planets and stellar spectral type. In Sect.~\ref{K2-157}, we contextualise K2-157~c and K2-157~d in the period-mass diagram of known neighbours of USP planets and discuss their unusual location in this region of the parameter space. In Sect.~\ref{sec:comparison_theories}, we compare the fundamental properties of K2-157 to different formation and evolution theories of USP planets. 

\subsection{K2-157~b in context}
\label{sec:K2-157b}

\begin{figure*}
    \centering
    \includegraphics[width=\textwidth]{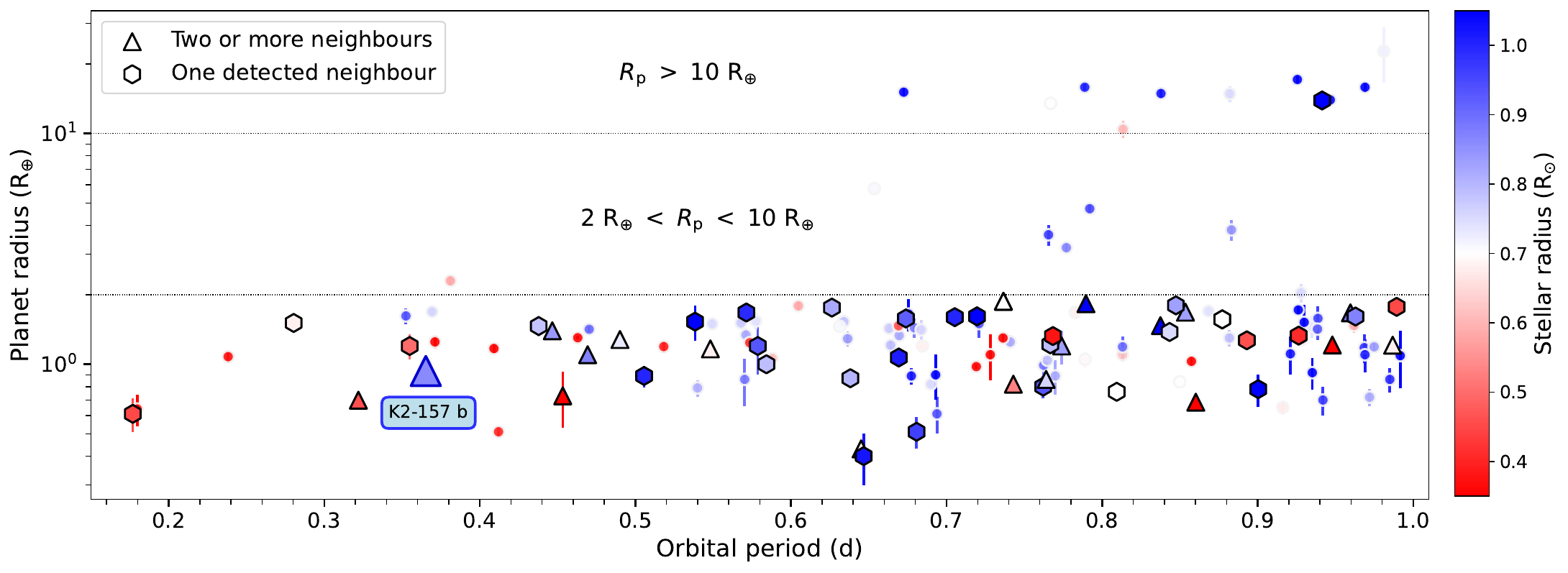}
    \caption{Period-radius diagram of all known USP planets with radii constrained to a precision better than 30$\%$ (source: NASA Exoplanet Archive, accessed on 24/03/2025). The colour code indicates the radii of the stellar hosts, with the blueish and reddish tones indicating radii above and below $R_{\star}$ $\simeq$ $0.7$ $\rm R_{\odot}$, respectively. USP planets with one detected neighbour are highlighted with hexagonal symbols, while those with two or more neighbours are shown with triangular symbols. }
    \label{fig:R_vs_P}
\end{figure*}

\subsubsection{Period-radius diagram}
\label{sec:P_R_diagram}

In Fig.~\ref{fig:R_vs_P}, we show the period-radius diagram of USP planets with radii constrained to precisions better than 30$\%$. Among the 135 planets composing the sample, 87$\%$ (i.e. 117 planets) show radii lower than 2 $\rm R_{\oplus}$, while the remaining 13$\%$ (i.e. 18 planets) show a wide range of larger radii (i.e. 2  $\rm R_{\oplus}$ $<$ $R_{p}$ $<$ 23 $\rm R_{\oplus}$). Interestingly, terrestrial USP planets ($R_{p}$ $<$ 2~$\rm R_{\oplus}$) have been detected with orbital periods as short as 0.18~d \citep[4.3~h;][]{2013ApJ...773L..15R,2018MNRAS.474.5523S}, while larger USP planets have been only found with periods as short as 0.65 d \citep[15.6~h;][]{2016ApJ...822...86M,2021AJ....162..256W}. In this context, having an orbital period of 0.365 d (8.8 h), K2-157~b is the ninth shortest-period transiting planet known to date\footnote{In Fig.~\ref{fig:R_vs_P}, K2-157~b appears to be the eighth shortest-period planet since EPIC 206042996~c and Kepler-78 b are perfectly overlaid with orbital periods of 0.355 d and radii of 1.20 $\rm R_{\oplus}$.}, only after KOI-1843.03 \citep{2013ApJ...773L..15R}, K2-137~b \citep{2018MNRAS.474.5523S}, TOI-6255~b \citep{2024AJ....168..101D}, K2-141~b \citep{2018A&A...612A..95B,2018AJ....155..107M}, GJ 367~b \citep{2021Sci...374.1271L}, TOI-2260~b \citep{2022AJ....163...99G}, EPIC 206042996~c \citep{2021PSJ.....2..152A}, and Kepler-78~b \citep{2013ApJ...774...54S}. Among these nine planets, four have been previously found to have additional planetary companions \citep{2016ApJ...822...86M,2018AJ....155..107M,2019A&A...627A..66H,2023A&A...677A..33B,2023ApJ...955L...3G}. The discovery of K2-157~c and K2-157~d thus implies that more than half the shortest-period planets (five out of nine) have neighbours. Among these systems, only GJ~367~b is known to have more than one neighbour \citep{2023ApJ...955L...3G}, so K2-157~b becomes the second shortest-period planet known to have at least two neighbours. In the terrestrial region ($R_{p}$ $<$ 2 $\rm R_{\oplus}$), 40$\%$ of USP planets (41$\%$ accounting for K2-157~b) have neighbours, while only 6~$\%$ (i.e. WASP-18~c; \citealt{2019AJ....158..243P}) of larger planets ($R_{p}$ $>$ 2~$\rm R_{\oplus}$) are known to have neighbours\footnote{In addition, WASP-18~c is flagged as a `controversial' planet in the NASA Exoplanet Archive \citep{2013PASP..125..989A,2022NatAs...6..516C}.}. Therefore, from a simple inspection of Fig.~\ref{fig:R_vs_P}, we can infer the existence of at least two classes of USP planets with a well-defined division line at $R_{\rm p}$ $\simeq$ 2 $\rm R_{\oplus}$. In addition to the different occurrence rates, orbital distribution, and fraction of systems with neighbours, small USP planets are known to be hosted by solar-metallicity stars, while hot Neptunes and hot Jupiters prefer higher metallicity hosts \citep[see][for a comparative study of USP planets]{2021AJ....162...62D}. Given these differences, which suggest well-differentiated formation or evolution pathways, we focus our discussion on USP planets with $R_{\rm p}$ $<$ 2 $\rm R_{\oplus}$ (i.e. the population to which K2-157~b belongs).

\begin{figure}
    \centering
    \includegraphics[width=0.48\textwidth]{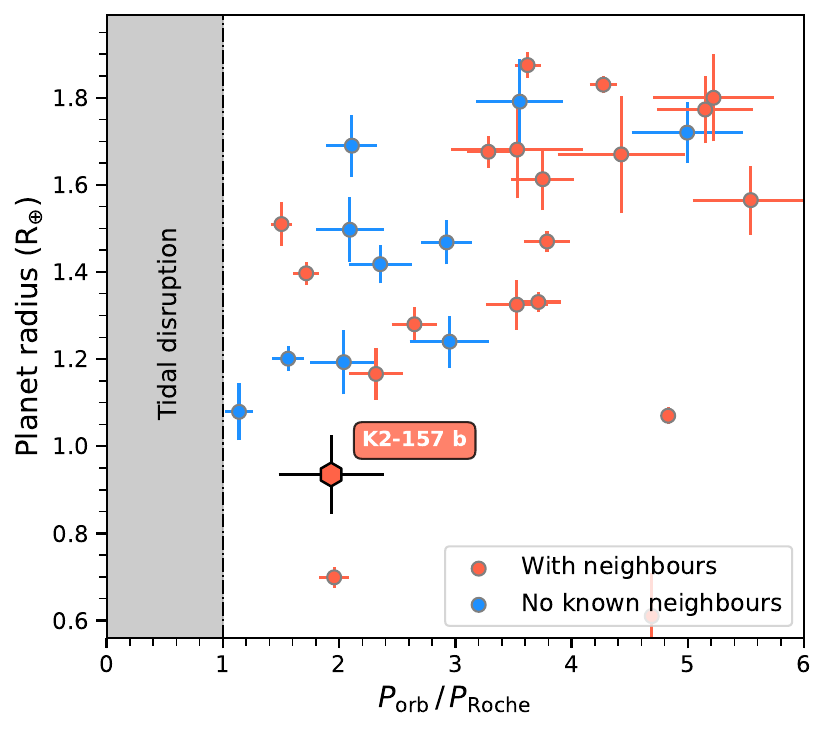}
    \caption{Planet radius versus orbital period scaled to the Roche limit for USP rocky planets. The vertical line indicates the tidal disruption limit. Data: NASA Exoplanet Archive (24/03/2025).}
    \label{fig:R_vs_Roche}
\end{figure}

The ultra-short orbital period of K2-157~b may place it near the Roche limit; that is, the star-planet separation where tidal forces exerted by the star are stronger than the planet's gravity. Below this threshold, a planet will start to disintegrate if the material strength is negligible. Following \citet{2013ApJ...773L..15R}, we estimated the orbital period of the Roche limit to be\footnote{This equation is only valid for bodies in circular orbits composed of
incompressible fluids with negligible bulk tensile strength. However, it serves as a good approximation for planets composed of iron and silicates since they are not highly compressible. According to \citet{1993ApJS...88..205L} and \citet{2013ApJ...773L..15R}, an extra multiplicative factor $(\rho_{0, \rm p}  / \rho_{\rm p})^{-0.16}$ should be included in the expression to account for this effect. $\rho_{0, \rm p}$ is the central density of the planet and cannot be easily inferred, but the $(\rho_{0, \rm p}  / \rho_{\rm p})^{-0.16}$ factor is only expected to vary from $\simeq$~0.85 to 1 for rocky planets \citep{2013ApJ...773L..15R}, so we can neglect it.}
\begin{equation} 
    P_{\rm Roche}  \approx 12.6 \, \textrm{h} \left( \frac{\rho_{\rm p}}{1 \textrm{g} \, \textrm{cm}^{-3}}  \right)^{-\frac{1}{2}}.
    \label{ec:roche}
\end{equation}
We obtain $P_{\rm Roche}$ = 4.5 $\pm$ 1.1 h, so that the orbital period of K2-157~b is approximately twice that of the tidal disruption limit, $P_{\rm orb}$/$P_{\rm Roche}$ = 1.93 $\pm$ 0.45. Interestingly, the tidal love theory \citep{love:hal-01307751} predicts that rocky planets with $P_{\rm orb}$/$P_{\rm Roche}$ $\simeq$ 2 such as K2-157~b may undergo a certain degree of tidal distortion, with a long axis of about $\simeq$ 2$\%$ longer than its short axis \citep[see][]{1980esvr.book.....L,2014A&A...570L...5C,2024AJ....168..101D}, which would decrease its bulk density by $\simeq$ 6$\%$ \citep[e.g.][]{1995geph.conf....1Y,2014A&A...570L...5C}.  We applied Eq.~(\ref{ec:roche}) to the sample of known USP rocky planets with measured masses and radii, and find that K2-157~b has one of the lowest $P_{\rm orb}$/$P_{\rm Roche}$ ratios, making it an excellent target to search for signatures of orbital decay (see Fig.~\ref{fig:R_vs_Roche}). Unfortunately, the long cadence of the K2 light curve combined with the low S/N of the individual transits and short time span of the observations prevent a meaningful search for orbital decay.

Going back to the period-radius parameter space, in Fig.~\ref{fig:R_vs_P} we colour the planets according to the stellar radii to illustrate an interesting trend that emerged over the last years: the shortest-period USP planets tend to orbit late-type stars \citep[e.g.][]{2013A&A...555A..58O,2013ApJ...773L..15R,2018MNRAS.474.5523S,2018A&A...612A..95B,2021PSJ.....2..152A,2021Sci...374.1271L,2024AJ....168..101D}. Interestingly, K2-157, with a stellar radius of $R_{\star}$ = 0.860 $\pm$ 0.019~$\rm R_{\odot}$, apparently contrasts with this trend. Other similar early-type stars hosting particularly close-in USP planets are K2-141~b \citep[$P_{\rm orb}$ = 0.28 d, $R_{\star}$ = 0.68 $\rm R_{\odot}$;][]{2023A&A...677A..33B}, TOI-2260~b \citep[$P_{\rm orb}$ = 0.35 d, $R_{\star}$ = 0.94 $\rm R_{\odot}$;][]{2022AJ....163...99G}, and Kepler-78~b \citep[$P_{\rm orb}$ = 0.36 d, $R_{\star}$ = 0.75~$\rm R_{\odot}$;][]{2023A&A...677A..33B}. In the following, we study this trend and contextualise K2-157~b in this parameter space.

\subsubsection{The semi-major axis dependency on spectral type}

\begin{figure*}
    \centering
    \includegraphics[width=0.48\textwidth]{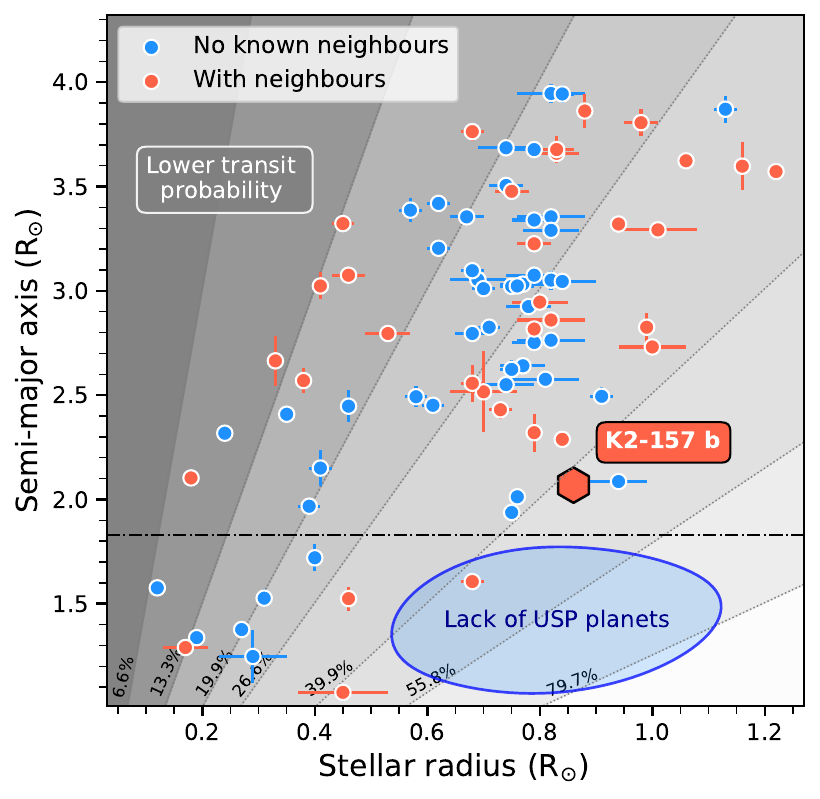}
\includegraphics[width=\columnwidth]{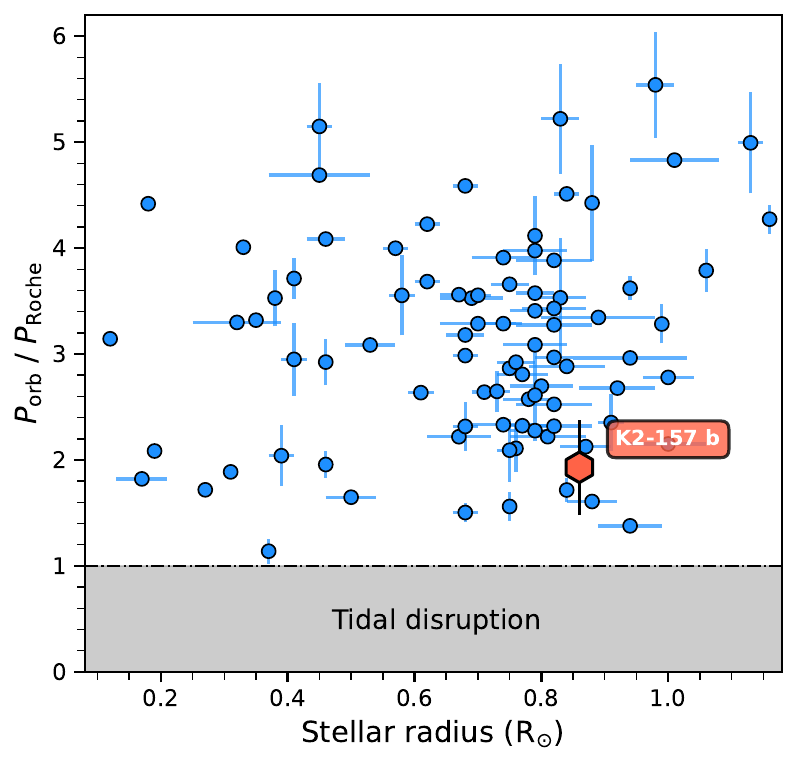}

    \caption{Left: Semi-major axis versus stellar radius of small USP planets ($R_{\rm p}$ $<$ 2 $\rm R_{\oplus}$), where we highlight a lack of planets that cannot be explained through observational bias, $a\lessapprox 0.0085$ au (1.8 $\rm R_{\odot}$) and $R_{\star}$ $\gtrapprox$ 0.5 $\rm R_{\odot}$. The dotted grey lines trace iso-probability transit regions. Right: Orbital period scaled to the Roche limit for USP rocky planets versus the stellar radius of their host stars, where we can appreciate how the scaled orbital locations do not depend on the stellar spectral types. The horizontal line indicates the tidal disruption limit. Data: NASA Exoplanet Archive (24/03/2025).}

    \label{fig:a_vs_Rs_main}
\end{figure*}

\citet{2021PSJ.....2..152A} found that the semi-major axis of USP planets and the radius of their host stars are linearly correlated. We here aim to contextualize K2-157~b in the $a-R_{\rm \star}$ space, revisit the correlation by considering the current planet sample, study whether it still holds when restricting to the small ($R_{\rm p}$ $<$ 2 $\rm R_{\oplus}$) USP planet population, try to quantify how much it is affected by observational bias, and discuss its possible origin. 

We first used a Markov chain Monte Carlo (MCMC) ensemble sample \citep{2010CAMCS...5...65G} as implemented in \texttt{emcee}\footnote{Available at \url{https://github.com/dfm/emcee}} \citep{2013PASP..125..306F} to sample the posterior probability density function of the coefficients A and B of a linear model, $y$ = $A$$x$+$B$\footnote{$y$ corresponds to the semi-major axis and $x$ to the stellar radius. $A$ has units of $\rm R_{\odot}^{-1}$ and $B$ has units of au.}. For the complete USP sample, we obtain $A = 0.00918^{+0.00088}_{-0.00091}$ and $B = 0.00692^{+0.00068}_{-0.00064}$. For the small USP planet sample ($R_{\rm p}$ $<$ 2 $\rm R_{\oplus}$), we obtain $A = 0.00907^{+0.00098}_{-0.00102}$ and $B = 0.00682^{+0.00073}_{-0.00068}$. The inferred slopes are compatible and differ from zero at the 10.3$\sigma$ and 9.1$\sigma$ levels. Therefore, we conclude that the $a-R_{\star}$ correlation holds for small USP planets. In Fig.~\ref{fig:a_vs_Rs}, we show the $a-R_{\star}$ distribution of the two samples together with a set of posterior models. We also tested separately fitting the samples of USP planets with and without neighbours, and found no statistical differences. 

We aim to examine whether the $a-R_{\star}$ correlation for small USP planets reflects a true feature of the exoplanet distribution, or if it could be driven by observational bias. Most USP planets have been detected through space-based photometers, so they are mostly affected by transit bias. Non-transiting configurations are a major bias of transit surveys, and it is particularly relevant for the USP population given their short orbital distances in comparison with the stellar radii. In Fig.~\ref{fig:a_vs_Rs_main} (left panel), we plot the $a-R_{\star}$ distribution of rocky USP planets together with different transit iso-probability lines. While there are several USP detections in the lower-left region of the diagram, where the transit probabilities range from $\simeq$7$\%$ to $\simeq$$30\%$, there is a clear scarcity of USP planets in the lower-right region, where the probabilities range from $\simeq$$40\%$ to $\simeq$$80\%$. We note that the transit S/N around larger hosts are lower. However, this can be arguably discarded as the cause of this lack of planets, given the large number of detections at larger orbital separations, which, in addition, statistically generate lower S/N signals given the lower number of detectable transit events. Therefore, we conclude that the lack of planets in the highlighted region of the left panel of Fig.~\ref{fig:a_vs_Rs_main},  $a\lessapprox 0.0085$~au (1.8~$\rm R_{\odot}$) and $R_{\star}$ $\gtrapprox$ 0.5~$\rm R_{\odot}$, cannot be explained through observational bias, thus requiring a physical explanation. On the opposite side of the diagram (i.e. upper-left region), there is also a lack of planets. This region has a low transit probability when compared to the whole USP population (i.e. $<$ 10~$\%$), and more distant planets produce lower S/N transit signals, so we cannot directly assess whether this lack of planets reflects a true property of the exoplanet distribution. A more comprehensive study of planet occurrences would be necessary.

We studied whether tidal disruption could instead determine the closest orbits of USP planets across stellar types. The Roche limit scales linearly with the stellar radius, so it is reasonable to think that the closest USP planets orbit at larger orbital distances around earlier-type stars. We computed $P_{\rm Roche}$\footnote{We use $P_{\rm Roche}$ instead of $a_{\rm Roche}$ since it allows us to decrease the error budget by cancelling out the stellar mass and radius, but we note that exchanging these magnitudes has no effect in this analysis.} (Eq.~\ref{ec:roche}) for the known planet population to test whether the $a-R_{\rm \star}$ trend holds after accounting for the minimum possible orbital distances in different stellar regimes. As we can see in Fig.~\ref{fig:a_vs_Rs_main} (right panel), the trend does not hold. There is not a lack of small  $P_{\rm orb}$/$P_{\rm Roche}$ ratios in early-type stars when compared to late-type stars, and instead we identify several ratios as short as $P_{\rm orb}$/$P_{\rm Roche}$ $\sim$ 1.5$-$2 distributed across a wide range of stellar radii, from $\sim$ 0.2 $\rm R_{\rm \odot}$ to $\sim$ 1~$\rm R_{\rm \odot}$. Hence, we interpret the $a-R_{\rm \star}$ trend in the shortest orbits of rocky planets as a direct consequence of the dependency between the Roche limit and the stellar radius. 

In all, the analyses presented above suggest that the distribution of the closest USP planets is not necessarily determined by the inner edges of the original proto-planetary discs, but simply by the present-day separation from the Roche limit, regardless of the spectral type. This aligns very well with the main formation theories of USP planets, which propose that their present-day orbits are primarily determined by their original migration triggered by interactions with outer companions and a subsequent orbital shrinkage that takes place before their final disruption. 

\subsubsection{Mass-radius diagram}

\begin{figure*}
    \centering
    \includegraphics[width=0.47\textwidth]{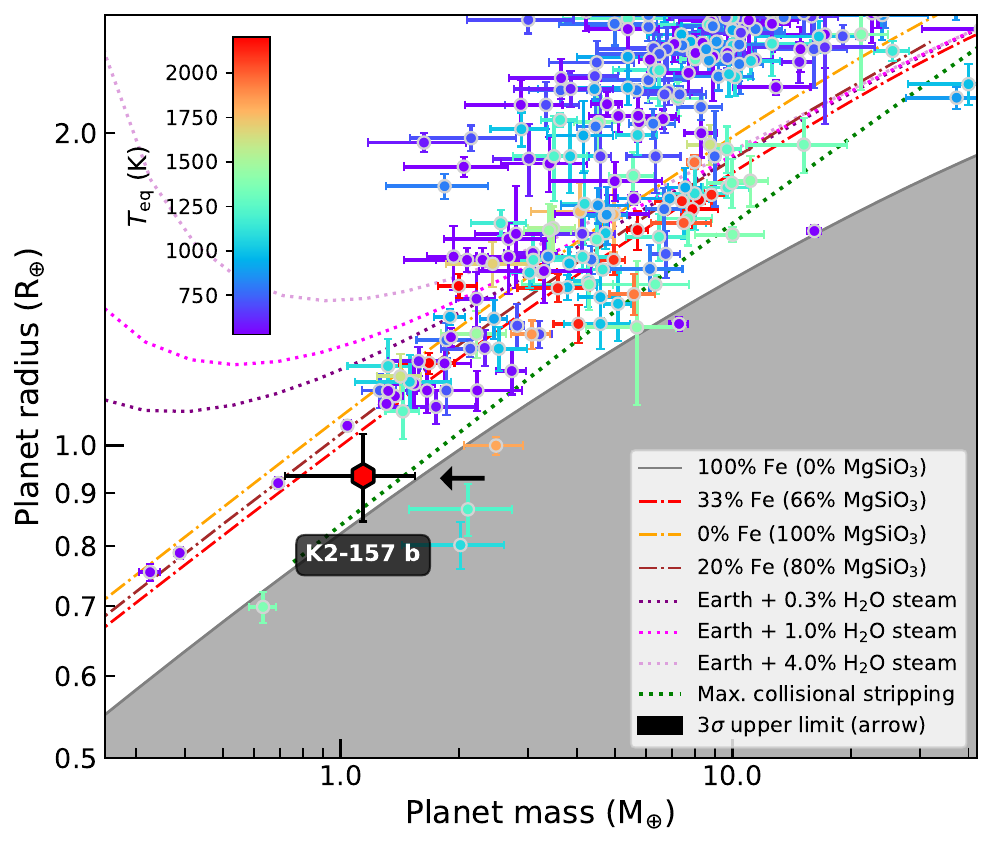}
     \includegraphics[width=0.47\textwidth]{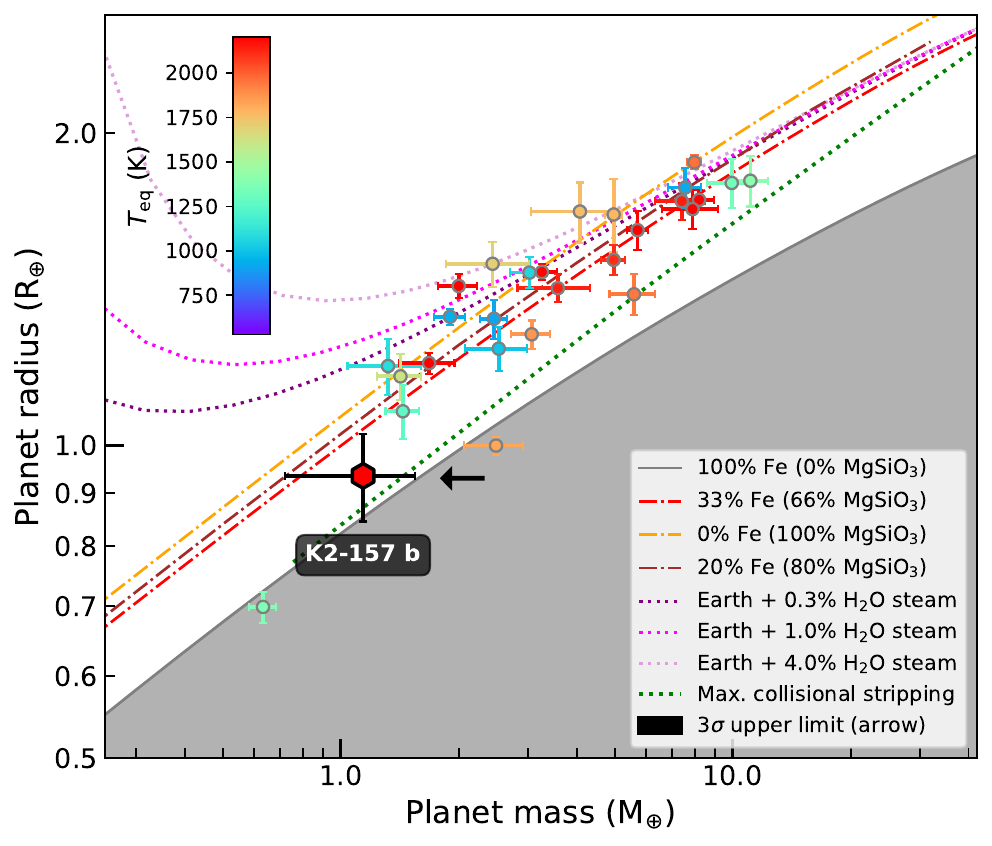}
    \caption{Mass-radius diagram of all small planets (left) and USP planets (right) with masses and radii constrained to a precision better than 35$\%$ and 20$\%$, respectively. The colour code represents the measured equilibrium temperatures. In both panels, we represent different composition curves for rocky planets derived in \citet{2010ApJ...712L..73M,2016ApJ...819..127Z,2019PNAS..116.9723Z,2020A&A...638A..41T}. The black arrow indicates the 3$\sigma$ upper limit of K2-157~b's mass: 2.3~$\rm M_{\oplus}$. Data: NASA Exoplanet Archive (24/03/2025). This plot was generated with \texttt{mr-plotter} \citep[\url{https://github.com/castro-gzlz/mr-plotter};][]{2023A&A...675A..52C}.}
    \label{fig:mr_main}
\end{figure*}

In Fig.~\ref{fig:mr_main}, left panel, we contextualise K2-157~b (both the 1$\sigma$ error bars and 3$\sigma$ mass upper limit) in the mass-radius diagram of all small planets with masses and radii constrained to precisions better than 35$\%$ and 20$\%$, respectively. In the right panel, we show the same diagram restricted to small USP planets. At a population level, we can see that planets with $R_{\rm p} \lessapprox$ 1.8 $\rm R_{\oplus}$ are consistent with an Earth-like bulk composition \citep[i.e. an iron-rich core making up $\simeq$33$\%$ of the planetary mass and a silicate-dominated mantle composing the remaining $\simeq$66$\%$;][]{2019PNAS..116.9723Z}. However, the wealth of precise mass measurements over the last years has revealed a non-negligible fraction of rocky planets with higher densities \citep[e.g.][]{2021Sci...374.1271L,2018NatAs...2..393S,2022A&A...657A..68A}. These planets are thought to have a larger iron-to-silicate mass fraction (i.e. larger cores or smaller mantles) than Earth, which could be due to a high metal content in the original proto-planetary disc \citep[e.g.][]{2009ApJ...693..722G} or to catastrophic giant impacts \citep[e.g.][]{2004E&PSL.223..241C} that strip away a significant fraction of the mantle \citep[e.g.][]{2010ApJ...712L..73M}. 

K2-157~b lies between the Earth-like and pure iron composition curves \citep{2019PNAS..116.9723Z}, with 1$\sigma$ uncertainties compatible with both models. Therefore, our mass and radius measurements indicate that K2-157~b may have a larger iron-to-silicate mass fraction than Earth. \citet{2021Sci...374..330A} found that stellar abundances and planet densities of highly irradiated rocky planets are correlated, suggesting a star-planet connection that may reflect the composition of the original proto-planetary discs. We used the [Fe/H], [Mg/H], and [Si/H] stellar abundances (Sect.~\ref{sec:stellar_charact}, Table~\ref{sec:stellar_charact}) and followed the stoichiometric model from \citet{2015A&A...580L..13S,2017A&A...608A..94S} to predict the iron-to-silicate mass fraction present in the original proto-planetary disc. We find a value compatible with that of the proto-solar disc (i.e. $f^{\rm star}_{\rm iron}$ = 30.5 $\pm$ 2.8$\%$). Therefore, in case K2-157~b has a considerably higher density than the expected for an Earth-like composition, it will not be easily explained through a primordial iron enhancement, and would thus possibly require the existence of a large-scale mantle-stripping impact \citep[e.g.][]{2019NatAs...3..416B}.

Additional high-precision RV and photometric data are necessary to determine whether K2-157~b is denser than Earth, and to precisely estimate its iron-to-silicate mass fraction. In particular, our stellar characterization imposes a radius and mass uncertainty floor of 0.021~$\rm R_{\oplus}$ and 0.024~$\rm M_{\oplus}$, respectively, in contrast to our derived uncertainties of 0.090~$\rm R_{\oplus}$ and 0.42~$\rm M_{\oplus}$, so the mass of K2-157~b still has a large room for improvement. We note, however, that while imprecise in relative terms, the mass measurement presented here still has a considerable statistical value. As we can see in Fig.~\ref{fig:mr_main}, the current mass-radius diagram is practically devoid of planets at $R_{\rm p}$ $\lessapprox$ 1~$\rm R_{\oplus}$, even when considering a relaxed mass precision threshold (i.e. $\delta M_{\rm p} / M_{\rm p} < 35 \%$). It is widely assumed that Earth and sub-Earth-sized planets broadly follow the Earth-like composition model, similarly to larger planets, and in consequence typically have rocky compositions. However, this inference is still based on a very limited observational evidence (i.e. the Earth, Venus, Mars, and the TRAPPIST-1 system; \citealt{2017Natur.542..456G,2016Natur.533..221G,2021PSJ.....2....1A}). Hence, the mass constraint on K2-157~b, although not very precise, still brings further evidence on the likely rocky nature of the smallest planets in our Galaxy. Interestingly, the rocky surfaces of small planets with equilibrium temperatures ($T_{\rm eq}$) above $\simeq$ 1800 K are expected to be completely melted, forming a `magma ocean' (irrespective of whether they host an atmosphere or not; \citealt{2025TrGeo...7...51L}). K2-157~b, with a $T_{\rm eq}$ of 2432~$\pm$~42~K, is expected to have one of the hottest dayside temperatures of the current sample of USP with constrained masses, thus being an excellent candidate to be a true `lava world'. Future characterization of K2-157~b may thus be able to probe planetary thermodynamic regimes that are inaccessible in the present-day Solar System, providing a fantastic test bed for the transition from primary to secondary atmospheres presumably happening in young rocky planets cooling down from primordial heating \citep[e.g.][]{2020PNAS..11718264K,2024ApJ...963..157T}.

\subsection{K2-157~c and K2-157~d in context: Period-mass diagram of neighbours of USP planets}
\label{K2-157}

\begin{figure}
    \centering
    \includegraphics[width=0.48\textwidth]{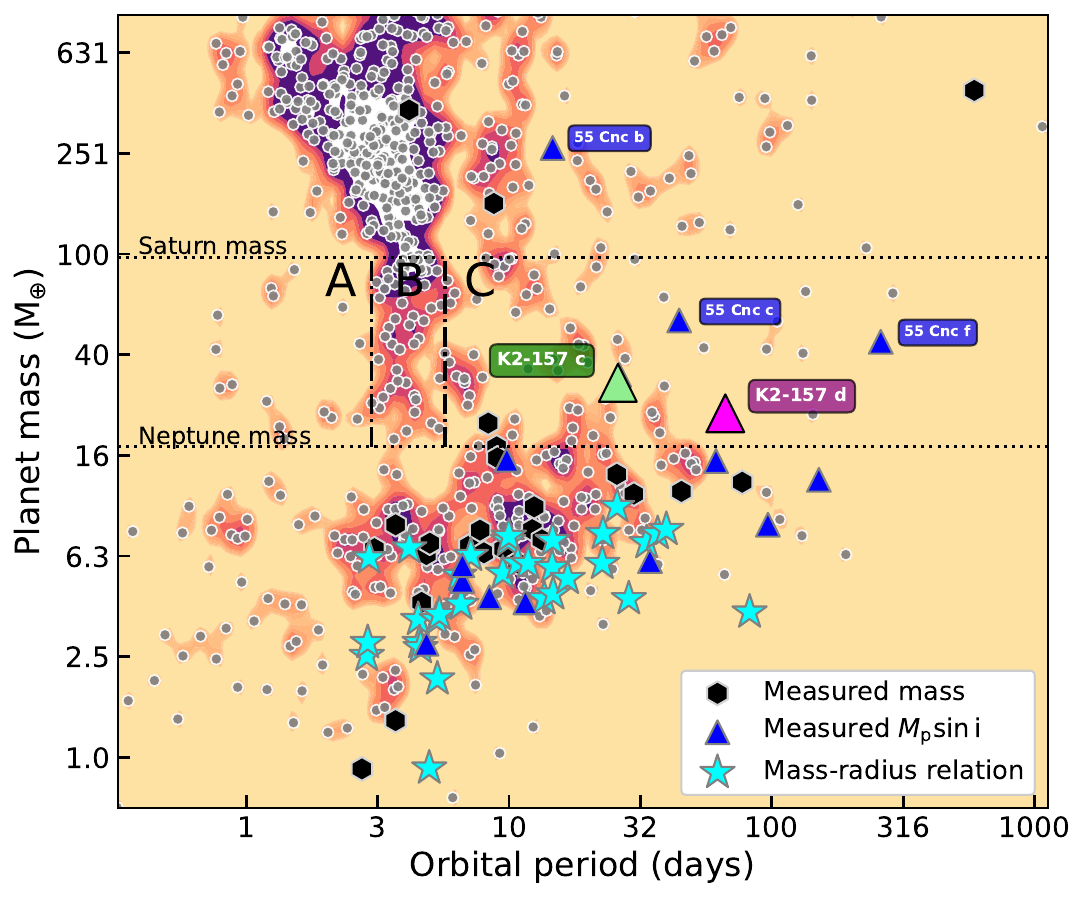}
    \caption{Contextualization of K2-157~c and K2-157~d in the period-mass diagram of known exoplanets with masses constrained to a precision better than 30$\%$. Black hexagons, blue triangles, and light-blue stars represent the current sample of known neighbours of USP planets. The delimited regions A, B, and C, represent the boundaries of the Neptunian desert, ridge, and savanna in the super-Neptune and sub-Jovian domains \citep{2024A&A...689A.250C}. Data: NASA Exoplanet Archive (24/03/2025). This plot was generated with \texttt{nep-des} (\url{https://github.com/castro-gzlz/nep-des}).}
    \label{fig:M_R_neighbourss}
\end{figure}

In Fig.~\ref{fig:M_R_neighbourss}, we contextualise K2-157~c and d in the period-mass diagram of known planets, where we highlight the current sample of neighbours of USP planets\footnote{Most neighbours of USP planets do not have mass measurements. For planets with only radius measurements, we estimated their masses through the empirical relations from \citet{2017ApJ...834...17C}. The massive and/or wide-orbit planets K2-187 e \citep{2018AJ....155..136M} and WASP-47 b, c, and d \citep{2012MNRAS.426..739H} fall outside the limits of the plot.}. These 78 planets occupy a wide region of the parameter space, showing masses from sub-Earths to super-Jovians and orbital periods from a few days to thousands of days. However, the vast majority of these planets (i.e. 83$\%$ of the current sample) lie within the sub-Neptune regime, showing a distribution in qualitative agreement with the overall planet distribution \citep[e.g.][]{2011arXiv1109.2497M,2012ApJS..201...15H,2019AJ....158..109H,2024AJ....167..288D}. The well-known planet occurrence drop occurring roughly at Neptune's mass (i.e. $M_{\rm p}$ $\simeq$ 17 $\rm M_{\oplus}$) is thus seemingly preserved within the observed distribution of companions to USP planets. In this context, K2-157~c and K2-157~d are both located in the poorly populated upper part of the period-mass space, in a warm region identified as the Neptunian savanna\footnote{The term `savanna' refers to the milder deficit of planets in this region compared to that in the Neptunian desert \citep{2011A&A...528A...2B,2011ApJ...727L..44S,2011ApJ...742...38Y}.} \citep{2023A&A...669A..63B}, which, as we can see in Fig.~\ref{fig:M_R_neighbourss}, is practically devoid of neighbours of USP planets. Interestingly, the only two known neighbours of USP planets in the warm savanna \citep[i.e. 55~Cnc~c and 55~Cnc~f;][]{2004ApJ...614L..81M,2008ApJ...675..790F,2018A&A...619A...1B} also belong to the same planetary system, 55 Cnc, which we highlight in Fig.~\ref{fig:M_R_neighbourss} together with K2-157~c and K2-157~d. Therefore, the system K2-157, having a USP planet accompanied by two neighbours in the warm savanna, of which the closer-in has a slightly higher minimum mass than the most distant one, and being possibly accompanied by an additional long-period giant planet (see Sect.~\ref{sec:blind_search}), becomes the most similar system to the iconic 55 Cnc system. We note, however, that 55 Cnc has an additional non-transiting Jupiter-mass planet between the USP planet and the two savanna neighbours (i.e. 55~Cnc~b, see Fig.~\ref{fig:M_R_neighbourss}), whose presence in the K2-157 system can be safely discarded based on our derived RV sensitivity limits (Sect.~\ref{sec:sensitivity_limits} and Fig.~\ref{fig:senslim}).

The unusual locations of K2-157~c and d in the period-mass parameter space can provide relevant clues on their evolution. Gas-rich and/or ice-rich giant planets such as K2-157~c and K2-157~d are widely thought to form at large orbital distances, beyond the ice line, and subsequently migrate inwards. Interestingly, the mere co-existence of two giant planets in relatively close-in orbits (i.e. within the ice line) favours the so-called disc-driven migration hypothesis\footnote{An early migration through the proto-planetary disc before the gas dissipates \citep[e.g.][]{1979ApJ...233..857G,1996Natur.380..606L}.} in contrast to the alternative HEM theory\footnote{An inward migration caused by a massive stellar or planetary companion that highly excites the planet eccentricity $-$making it lose orbital momentum$-$ and whose orbit is subsequently shrunken and circularized due to stellar tides $-$making it lose energy \citep[e.g.][]{2003ApJ...589..605W,2008ApJ...686..580C,2011CeMDA.111..105C}.} to explain the origins of these planets. Disc-driven migration can deliver giant planets with companions\footnote{These planets are expected to originally make up resonant configurations, but they also have the potential to evolve and escape the resonance \citep[][]{2014AJ....147...32G}, which could have been the case for K2-157~c and K2-157~d.} \citep[e.g.][]{1993Natur.365..819M,2002ApJ...567..596L,2006Sci...313.1413R}, while HEM is expected to remove any planet situated within the orbit of the migrating giant \citep[e.g.][]{2015ApJ...808...14M}, which would make it difficult to generate the observed configuration in K2-157. There are also increasing observational constraints that suggest that intermediate planets in the warm savanna were preferentially brought inwards through disc-driven migration, in contrast to closer-in hot Neptunes, which might have been primarily brought through HEM towards the Neptunian ridge \citep[see][]{2020A&A...635A..37C,2023A&A...669A..63B,2024A&A...689A.250C,2024A&A...691A.233C,2025AJ....169..117V}. 

\subsection{Orbital stability and constraints}
\label{sec:alexandre}

The K2-157 system is composed of an Earth-mass planet ($M_{\rm p,b} \approx 1.14$~$ \rm M_\oplus)$ with a USP orbit ($P_{\rm orb, b} \approx 8.8$~h), and two super-Neptune-mass planets ($M_{\rm p, c} \approx 30.8$~$\rm M_\oplus$, $M_{\rm p, d} \approx 23.3$~$\rm M_\oplus$) further away, in the vicinity of a 5:2 mean motion resonance ($P_{\rm orb, d} / P_{\rm orb, c} \approx 2.563$).
While the mass of the innermost planet is relatively well estimated from RV measurements, the masses of the other two planets correspond to minimal estimations, since the inclinations of their orbits are unknown.
The eccentricities are also loosely constrained from RV measurements.

To inspect the stability of the system and derive constraints for the unknown orbital parameters, we performed a dynamical analysis as in previous works \citep[e.g.][]{Correia_etal_2005, Correia_etal_2010}.
The full system was integrated around the best fit (Table~\ref{tab:parameters_joint}), varying two orbital parameters on a uniform grid of initial conditions.
Each initial condition was integrated for 10\,000~yr, using the symplectic integrator SABAC4 \citep{Laskar_Robutel_2001}, with a step size of $2.5 \times 10^{-5} $~yr and general relativity corrections.
We then performed a frequency analysis \citep{Laskar_1990, Laskar_1993PD} of the mean longitude of K2-157~d over two consecutive time intervals of 5\,000~yr, and determined the main frequency, $n$ and $n'$, respectively.
The stability was measured with $\Delta = |1-n'/n|$, which estimates the chaotic diffusion of the orbits.

\begin{figure*}
    \centering
	\includegraphics[width=.95\textwidth]{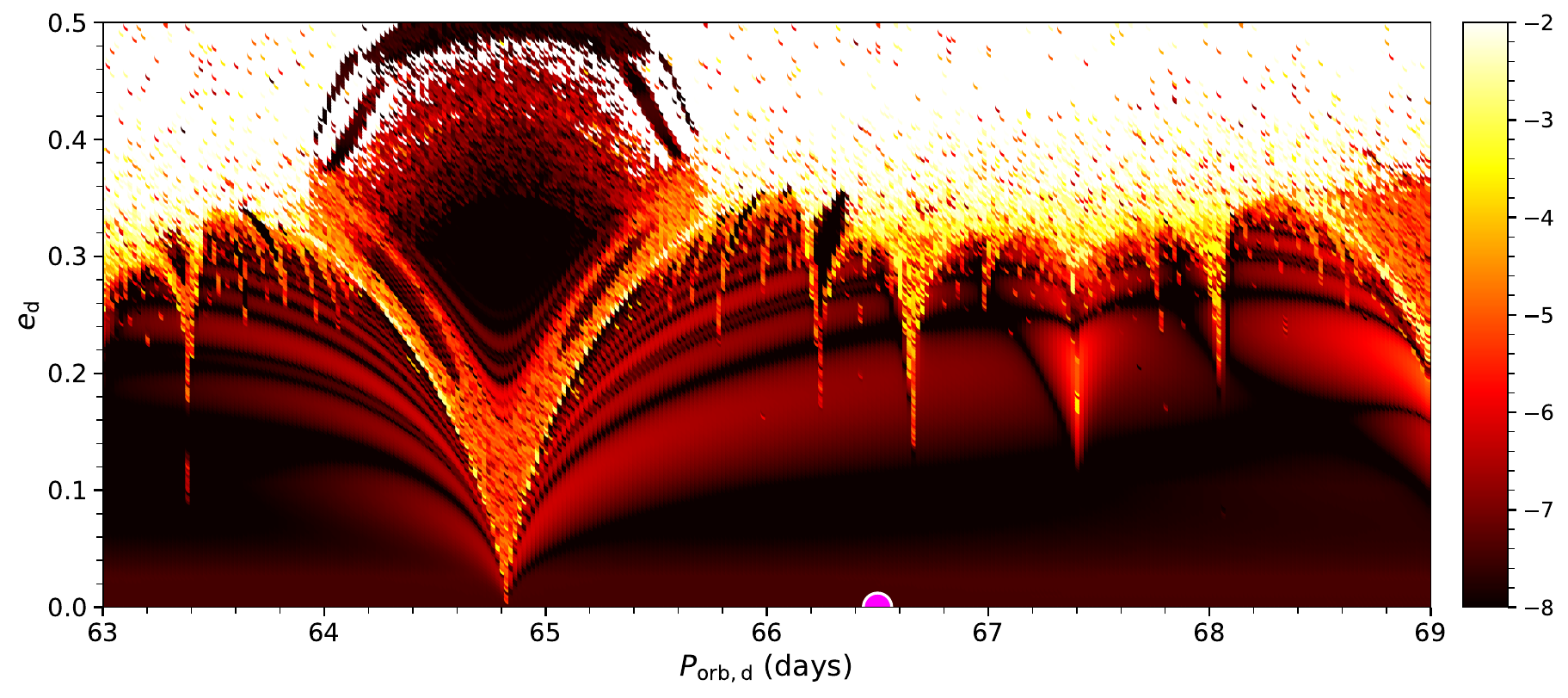}
    \caption{Stability analysis of the K2-157 planetary system. For fixed initial conditions (Table~\ref{tab:parameters_joint}), the parameter space of the system is explored by varying the orbital period and eccentricity of K2-157~d. The step size is $0.02$~d in orbital period and $0.0025$ in eccentricity. For each initial condition, the system was integrated over $10^4$~yr and a stability indicator was calculated, which involved a frequency analysis of the mean longitude of the inner planet. The chaotic diffusion is measured by the variation in the frequency. White-yellow points correspond to highly unstable orbits, while red points correspond to orbits which are stable on Gyr timescales. The magenta dot corresponds to the best-fit solution ($P_{\rm orb, d}$ = 66.50 d, $e_{\rm d}$ = 0).}
    \label{figPe}
\end{figure*}

We first explored the stability of the system by varying the orbital period and the eccentricity of K2-157~d (Fig.~\ref{figPe}). The results are reported in colour: white and yellow correspond to strongly chaotic unstable trajectories, while red and black give extremely stable quasi-periodic orbits on Gyr timescales. We observe that the best fit solution from Table~\ref{tab:parameters_joint} (marked with a blue dot) is completely stable, even for eccentricities up to 0.3.
We can also clearly identify a V-shape structure on the left-hand side of Fig.~\ref{figPe} corresponding to the 5:2 mean motion resonance with K2-157~c.
Since the current best fit is completely outside this region, we conclude that the two planets are not in resonance, even for higher eccentricity values.
We additionally note that the system is on the correct side of the resonance predicted by planetary migration models \citep[e.g.][]{Lissauer_etal_2011K}.
This feature strengthens the hypothesis that the USP planet K2-157~b also formed further away and then migrated inwards.

\begin{figure}
    \centering
	\includegraphics[width=0.99\columnwidth]{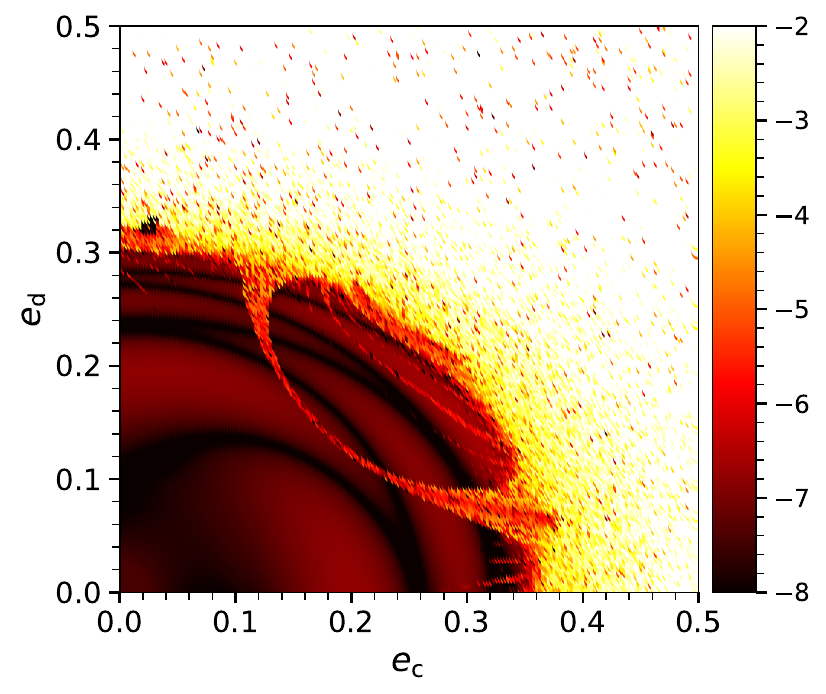}
    \caption{Stability analysis of the K2-157 system in the plane $(e_{\rm c}, e_{\rm d})$, assuming coplanar orbits. For fixed initial conditions (Table~\ref{tab:parameters_joint}), the parameter space is explored by varying the eccentricities of K2-157 c and K2-157 d with a step size of $0.0025$. The stability index and colour codes are obtained as for Fig.~\ref{figPe}.}   
    \label{figee}
\end{figure}

In a second experiment, we varied the eccentricities of the two super-Neptune-mass planets (Fig.~\ref{figee}).
These two eccentricities were not constrained by the best fit solution (Table~\ref{tab:parameters_joint}), but can play an important role in the stability of the system.
Indeed, we observe that high eccentricity values of both planets can lead to unstable orbits, although the system remains stable as long as $e_{\rm c} \lesssim 0.35$ and $e_{\rm d} \lesssim 0.3$.
We then conclude that the K2-157 planetary system is very resilient to the uncertainties in the determination of the eccentricities.

\begin{figure}
    \centering
	\includegraphics[width=0.99\columnwidth]{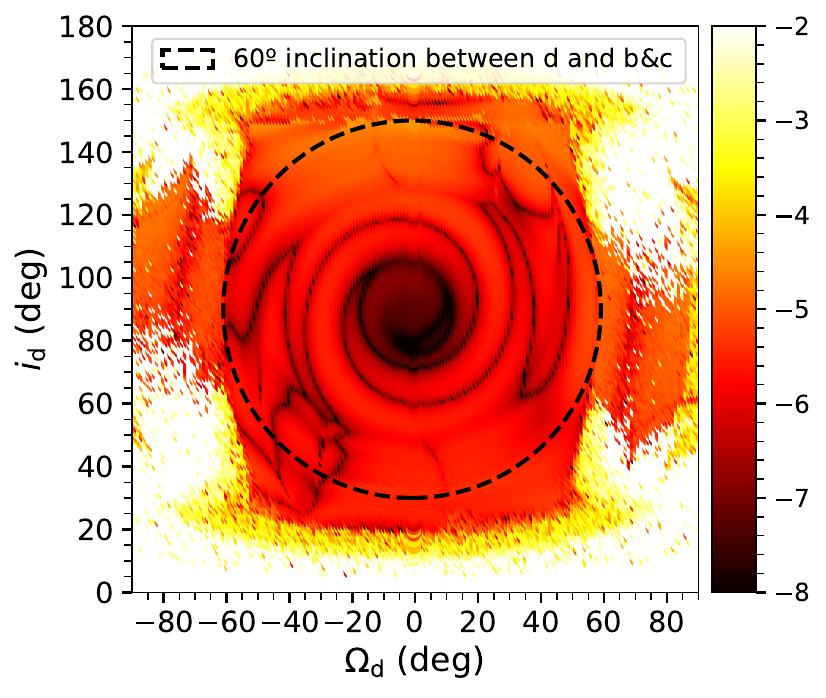}
    \caption{Stability analysis of the K2-157 system in the plane $(\Omega_{\rm d}, i_{\rm d})$. For fixed initial conditions (Table~\ref{tab:parameters_joint}), the parameter space is explored by varying the longitude of the node $\Omega_{\rm d}$ and the inclination $i_{\rm d}$ of the outer planet, with a step size of $1^\circ$. The stability index and colour codes are obtained as for Fig.~\ref{figPe}.}   
 \label{figOI}
\end{figure}

The RV technique alone is unable to constrain the inclination, $i$, and the longitude of the node, $\Omega$, of the planets.
Therefore, in a final experiment, we vary these two parameters for the outermost planet K2-157~d (Fig.~\ref{figOI}).
We observe that the system is stable within a circle centred at the coplanar solution, which corresponds to mutual inclinations smaller than $60^\circ$ between the orbit of K2-157~d and the orbits of the remaining two planets in the system\footnote{Additional stability regions also exists around ($\Omega_\mathrm{d} = 180^\circ, i_\mathrm{d}=90^\circ$), but they are not shown in Fig.~\ref{figOI} because they correspond to retrograde orbits, which are more unlikely from a formation point-of-view.}.
As we change the inclination, the mass of the outer planet increases.
At the boundary of stability circle, $i_\mathrm{d} = 90^\circ \pm 60^\circ$, we get a maximum mass for K2-157~d of about twice its minimum value (i.e. $M_{\rm d,\mathrm{max}} \approx 47$~$\rm M_\oplus$).
A similar constraint could be derived for the maximal mass of K2-157~c (i.e. $M_{\rm c,\mathrm{max}} \approx 62$~$\rm M_\oplus$).
Indeed, we can also get a maximal mutual inclination of $60^\circ$ with  $\Omega_{\rm d} = \pm 60^\circ$ and $i_{\rm d}= 90^\circ$ (Fig.~\ref{figOI}), for which the mass of K2-157~d is minimal ($M_{\rm d} = 23.3$~$\rm M_\oplus$).
That is, the main source of instability results from mutual inclinations above the $60^\circ$ threshold and not from the increase in the masses.

\subsection{Comparison to formation mechanisms of USP planets}
\label{sec:comparison_theories}

Formation theories of USP planets typically require the existence of outer planetary companions that brought them towards their present-day orbits. However, there is no consensus on the predominant mechanism triggering this likely migration. In this section, we briefly introduce three promising theories and compare their predictions to the properties of K2-157. 

\subsubsection{High-eccentricity migration}

\citet{2019AJ....157..180P} propose that USP planets were originally formed at orbital periods of $\sim$5-10~d ($\sim$0.05-0.1 au) and reached their current orbits through HEM. Chaotic interactions with outer planetary companions would significantly increase the eccentricities of these planets (e.g. 1$-$e $\ll$ 1). Subsequently, energy dissipation from tides raised by the star on the planets would shrink and circularise their orbits. This theory is closely related to the secular chaos theory proposed to explain the short orbits of hot Jupiters \citep[e.g.][]{2011ApJ...735..109W}. The initial conditions required to form USP planets through this channel are the existence of several neighbours (typically $\geq$ 3) and moderate eccentricities and/or inclinations ($e_{\rm rms}$, $i_{\rm rms}$ $\sim$ 0.1) able to drive chaotic diffusion. 

This mechanism predicts that neighbours of USP planets should have extended orbits, from tens of days (i.e $P_{\rm orb} >$ 10~d) up to beyond $\sim$~au distances, as well as relatively large period ratios between adjacent planets (typically $P_{\rm i}$ / $P_{\rm i+1}$ $\gtrsim$ 3). In addition, the outer neighbours need to be considerably more massive than the USP planets (typically $\gtrsim$ 10 $\rm M_{\oplus}$). This mechanism also predicts mutual spin-orbit angles and inclinations between the USP and the outer planets in the range of $\sim$10-50$^{\circ}$. When compared to K2-157, the relatively large orbits and masses of the neighbours K2-157~c and K2-157~d agree with the predictions of this mechanism. In addition, the measured ratio $P_{\rm c} / P_{\rm d}$ $\simeq$ 2.6 is near the typical value, and the subsequent ratio would be considerably larger in case the quadratic trend in the RVs corresponds to an outer massive companion. Regarding the mutual inclinations, our orbital stability analysis indicates that inclinations below $\sim$60$^{\circ}$ in K2-157 result in a dynamically stable system, regardless of the true masses of K2-157~c and K2-157~d  (Sect.~\ref{sec:alexandre}), which is compatible with the predicted range of inclinations from this mechanism. 

\subsubsection{Low-eccentricity migration}

\citet{2019MNRAS.488.3568P} propose that USP planets originally formed in closer-in orbits, at orbital periods of $\sim$1-3 d ($\sim$0.02 and 0.04 au), and were then brought inwards through a low-eccentricity migration mechanism (LEM). This mechanism relies upon the combination of secular interactions and tidal dissipation \citep[e.g.][]{2002ApJ...564.1024W,2007MNRAS.382.1768M,2010MNRAS.407.1048M}. Contrary to the HEM mechanism, LEM requires the existence of a multi-planet system with typically $\geq$ 2 neighbours, and requires the inner planet to achieve mild eccentricities ($e$ $\sim$ 0.05-0.2) before entering the tidal decay process. Therefore, regarding the initial conditions, LEM requires more constrained proto-USP periods, but requires fewer neighbours to trigger their migration.

The LEM mechanism allows for more relaxed properties of USP-hosting systems than the HEM mechanism. Exterior neighbours can be less massive (i.e. $\gtrsim$ 3 $\rm M_{\oplus}$) and there is no need for large values of $P_{\rm i}$ / $P_{\rm i+1}$. As was discussed before, the mass condition is well met in the K2-157 system. In addition, LEM predicts milder mutual inclinations than HEM, of about 20$^{\circ}$, which is compatible with our dynamical stability analysis.

From the comparisons discussed above, we conclude that K2-157 is perfectly compatible with the relaxed constraints of the LEM mechanism, but so it is with the more restrictive predictions of the HEM mechanism, thus converting these two theories into promising contenders to explain its origin. We also note that these two eccentricity-based mechanisms may expect some degree of eccentricity enhancements for the companion planets in USP systems, which is allowed by the RVs ($e_{\rm c}$ $<$ 0.2 and $e_{\rm d}$ $<$ 0.5 at 3$\sigma$) and stability analysis ($e_{\rm c}$,~$e_{\rm d}$ $\lesssim$ 0.3) of this system. 

\subsubsection{Obliquity-driven migration}

\citet{2020ApJ...905...71M} propose that USP initially come from orbits $\lesssim$ 0.04-0.05 au and arrive into their present-day locations through obliquity-driven tidal migration. This mechanism states that the spin of proto-USP planets first locks on an equilibrium configuration, called Cassini state 2, due to tidal dissipation. Then, the planetary obliquity is significantly enhanced and the planet starts migrating inwards, which would generate an even larger obliquity, entering a process of runaway orbital decay that stops when the Cassini state 2 is destabilised. 

Contrary to the HEM and LEM eccentricity-based mechanisms, the obliquity-driven tidal migration is most efficient when the planetary neighbours have close-in orbits, typically with periods $P_{\rm orb}$ $<$ 10 d. This observational prediction contrasts with the substantially larger periods of planets K2-157~c and K2-157~d. We note, however, that for $P_{\rm orb}$ $\lesssim$ 10 d, our RV sensitivity analysis only discards the existence of additional planets with $M_{\rm p}$ $\gtrsim$ 4-5~$\rm M_{\oplus}$ (Sect.~\ref{sec:sensitivity_limits}), so there is still room for the existence of non-transiting smaller planets in this close-in orbital region which could have been responsible for the obliquity-driven migration of K2-157~b.


\section{Summary and conclusions}
\label{sec:conclusions}

We acquired 49 ESPRESSO spectra to confirm and characterize the USP ($P_{\rm orb}$ = 8.8 h) Earth-sized ($R_{\rm p}$ = 0.935 $\pm$ 0.090~$\rm R_{\oplus}$) planet K2-157~b, as well as to constrain the presence of additional companions in the system. We first derived the stellar parameters of K2-157 through a high-resolution, high-S/N spectrum obtained from the co-adding of the individual ESPRESSO spectra. We find that K2-157 is a G9~V star with $T_{\rm eff}$ = $5334 \pm 64$~K, $R_{\star}$ = $0.860 \pm 0.019$ $\rm R_{\odot}$, and $M_{\star}$ = $0.890 \pm 0.029$~$\rm M_{\odot}$,  in agreement with previous estimations. A blind-search analysis of the ESPRESSO RVs revealed the existence of two additional Keplerian signals in the system, which, after discarding stellar activity as their possible origin, we confirm as planets: K2-157~c ($P_{\rm orb, c}$ = $25.942^{+0.045}_{-0.044}$~d,  $M_{\rm p, c} \, \textrm{sin} \, i$ = $30.8 \pm 1.9$ $\rm M_{\oplus}$), and K2-157~d ($P_{\rm orb, d}$ = $66.50^{+0.71}_{-0.59}$ d,  $M_{\rm p, d} \, \textrm{sin} \, i$ = $23.3 \pm 2.5$ $\rm M_{\oplus}$). This analysis also revealed the existence of a long-term quadratic trend that could be due to an additional long-period massive companion or, alternatively, reflect the magnetic cycle of the star. 

A joint analysis of the K2 and ESPRESSO data allowed us to infer the properties of the system. This analysis constrained the mass of K2-157~b at the 2.7$\sigma$ level, $M_{\rm p, b}$ = $1.14^{+0.41}_{-0.42}$~$\rm M_{\oplus}$ ($<$ 2.4~$\rm M_{\oplus}$ at 3$\sigma$), which makes the planet compatible with a rocky composition with a likely ($68\%$ confidence) higher iron-to-silicate mass fraction than Earth. K2 and TESS photometry did not allow us to determine whether K2-157~d transits its star, but K2 data allowed us to discard non-grazing transit configurations for K2-157~c ($i_{\rm c}$ $<$ 88.4$^{\circ}$ at 3$\sigma$). ESPRESSO data also allowed us to constrain the eccentricities of K2-157~c and d: $e_{\rm c}$ $<$ 0.2 and $e_{\rm d}$~$<$~0.5 at 3$\sigma$. In this regard, we performed a dynamical analysis that showed that the system is stable for eccentricities up to $e_{\rm c}$,~$e_{\rm d}$ $\sim$ 0.3 and mutual inclinations up to $\sim$ 60$^{\circ}$. Therefore, both the observations and stability analysis discard highly eccentric orbits for the outer planetary companions but are still compatible with moderate eccentricities up to $\sim$ 0.2$-$0.3.

The orbital architecture of K2-157 (a USP planet accompanied by massive neighbours in the warm Neptunian savanna) is infrequent, with only one other similar case reported to date: the iconic 55 Cnc system. Whether these systems are intrinsically unusual or the low observed occurrence is dominated by observational bias needs to be studied. Interestingly, the USP planets of these systems, which have massive ($M_{\rm p}$ $>$ 10 $\rm M_{\oplus}$), long-period ($P_{\rm orb}$ $>$ 10 d), relatively spaced ($P_{\rm i}$ / $P_{\rm i+1}$ $\gtrsim$ 3), and possibly misaligned planetary neighbours, could
have migrated inwards through a HEM process triggered by chaotic secular interactions. A LEM process could have also played a role.

K2-157~b has one of the shortest orbits known to date. This characteristic makes the K2-157 system particularly interesting when considering the early G9 V spectral type of the host star, given that the rocky planets with the shortest orbits are preferentially found around late-type stars. We further studied this trend and found that it cannot be explained through observational bias, thus requiring a physical explanation. We also found that this trend does not hold when scaling the orbital separation to the Roche limit. Instead, we identify a featureless distribution across a wide range of spectral types, which leads us to interpret the trend as a direct consequence of the dependency between the Roche limit and the stellar radius. We thus conclude that the distribution of the closest USP planets is not necessarily determined by
the inner edges of the original proto-planetary discs, as was previously suggested, but simply
by the present-day separation from the Roche limit, regardless of the spectral type of the host star. This explanation aligns very well with the main formation
theories of USP planets, according to which their present-day orbits are primarily determined by the original interactions with outer companions and subsequent tidal decay, with no preservation of signatures of their primordial distribution across the proto-planetary discs. 

\begin{acknowledgements}
We are very grateful to the reviewer, who provided insightful suggestions that improved the quality of this work.
A.C.-G. is funded by the Spanish Ministry of Science through MCIN/AEI/10.13039/501100011033 grant PID2019-107061GB-C61. 
A.C.M.C. acknowledges support from the FCT, Portugal, through the CFisUC projects UIDB/04564/2020 and UIDP/04564/2020, with DOI identifiers 10.54499/UIDB/04564/2020 and 10.54499/UIDP/04564/2020, respectively.
This research is co-funded by the European Union (ERC, FIERCE, 101052347). Views and opinions expressed are however those of the author(s) only and do not necessarily reflect those of the European Union or the European Research Council. Neither the European Union nor the granting authority can be held responsible for them. This work was supported by FCT - Fundação para a Ciência e a Tecnologia through national funds by these grants: UIDB/04434/2020 DOI: 10.54499/UIDB/04434/2020, UIDP/04434/2020 DOI: 10.54499/UIDP/04434/2020.
JIGH and ASM acknowledge financial support from the Spanish Ministry of Science, Innovation and Universities (MICIU) projects PID2020-117493GB-I00 and PID2023-149982NB-I00.
The INAF authors acknowledge financial support of the Italian Ministry of Education, University, and Research
with PRIN 201278X4FL and the "Progetti Premiali" funding scheme.
This work was financed by Portuguese funds through FCT (Funda\c c\~ao
para a Ci\^encia e a Tecnologia) in the framework of the project
2022.04048.PTDC (Phi in the Sky, DOI 10.54499/2022.04048.PTDC).
CJM also acknowledges FCT and POCH/FSE (EC) support through
Investigador FCT Contract 2021.01214.CEECIND/CP1658/CT0001
(DOI 10.54499/2021.01214.CEECIND/CP1658/CT0001).  
We acknowledge financial support from the Agencia Estatal de Investigaci\'on of the Ministerio de Ciencia e Innovaci\'on MCIN/AEI/10.13039/501100011033 and the ERDF “A way of making Europe” through project PID2021-125627OB-C32, and from the Centre of Excellence “Severo Ochoa” award to the Instituto de Astrofisica de Canarias.
FPE would like to acknowledge the Swiss National Science Foundation (SNSF) for supporting research with ESPRESSO through the SNSF grants nr. 140649, 152721, 166227, 184618 and 215190. The ESPRESSO Instrument Project was partially funded through SNSF’s FLARE Programme for large infrastructures.
This publication made use of \texttt{TESS-cont} (\url{https://github.com/castro-gzlz/TESS-cont}), which also made use of \texttt{tpfplotter} \citep{2020A&A...635A.128A} and \texttt{TESS\_PRF} \citep{2022ascl.soft07008B}.
This work made use of \texttt{mr-plotter} (available in \url{https://github.com/castro-gzlz/mr-plotter}).

\end{acknowledgements}

\bibliographystyle{aa} 
\bibliography{biblio_K2-157} 

%


\begin{appendix}

\section{Additional figures}

\begin{figure*}
    \centering
    \includegraphics[width=\textwidth]{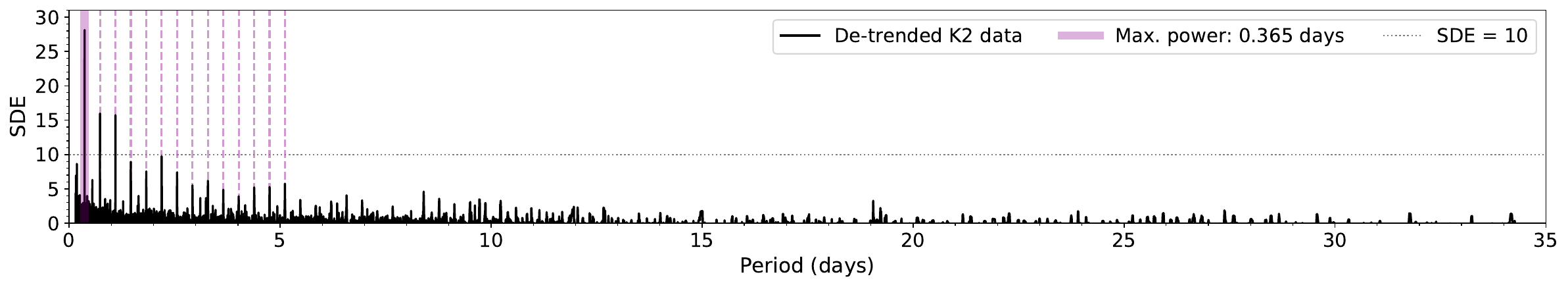}
    \includegraphics[width=0.995\textwidth]{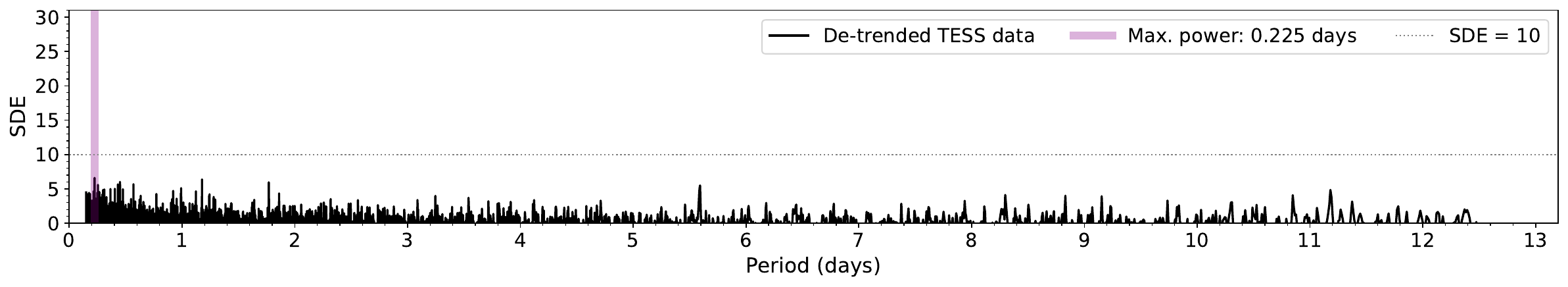}
    \caption{Transit least squares (\texttt{TLS}) periodogram of the K2 (C10) and TESS (S46) photometry of K2-157 detrended with the bi-weight technique considering a window length of 0.5~d. The horizontal dotted lines indicate a conservative threshold for considering a significant transit detection. The vertical dashed lines indicate the first harmonics of the maximum power period (0.365~d). }
    \label{fig:TLS}
\end{figure*}

\begin{figure*}
    \centering
    \includegraphics[width=0.98\textwidth]{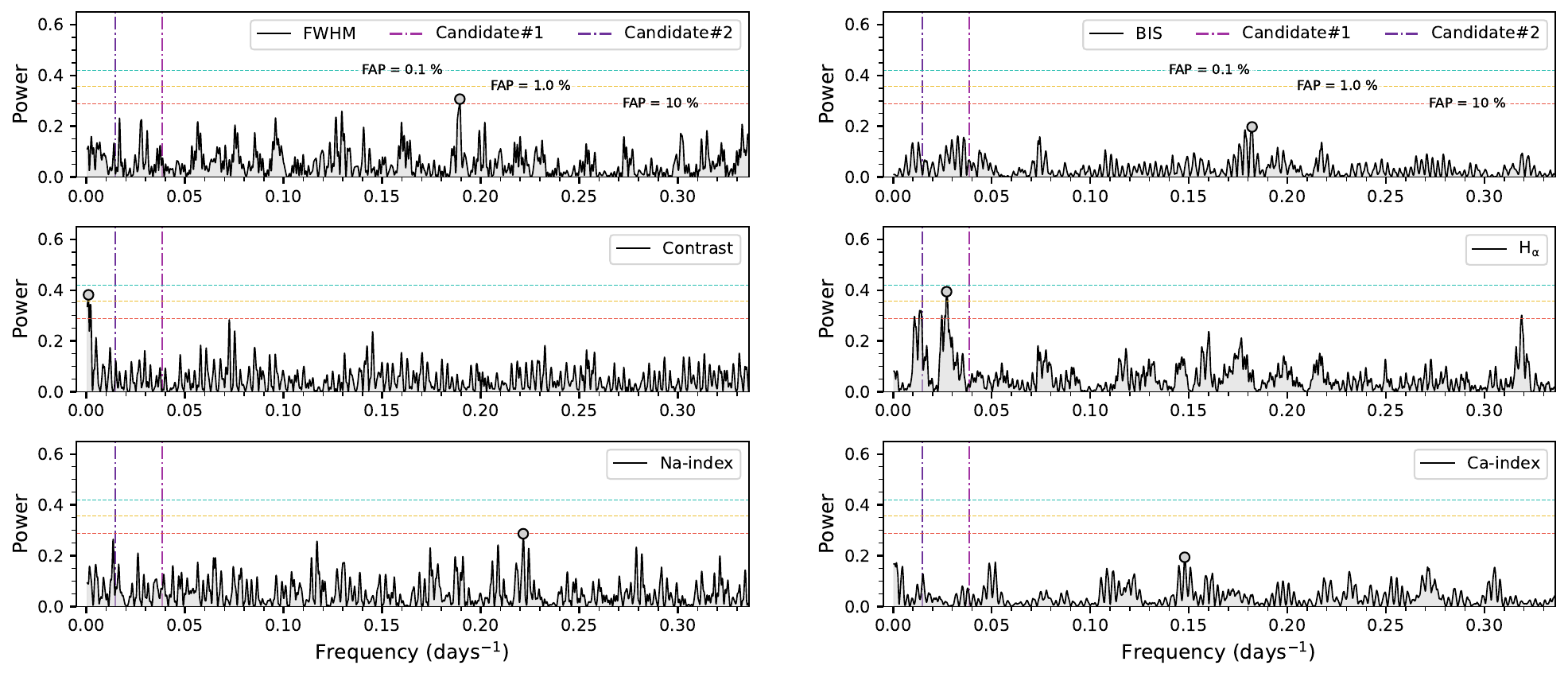}
    \caption{Generalized Lomb-Scargle periodograms (\texttt{GLS}) of the ESPRESSO activity indicators of K2-157. The grey circles indicate the maximum power frequencies. The vertical dash-dotted lines indicate the frequencies of the sinusoidal periodicities of Candidate$\#1$ and Candidate$\#2$. The horizontal dotted lines show the 10$\%$, 1$\%$, and 0.1$\%$ FAP levels. }
    \label{fig:gls_to_indicators}
\end{figure*}

\begin{figure*}
    \centering
    \includegraphics[width=0.98\textwidth]{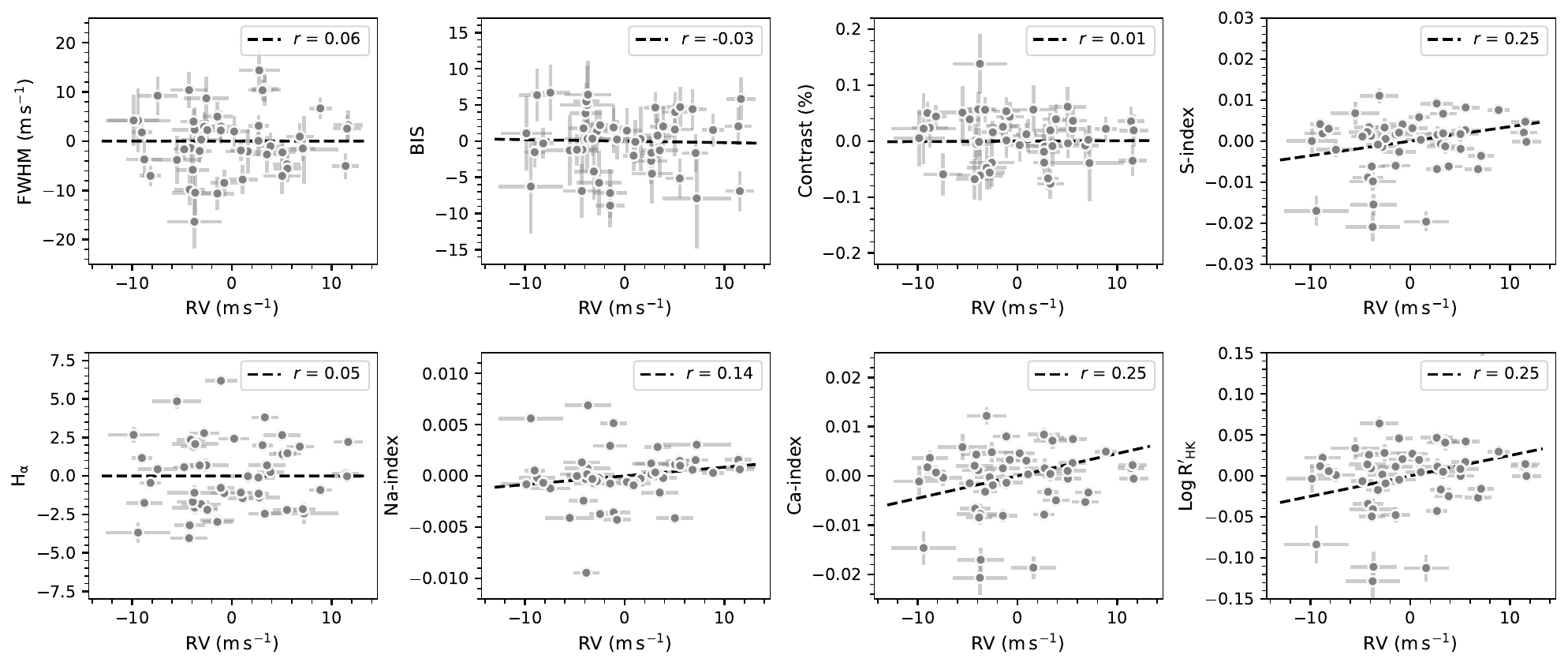}
    
    \caption{ESPRESSO activity indicators as a function of the RVs of K2-157. The dashed lines represent linear trends fitted to the data. The legends indicate the Pearson product-moment correlation coefficients $r$.}
    \label{fig:person_r}
\end{figure*}

\begin{figure*}
    \centering
    \includegraphics[width=0.65\textwidth]{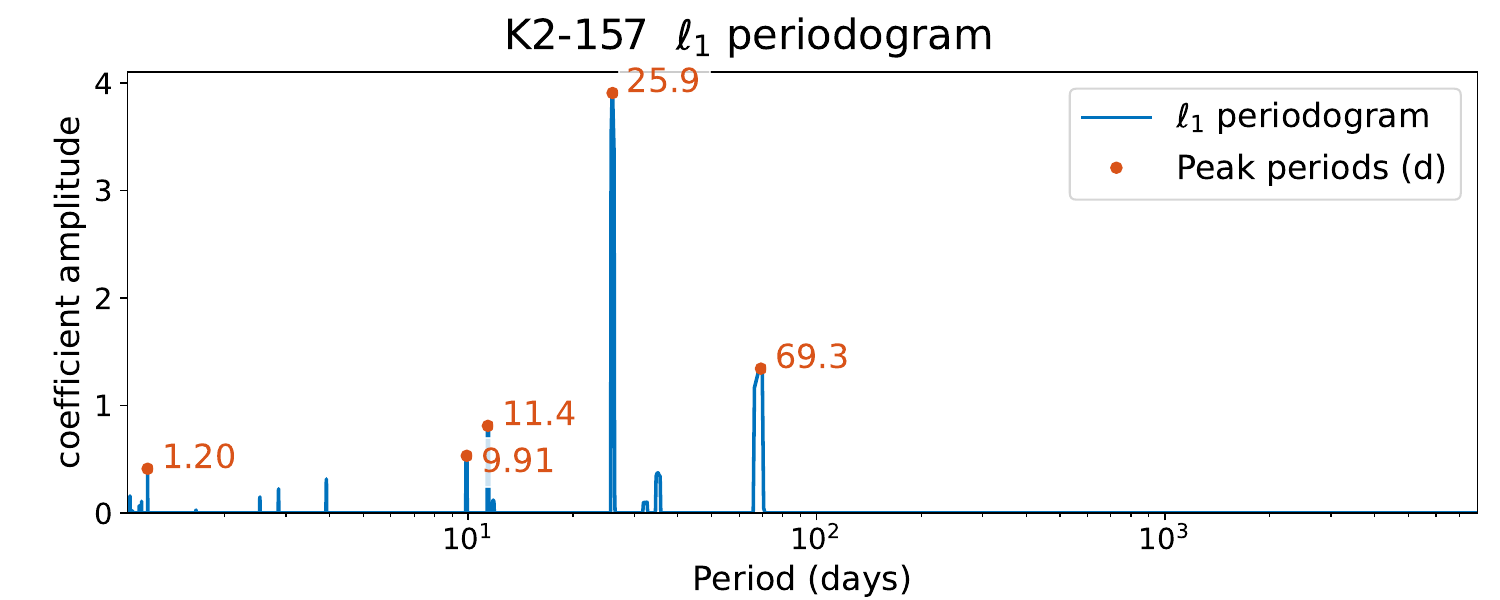}    \caption{$\ell_1$ periodogram of the ESPRESSO RVs, assuming a quadratic trend, and a covariance model selected with highest Bayesian evidence, computed with the Laplace approximation.}
    \label{fig:l1per}
\end{figure*}

\begin{figure*}
\noindent
\centering
\begin{tikzpicture}		
\path (0,0) node[above right]{\includegraphics[width=0.97\textwidth]{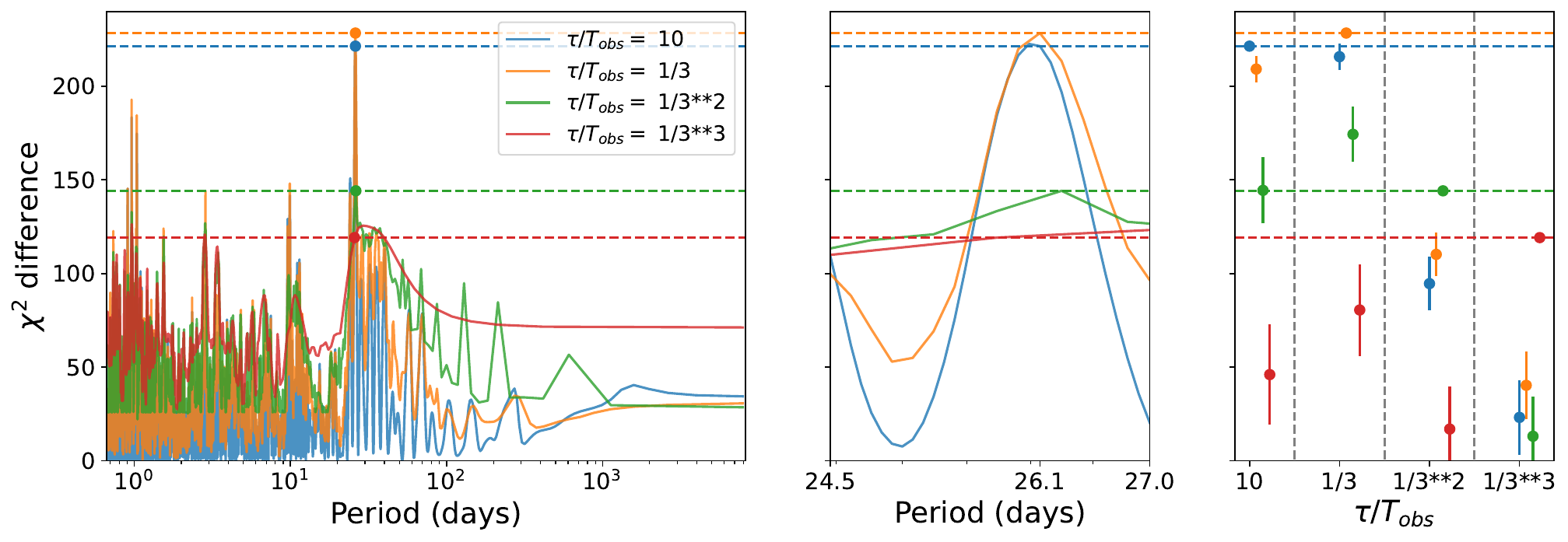}};
\path (1.15,4.5);
\begin{scope}[yshift=-7cm]
\path (0,0) node[above right]{\includegraphics[width=0.97\textwidth]{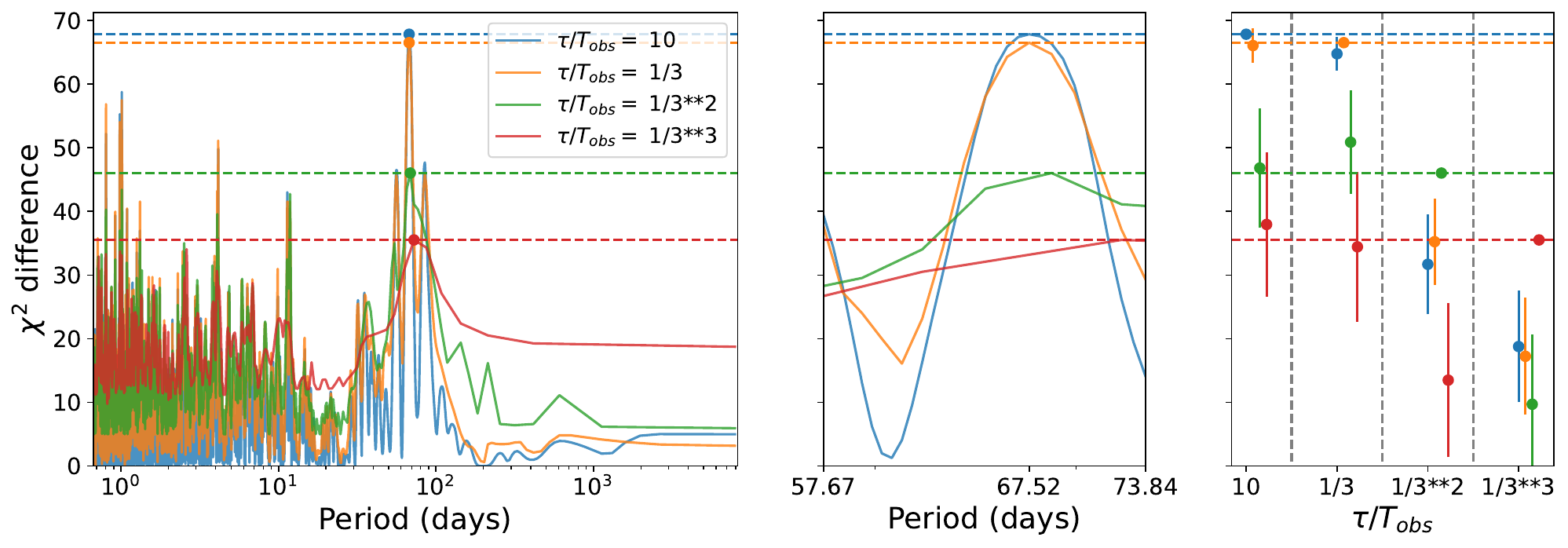}};
\path (1.15,4.5) ;
\end{scope}
\begin{scope}[yshift=-14cm]
\path (0,0) node[above right]{\includegraphics[width=0.97\textwidth]{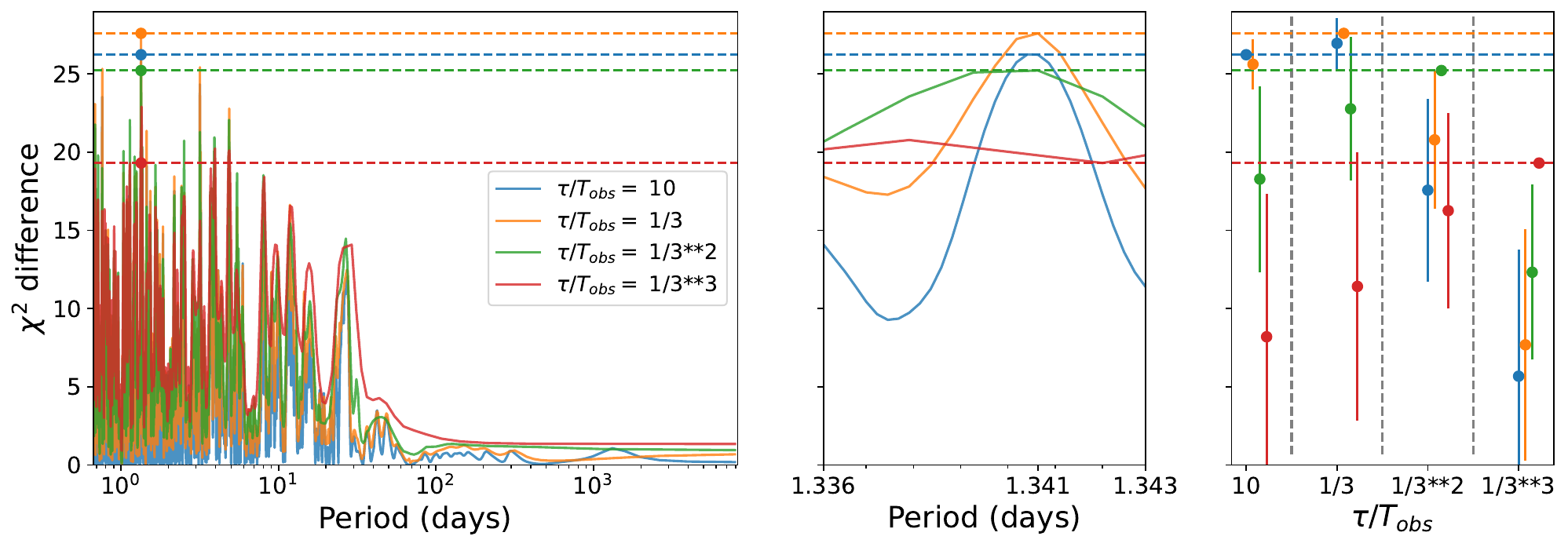}};
\path (1.15,4.5);
\end{scope}
\end{tikzpicture}
\caption{First three iterations of the apodized sine periodograms (ASPs) method. Models correspond to the maximum of the ASPs. $T_\mathrm{obs}$ designates the timespan between the first and last RV observation, and $\tau$ has the same meaning as in Eq.~\eqref{eq:asp}. The left panels represent the periodograms, the middle panels show a zoom in on the highest peak, and the right-hand panels represent a statistical significance test \citep[see][for details]{hara2022}. We note that the best fitting $\tau$ for the 26 d signal is not the longest. We attribute this to the sequential fitting. In the first iteration, the 69 d signal is not removed. Because there are only 49 RV measurements, the aliasing is strong, and the presence of the 69 d signal in the first iteration creates interferences. If we first remove the 69 d signal, the longest timescale has the same statistical significance as $\tau/T_\mathrm{obs} =3$. }
\label{fig:ASP}
\end{figure*}
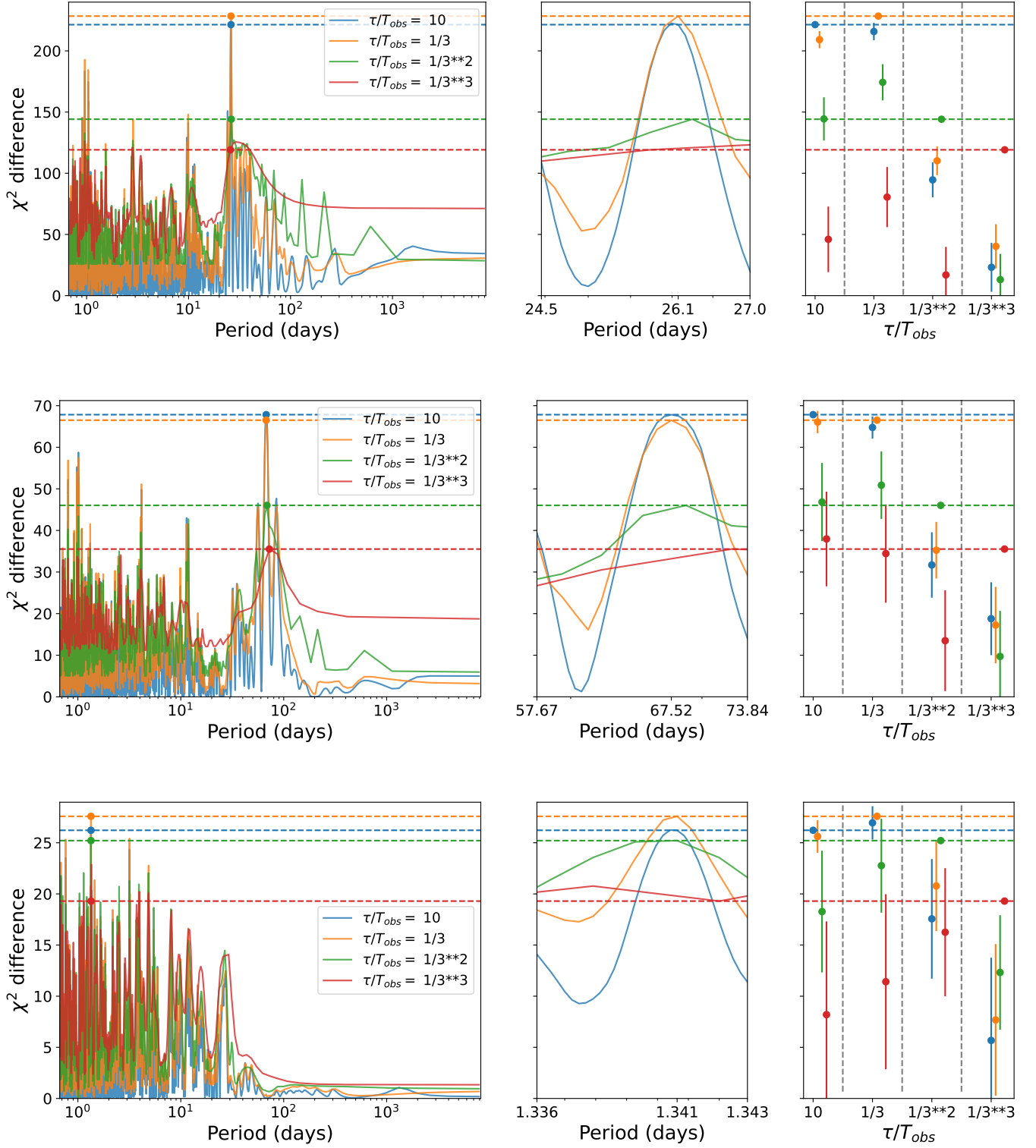

\begin{figure*}
    \centering

    \includegraphics[width=0.31\textwidth]{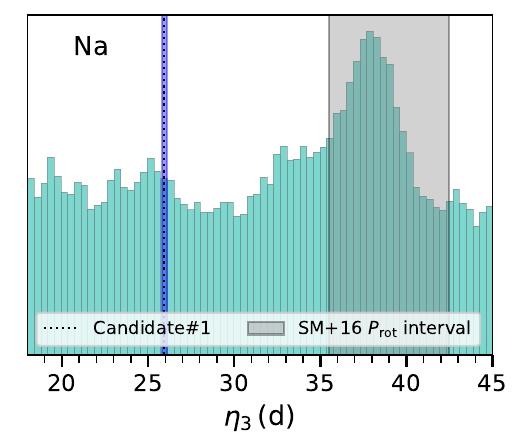}
    \includegraphics[width=0.31\textwidth]{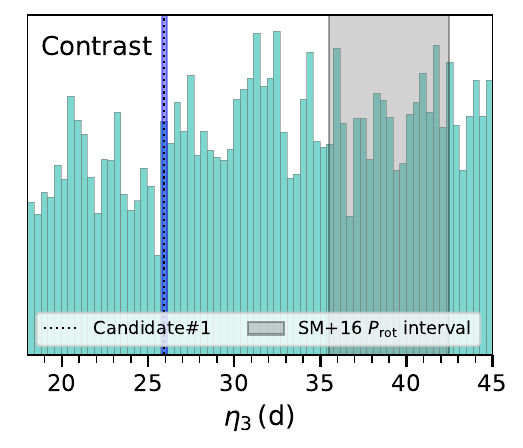}
    \includegraphics[width=0.31\textwidth]{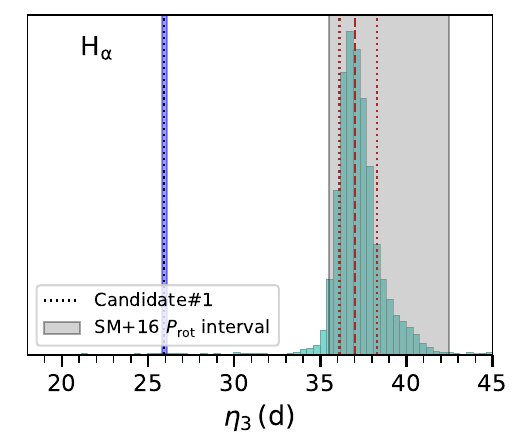}
    \includegraphics[width=0.31\textwidth]{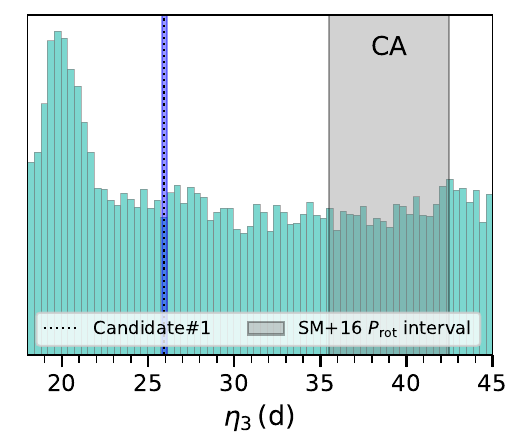}
    \includegraphics[width=0.31\textwidth]{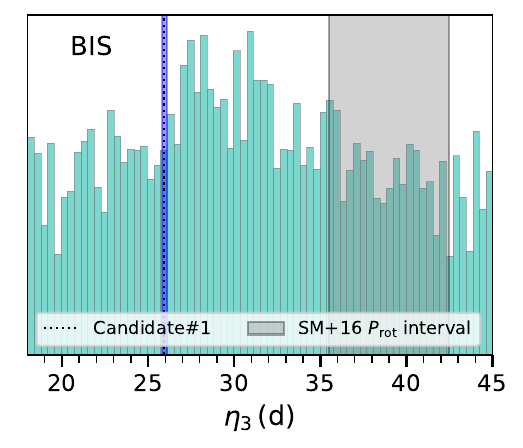}
    \includegraphics[width=0.31\textwidth]{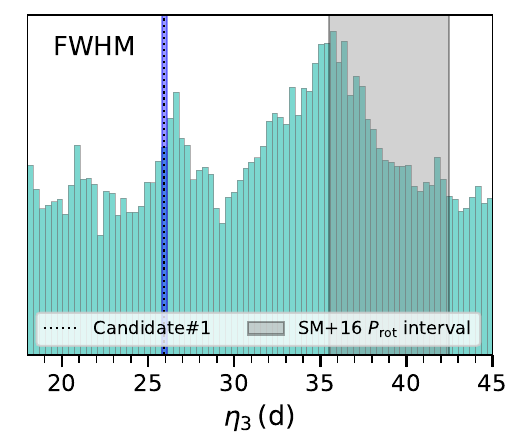}

    \caption{Posterior distribution of the $\eta_{3}$ hyperparameter of the quasiperiodic GP kernel used to describe the ESPRESSO activity indicators (Sect.~\ref{sec:stellar_activity}). The vertical dash-dotted black line indicates the orbital period of Candidate$\#$1 and the vertical blue shade indicates its 3$\sigma$ confidence interval. The vertical grey shade represents the 1$\sigma$ confidence interval of the expected $P_{\rm rot}$, inferred through the $\textrm{log}(R'_{\rm HK }$)-$P_{\rm rot}$ relations from \citet{2016A&A...595A..12S}. In the $H_{\rm \alpha}$ panel, the vertical dashed and dotted brown lines indicate the median value and the 1$\sigma$ confidence intervals of the distribution, respectively ($\eta_{\rm 3, H_{\alpha}}$ = $37.03^{+1.25}_{- 0.92}$ d).}
    \label{fig:eta3_posteriors}
\end{figure*}

\begin{figure*}
    \centering
    \includegraphics[width=0.65\textwidth]{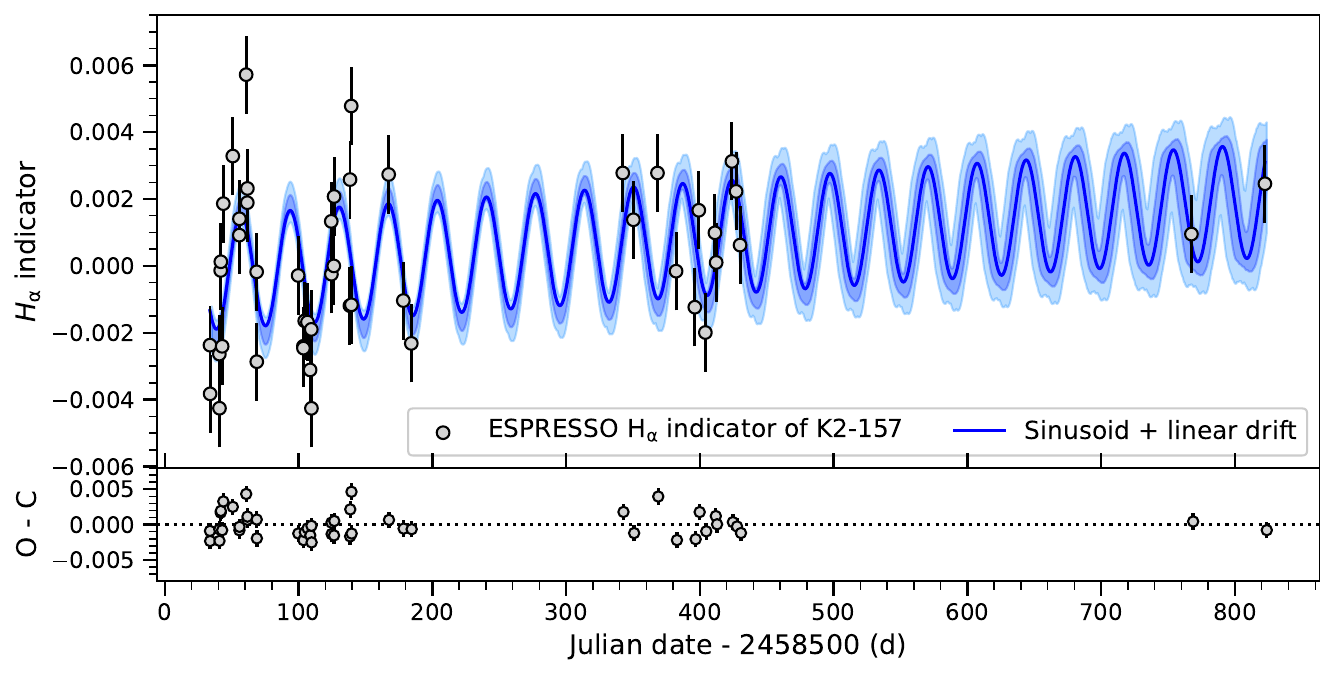}
    \includegraphics[width=0.3\textwidth]{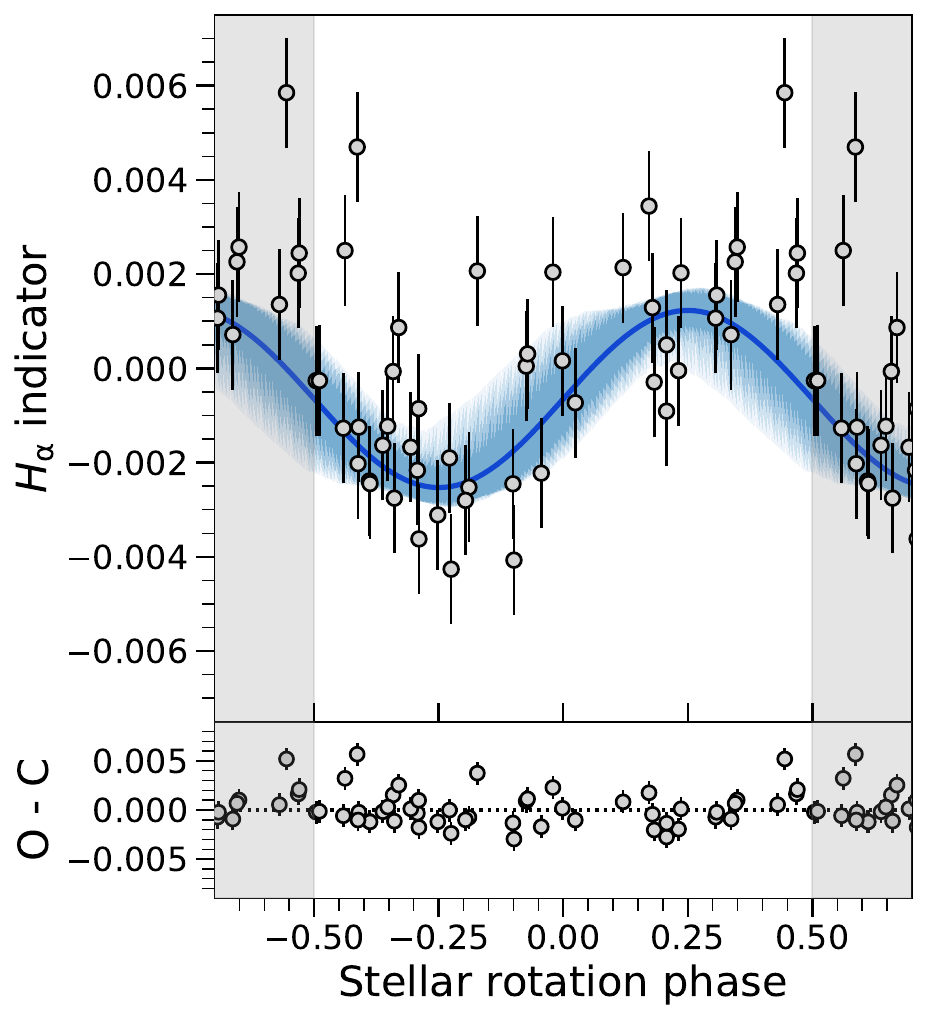}
    \caption{ESPRESSO $H_{\rm \alpha}$ indicator of K2-157. The solid line represents the median posterior sinusoidal model, and the dark and light shades represent the 1$\sigma$ and 3$\sigma$ confidence intervals, respectively. The left panel contains the full data set, and the right panel contains the $H_{\rm \alpha}$ data folded to the inferred periodicity ($P_{\rm rot, H_{\alpha}}$ = $36.76^{+0.40}_{-0.44}$ d). }
    \label{fig:ha_timeseries}
\end{figure*}

\begin{figure*}
    \centering
    \includegraphics[width=0.48\textwidth]{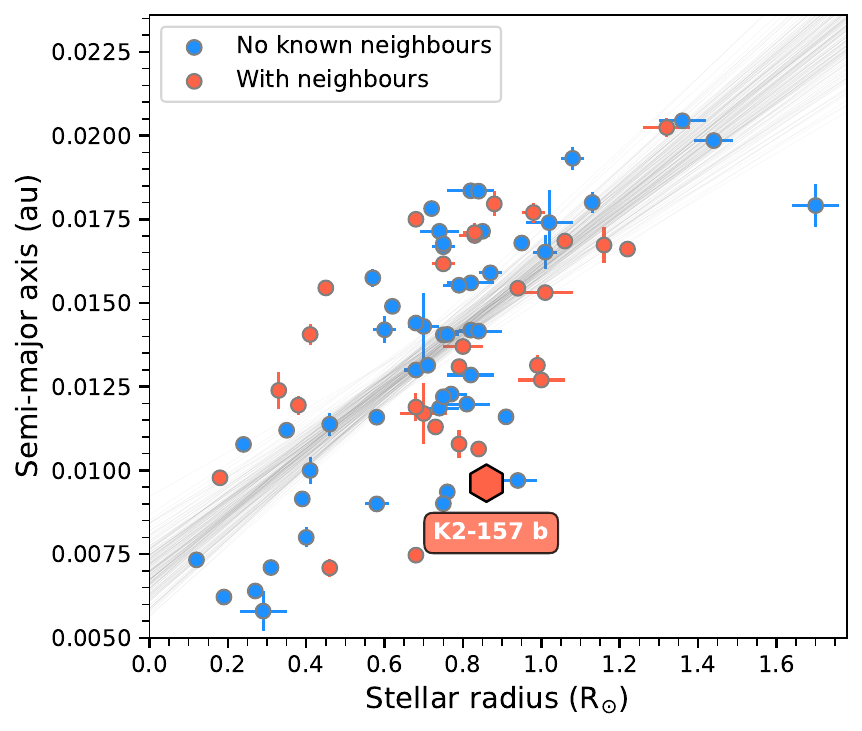}
    \includegraphics[width=0.48\textwidth]{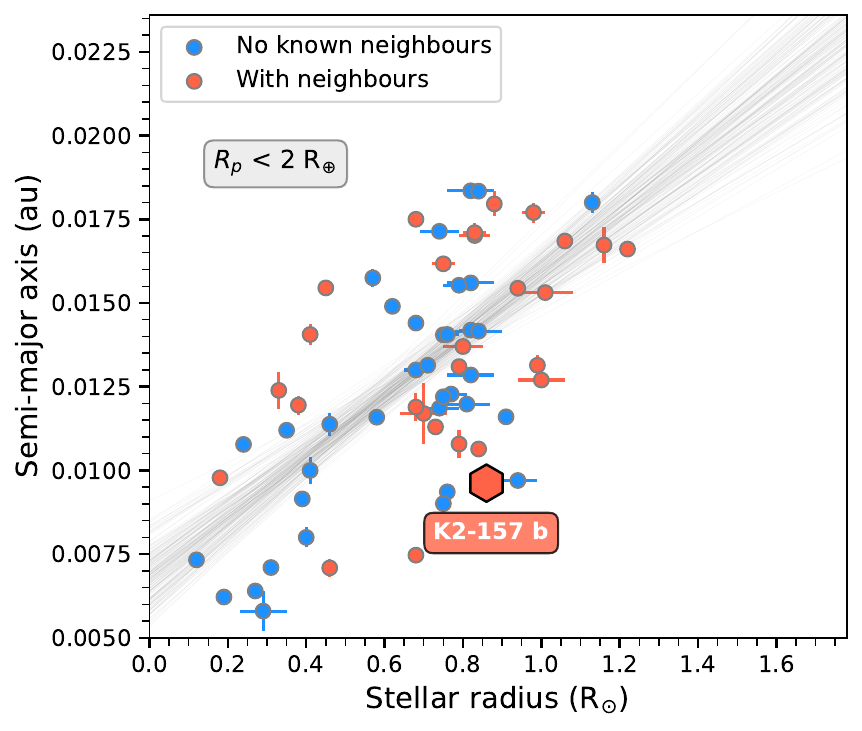}
    \caption{Semi-major axis versus stellar radius of all USP planets (left) and of a restricted sample with $R_{\rm p}$ $<$ 2 $\rm R_{\oplus}$ (right). Planets without reported neighbours are coloured in blue, while those with known neighbours a coloured in orange. The straight grey lines are 200 random posterior linear models (i.e. $A$x + $B$). Data: NASA Exoplanet Archive (24/03/2025). }
    \label{fig:a_vs_Rs}
\end{figure*}

\begin{figure*}
    \centering
    \includegraphics[width=\textwidth]{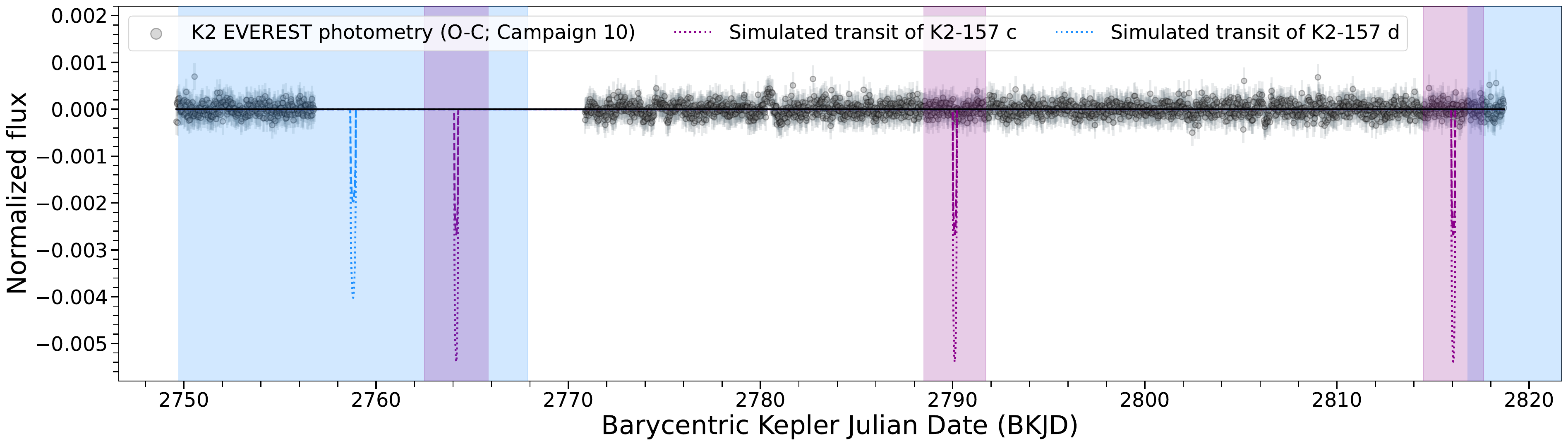}
    \includegraphics[width=\textwidth]{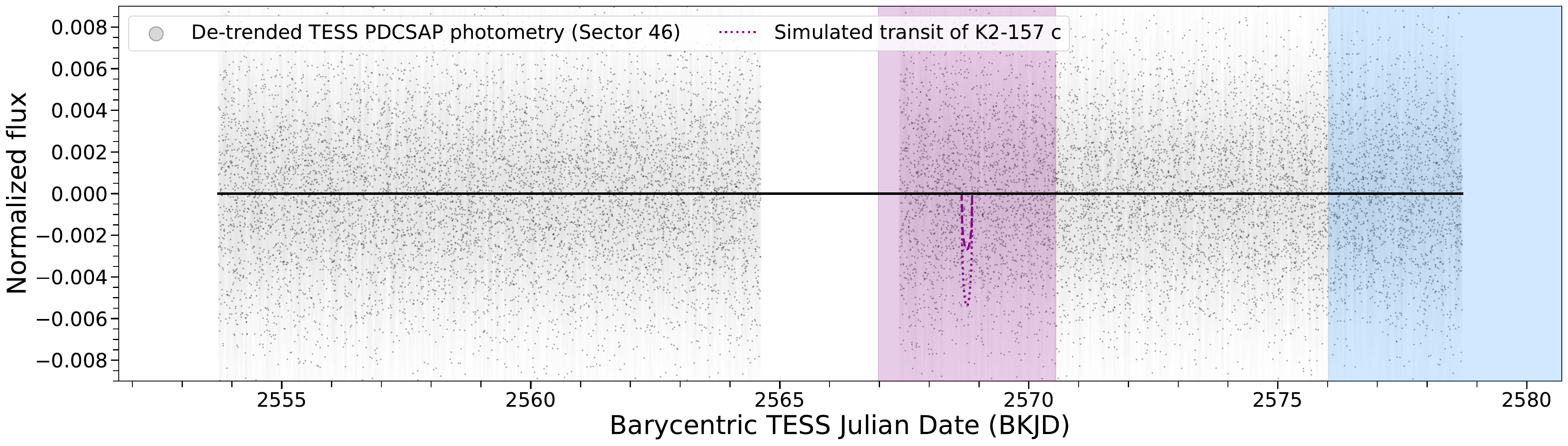}
    \caption{Predicted transit times of  K2-157~c and K2-157~d along the K2 (C10, upper panel) and TESS (S46, lower panel) observing baselines, obtained from our RV-derived ephemeris (Table~\ref{tab:parameters_joint}). Shaded regions represent the 1$\sigma$ confidence intervals. The dotted (dashed) simulated transits assume the median (1$\sigma$ lower limit) $R_{\rm p} / R_{\star}$ as estimated by \texttt{forecaster} \citep{2017ApJ...834...17C}.}
    \label{fig:tr_simul}
\end{figure*}

\section{Additional tables}


\onecolumn

\begin{table}
\renewcommand{\arraystretch}{1.23}
\setlength{\tabcolsep}{9.2pt}
\parbox{.47\linewidth}{
\caption{K2 \texttt{everest} photometry of K2-157. The complete table is available at the CDS.}
\begin{tabular}{cccc}
\hline \hline
BJD (days)  & FLUX & FLUX\_ERR  & Campaign \\ \hline
2457582.6106 & 0.9997     & 0.0003   & C10 \\
2457582.6310      & 1.0001   & 0.0003 & C10    \\
...         & ...                     & ... & ...    \\ \hline
\end{tabular}
\label{tab:K2_data}
}
\hfill
\parbox{0.47\linewidth}{
\caption{TESS PDCSAP photometry of K2-157. The complete table is available at the CDS.}
\begin{tabular}{cccc}
\hline \hline
BJD (days)  & FLUX & FLUX\_ERR  & Sector \\ \hline
2459553.7278 & 0.9974 & 0.0044 & S46 \\
2459553.7291 & 0.9995 & 0.0044 & S46 \\
...         & ...                     & ... & ...    \\ \hline
\end{tabular}
\label{tab:TESS_data}
}
\end{table}


\begin{table}[]
\tiny
\caption{ESPRESSO RVs and activity indicators of K2-157 acquired between 19 February 2019 and 19 April 2021 under the programmes with IDs 1102.C-0744, 1102.C-0958, 1104.C-0350, and 106.21M2.004. The complete table is available at the CDS.}
\renewcommand{\arraystretch}{1.4}
\setlength{\tabcolsep}{3.6pt}
\begin{tabular}{ccccccccc}
\hline \hline
BJD (days)   & $\rm RV_{DRS}$ ($\rm m \, s^{-1}$) & $\rm RV_{\texttt{sbart}}$ ($\rm m \, s^{-1}$) & FWHM & BIS             & Contrast      & $\rm H_{\alpha}$      & Na-index              & Ca-index            \\ \hline
2458533.7351 & 43436.9 $\pm$ 1.8                  & 43439.7 $\pm$ 1.1                             & 6413.4 $\pm$ 3.7         & -60.5 $\pm$ 3.7 & 62.820 $\pm$ 0.036 & 0.20662 $\pm$ 0.00036 & 0.22297 $\pm$ 0.00026 & 0.1372 $\pm$ 0.0015 \\
2458533.8509 & 43441.4 $\pm$ 1.5                  & 43438.8 $\pm$ 1.0                             & 6407.2 $\pm$ 3.1         & -68.0 $\pm$ 3.1 & 62.797 $\pm$ 0.030 & 0.20517 $\pm$ 0.00030 & 0.22328 $\pm$ 0.00021 & 0.1269$\pm$ 0.0011  \\
...          & ...                                & ...                                           & ...                      &       ...          & ...                & ...                   & ...                   &        ...             \\ \hline
\end{tabular}
\label{tab:ESPRESSO_data}
\end{table}



\begin{table*}[]
\centering
\renewcommand{\arraystretch}{1.38}
\setlength{\tabcolsep}{20pt}
\caption{Parameters of K2-157~c and K2-157~d based on the ESPRESSO blind-search analysis described in Sect.~\ref{sec:blind_search}.}
\label{tab:parameters_ESPRESSO}
\begin{tabular}{lll}
\hline \hline
Parameter                                                                 & Priors                          & Posteriors                    \\ \hline 
\multicolumn{3}{l}{Orbital and physical parameters of K2-157 c}                                                                             \\ \hline 
Orbital period, $P_{\rm orb, c}$ (days)                                   & $\mathcal{U}(0, 400)$           & $25.968^{+0.051}_{-0.052}$    \\
Time of mid-transit, $T_{\rm 0, c}$ (JD)                                  & $\mathcal{U}(2458533, 2458933)$ & $2458557.03 \pm 0.36$         \\
RV semi-amplitude, $K_{\rm c} \, (\rm m\,s^{-1})$                         & $\mathcal{U}(0, 20)$            & $7.10 \pm 0.49$               \\
Minimum mass, $M_{\rm p, c} \, \textrm{sin} \, i \, (\rm M_{\rm \oplus})$ & (derived)                       & $30.4 \pm 2.2$                \\
Minimum mass, $M_{\rm p, c} \, \textrm{sin} \, i \, (\rm M_{\rm J})$      & (derived)                       & $0.0958 \pm 0.0069$             \\
Relative orbital separation, $a_{\rm c} / R_{\star}$                      & (derived)                       & $41.3 \pm 1.0$                \\
Orbit semi-major axis, $a_{\rm c}$ (au)                                   & (derived)                       & $0.1651 \pm 0.0018$           \\
Incident flux, $F_{\rm inc, c} \, (\rm F_{\oplus})$                       & (derived)                       & $22.62 \pm 0.68$              \\
Equilibrium temperature {[}A=0{]}, $T_{\rm eq, c} \, (\rm K)$             & (derived)                       & $587 \pm 10$                  \\ \hline
\multicolumn{3}{l}{Orbital and physical parameters of K2-157 d}                                                                             \\ \hline
Orbital period, $P_{\rm orb, d}$ (days)                                   & $\mathcal{U}(0, 400)$           & $66.62^{+0.99}_{-0.75}$       \\
Time of mid-transit, $T_{\rm 0, d}$ (JD)                                  & $\mathcal{U}(2458533, 2458933)$ & $2458588.9^{+2.2}_{-2.8}$     \\
RV semi-amplitude, $K_{\rm d} \, (\rm m\,s^{-1})$                         & $\mathcal{U}(0, 20)$            & $3.79^{+0.49}_{-0.48}$        \\
Minimum mass, $M_{\rm p, d} \, \textrm{sin} \, i \, (\rm M_{\rm \oplus})$ & (derived)                       & $22.3 \pm 2.8$                \\
Minimum mass, $M_{\rm p, d} \, \textrm{sin} \, i \, (\rm M_{\rm J})$      & (derived)                       & $0.0701 \pm 0.0089$             \\
Relative orbital separation, $a_{\rm d} / R_{\star}$                      & (derived)                       & $77.4 \pm 2.0$                \\
Orbit semi-major axis, $a_{\rm d}$ (au)                                   & (derived)                       & $0.3093\pm 0.0041$            \\
Incident flux, $F_{\rm inc, d} \, (\rm F_{\oplus})$                       & (derived)                       & $6.44 \pm 0.22$               \\
Equilibrium temperature {[}A=0{]}, $T_{\rm eq, d} \, (\rm K)$             & (derived)                       & $429 \pm 8$                   \\ \hline
RV jitter and quadratic trend                                             &                                 &                               \\ \hline
RV jitter, $\sigma_{\rm jit}$ $(\rm m\,s^{-1})$               & $\mathcal{U}(0, 10)$            & $1.43^{+0.35}_{-0.39}$        \\
Systemic velocity, $v_{\rm sys}$ $(\rm m\,s^{-1})$                        & $\mathcal{U}(43430, 43460)$     & $43448.00^{+0.79}_{-0.83}$    \\
Linear term, $\gamma$ $(\rm m\,s^{-1}\,day^{-1})$                         & $\mathcal{U}(-0.1, 0.1)$        & $-0.0498^{+0.0062}_{-0.0061}$ \\
Quadratic term, $\delta$ $(\rm m\,s^{-1}\,day^{-2})$                      & $\mathcal{U}(-0.01, 0.01)$      & $6.5 \pm 1.1 \times 10^{-5}$  \\ \hline
\end{tabular}
\end{table*}



\begin{table*}[]
\centering
\renewcommand{\arraystretch}{1.14}
\setlength{\tabcolsep}{20pt}
\caption{Parameters of K2-157~b,  K2-157~c, and K2-157~d based on the joint fit analysis described in Sect.~\ref{sec:joint_fit}.}
\label{tab:parameters_joint}
\begin{tabular}{lll}
\hline \hline
Parameter                                                                 & Priors                              & Posteriors                            \\ \hline
Orbital and physical parameters of K2-157 b                               &                                     &                                       \\ \hline
Orbital period, $P_{\rm orb, b}$ (days)                                   & $\mathcal{U}(0.3, 0.4)$             & $0.3652575^{+0.0000078}_{-0.0000088}$ \\
Time of mid-transit, $T_{\rm 0, b}$ (JD)                                  & $\mathcal{U}(2457582.6, 2457583.0)$ & $2457582.8220^{+0.0041}_{-0.0032}$    \\
Orbital inclination, $i_{\rm b}$ (degrees)                                & $\mathcal{U}(50, 150)$              & $89.2^{+9.3}_{-9.1}$                  \\
Scaled planet radius, $R_{\rm p, b}/ R_{\star}$                           & $\mathcal{U}(0.00, 0.05)$           & $0.00996 \pm 0.00094$                 \\
RV semi-amplitude, $K_{\rm b} \, (\rm m\,s^{-1})$                         & $\mathcal{U}(0, 20)$                & $1.10^{+0.39}_{-0.41}$                \\
Planet radius, $R_{\rm p, b} \, (\rm R_{\rm \oplus})$                     & (derived)                           & $0.935 \pm 0.090$                     \\
Planet radius, $R_{\rm p, b} \, (\rm R_{\rm J})$                          & (derived)                           & $0.0834 \pm 0.0081$                   \\
Planet mass, $M_{\rm p, b} (\rm M_{\rm \oplus})$                          & (derived)                           & $1.14^{+0.41}_{-0.42}$                       \\
Planet mass, $M_{\rm p, b} (\rm M_{\rm J})$                               & (derived)                           & $0.0036 \pm 0.0013$                   \\
Planet density, $\rho_{\rm p, b}$ ($\rm g \, cm^{-3}$)                    & (derived)                           & $7.7 \pm 3.6$                         \\
Transit depth, $\Delta_{\rm b}$ (ppm)                                     & (derived)                           & $99 \pm 19$                           \\
Transit duration, $T_{\rm 14, b}$ (hours)                                 & (derived)                           & $1.161 \pm 0.032$                     \\
Relative orbital separation, $a_{\rm b} / R_{\star}$                      & (derived)                           & $2.405 \pm 0.059$                     \\
Orbit semi-major axis, $a_{\rm b}$ (au)                                   & (derived)                           & $0.00962 \pm 0.00010$                 \\
Planet surface gravity, $g_{\rm b}$ $(\rm m\,s^{-2})$                     & (derived)                           & $12.8 \pm 5.3$                        \\
Impact parameter, $b_{\rm b}$                                             & (derived)                           & $0.03 \pm 0.38$                       \\
Incident flux, $F_{\rm inc, b} \, (\rm F_{\oplus})$                       & (derived)                           & $(6.66 \pm 0.20) \times 10^{3}$       \\
Equilibrium temperature [A=0], $T_{\rm eq, b} \, (\rm K)$                 & (derived)                           & $2432 \pm 42$                         \\ \hline
Orbital and physical parameters of K2-157 c                               &                                     &                                       \\ \hline
Orbital period, $P_{\rm orb, c}$ (days)                                   & $\mathcal{U}(0, 400)$               & $25.942^{+0.045}_{-0.044}$            \\
Time of mid-transit, $T_{\rm 0, c}$ (JD)                                  & $\mathcal{U}(2458533, 2458933)$     & $2458557.02 \pm 0.30$                 \\
RV semi-amplitude, $K_{\rm c} \, (\rm m\,s^{-1})$                         & $\mathcal{U}(0, 20)$                & $7.17 \pm 0.42$                       \\
Minimum mass, $M_{\rm p, c} \, \textrm{sin} \, i \, (\rm M_{\rm \oplus})$ & (derived)                           & $30.8 \pm 1.9$                        \\
Minimum mass, $M_{\rm p, c} \, \textrm{sin} \, i \, (\rm M_{\rm J})$      & (derived)                           & $0.0967 \pm 0.0060$                   \\
Relative orbital separation, $a_{\rm c} / R_{\star}$                      & (derived)                           & $41.2 \pm 1.0$                        \\
Orbit semi-major axis, $a_{\rm c}$ (au)                                   & (derived)                           & $0.1650 \pm 0.0018$                   \\
Incident flux, $F_{\rm inc, c} \, (\rm F_{\oplus})$                       & (derived)                           & $22.65 \pm 0.68$                      \\
Equilibrium temperature {[}A=0{]}, $T_{\rm eq, c} \, (\rm K)$             & (derived)                           & $587 \pm 10$                          \\ \hline
\multicolumn{3}{l}{Orbital and physical parameters of K2-157 d}                                                                                         \\ \hline
Orbital period, $P_{\rm orb, d}$ (days)                                   & $\mathcal{U}(0, 400)$               & $66.50^{+0.71}_{-0.59}$               \\
Time of mid-transit, $T_{\rm 0, d}$ (JD)                                  & $\mathcal{U}(2458533, 2458933)$     & $2458589.3^{+1.7}_{-2.0}$             \\
RV semi-amplitude, $K_{\rm d} \, (\rm m\,s^{-1})$                         & $\mathcal{U}(0, 20)$                & $3.96^{+0.42}_{-0.41}$                \\
Minimum mass, $M_{\rm p, d} \, \textrm{sin} \, i \, (\rm M_{\rm \oplus})$ & (derived)                           & $23.3 \pm 2.5$                        \\
Minimum mass, $M_{\rm p, d} \, \textrm{sin} \, i \, (\rm M_{\rm J})$      & (derived)                           & $0.0732 \pm 0.0077$                   \\
Relative orbital separation, $a_{\rm d} / R_{\star}$                      & (derived)                           & $77.3 \pm 2.0$                        \\
Orbit semi-major axis, $a_{\rm d}$ (au)                                   & (derived)                           & $0.3090 \pm 0.0038$                   \\
Incident flux, $F_{\rm inc, d} \, (\rm F_{\oplus})$                       & (derived)                           & $6.46 \pm 0.21$                       \\
Equilibrium temperature {[}A=0{]}, $T_{\rm eq, d} \, (\rm K)$             & (derived)                           & $429.1 \pm 7.5$                       \\ \hline
RV quadratic trend                                                        &                                     &                                       \\ \hline
Systemic velocity, $v_{\rm sys}$ $(\rm m\,s^{-1})$                        & $\mathcal{U}(43430, 43460)$         & $43448.11^{+0.64}_{-0.66}$            \\
Linear term, $\gamma$ $(\rm m\,s^{-1}\,day^{-1})$                         & $\mathcal{U}(-0.1, 0.1)$            & $-0.0491^{+0.0052}_{-0.0051}$         \\
Quadratic term, $\delta$ $(\rm m\,s^{-1}\,day^{-2})$                      & $\mathcal{U}(-0.01, 0.01)$          & $6.20^{+0.92}_{-0.94} \times 10^{-5}$ \\ \hline
Instrument-dependent parameters                                           &                                     &                                       \\ \hline
K2 LC level C10, $F_{\rm 0, C10}$                                         & $\mathcal{U}(-0.1, 0.1)$            & $-0.000042^{+0.000216}_{-0.000233}$   \\
K2 LC jitter C10, $\sigma_{\rm K2, C10}$                                  & $\mathcal{U}(0, 0.005)$             & $4.6^{+4.8}_{-3.2} \times 10^{-6}$    \\
RV jitter, $\sigma_{\rm jit}$                          & $\mathcal{U}(0, 10)$                & $1.06^{+0.38}_{-0.43}$                \\ \hline
Limb-darkening coefficients                                               &                                     &                                       \\ \hline
Limb-darkening coefficient, $q_{\rm 1, Kepler}$                           & $\mathcal{U}(0, 1)$                 & $0.62^{+0.26}_{-0.34}$                \\
Limb-darkening coefficient, $q_{\rm 2, Kepler}$                           & $\mathcal{U}(0, 1)$                 & $0.45^{+0.33}_{-0.30}$                \\ \hline
GP hyperparameters                                                        &                                     &                                       \\ \hline
$\eta_{\rm \sigma}$                                                       & $\mathcal{U}(0, 0.5)$               & $0.00048^{+0.00031}_{-0.00015}$       \\
$\eta_{\rm \rho}$ (days)                                                  & $\mathcal{U}(0, 30)$                & $6.3^{+3.1}_{-1.8}$  \\

\hline 

\end{tabular}
\end{table*}


\section{Corner plots}

\begin{figure*}
    \centering
    \includegraphics[width=\textwidth]{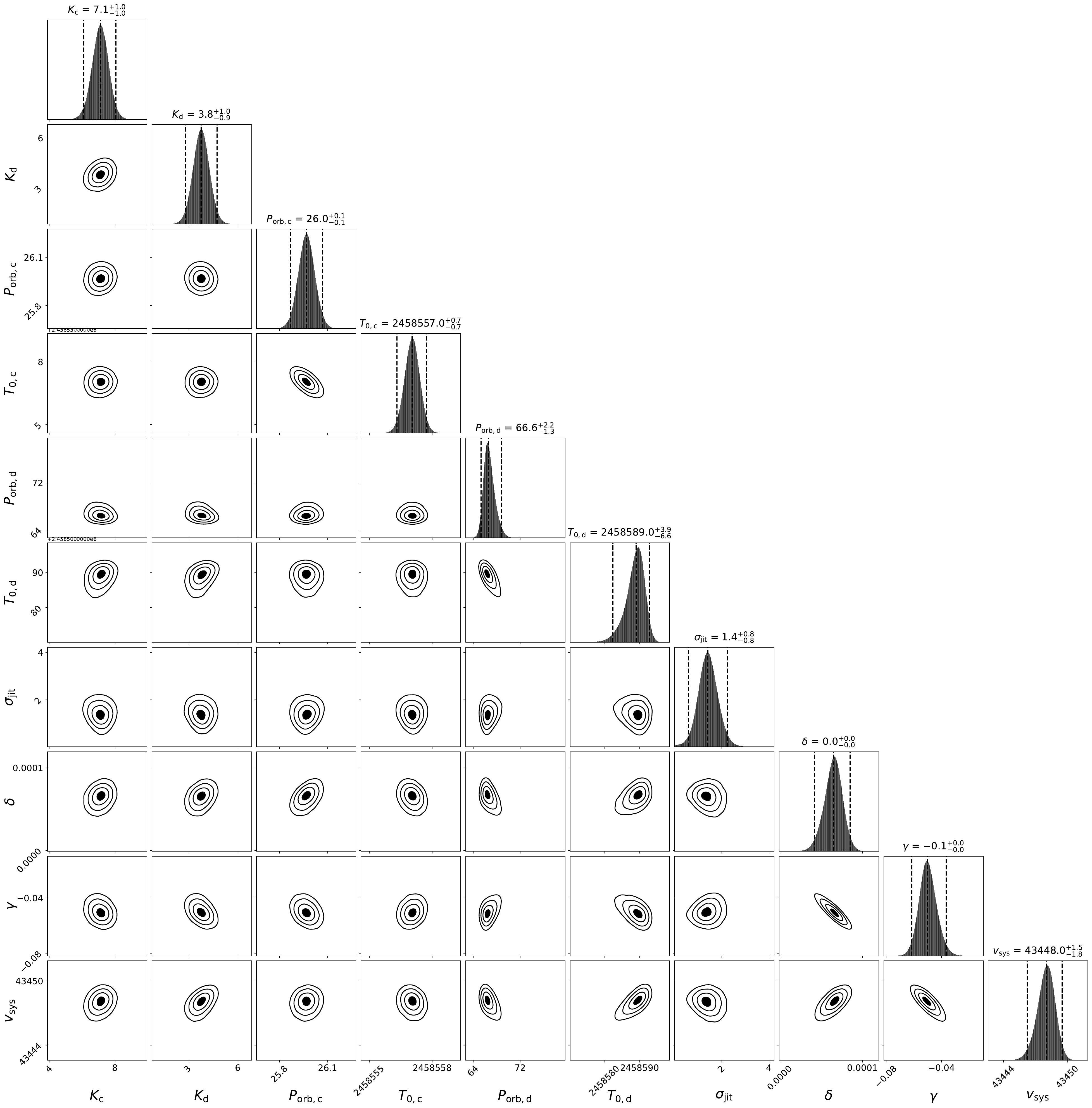}
    \caption{Corner plot with the 1D and 2D posterior distributions of the parameters of the model that best describes the RVs: 2p1c2cQ (i.e. 2 planets with circular orbits plus a quadratic trend; see Sect.~\ref{sec:blind_search}). The vertical lines indicate the median and 2$\sigma$ intervals.}
    \label{fig:corner_ESPRESSO}
\end{figure*}

\begin{figure*}
    \centering
    \includegraphics[width=\textwidth]{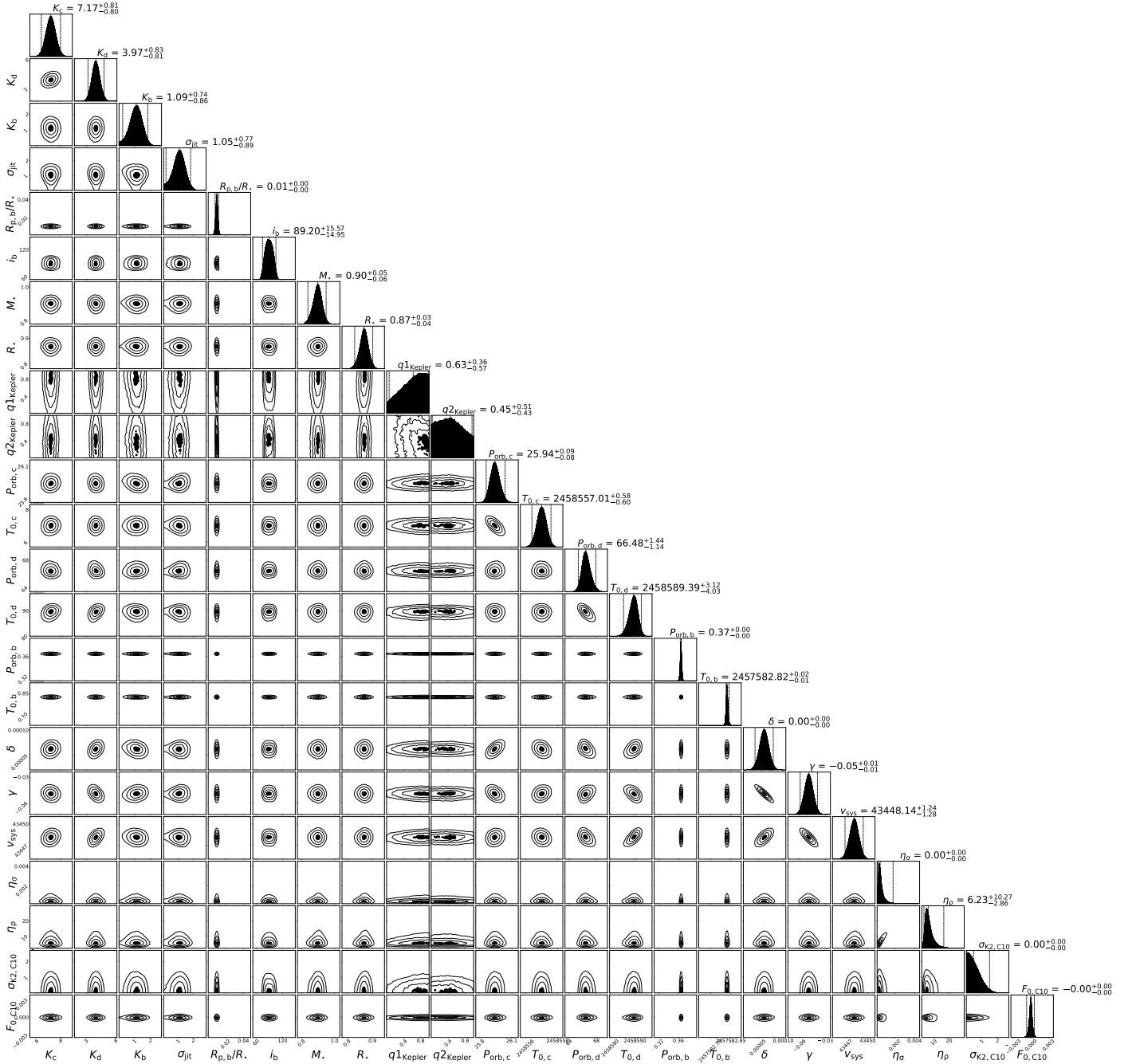}
    \caption{Corner plot with the 1D and 2D posterior distributions of the parameters of the global model (transit + RV) used in the joint fit analysis (see Sect.~\ref{sec:joint_fit}). The vertical lines indicate the median and 2$\sigma$ intervals.}
    \label{fig:corner_joint}
\end{figure*}






\end{appendix}

\end{document}